\documentclass[a4paper,11pt]{article}
\pdfoutput=1

\usepackage{jcappub}
\usepackage[normalem]{ulem}
\usepackage[utf8]{inputenc}
\usepackage{amsmath, amssymb, amsthm, graphicx, epsfig, fancyhdr,epsfig, slashed}
\let\oldcdot\cdot
\usepackage{breqn}
\let\cdot\oldcdot
\usepackage{enumerate}
\usepackage{epsfig}  
\usepackage{subcaption} % for subplot
\usepackage{graphicx}   
\usepackage{slashed}       
\usepackage{tikz}
\usepackage{url}
\usepackage{color,soul} % soul pakage for highlighting (to hightlight use \hl{}) 
\usepackage{multirow}
\usepackage{comment}
\usepackage{mathtools}
\usepackage{bm}
\usepackage{mathrsfs}
\usepackage{calligra}
\usepackage{hyperref}
\usepackage[all]{hypcap}
\hypersetup{  % set up for colour hyperlink
    colorlinks=true,
    linkcolor=blue,
    filecolor=green,      
    urlcolor=cyan,
    citecolor=red,
    }

\newcommand{\Trh}{T_{\rm RH}}

\newcommand{\arh}{a_{\rm RH}}

\newcommand{\aend}{a_{\rm end}}
\newcommand{\krh}{k_{\rm RH}}
\newcommand{\kend}{k_{\rm end}}
\newcommand{\nt}{n_{\rm T}}
\newcommand{\ax}{\textsl{a}}

\begin{document}

\title{\sf{Primordial Gravitational Waves as Probe of Dark Matter in Interferometer Missions:} \\ \it{Fisher Forecast and MCMC}}

\author[a]{Anish Ghoshal,}
\emailAdd{anish.ghoshal@fuw.edu.pl}

\author[b,1]{ Debarun Paul,\note{Corresponding author.}}
\emailAdd{debarun31paul@gmail.com}

\author[b,c]{ Supratik Pal}
\emailAdd{supratik@isical.ac.in}

\affiliation[a]{\,Institute of Theoretical Physics, Faculty of Physics, University of Warsaw, ul. Pasteura 5, 02-093 Warsaw, Poland }
\affiliation[b]{\,Physics and Applied Mathematics Unit, Indian Statistical Institute, 203 B.T. Road, Kolkata 700108, India}
\affiliation[c]{\,Technology Innovation Hub on Data Science, Big Data Analytics and Data Curation, Indian Statistical Institute, 203 B.T. Road, Kolkata 700108, India}

\smallskip

%%%%%%%%%%%%%%%%%%%%%%%%%%%
\abstract{We propose novel inflationary primordial gravitational wave (GW) spectral shapes at interferometer-based current and future GW missions to test dark matter (DM) production via gravity-portal. We consider three right-handed neutrinos (RHNs), required to generate Standard Model (SM) neutrino masses via seesaw mechanism, are produced via gravity-portal in early universe. The lightest among them is stable and is the DM candidate of the Universe. The other two RHNs decay and generate matter-antimatter asymmetry due to baryogenesis via leptogenesis. 
We find that future GW detectors BBO, DECIGO, ET, for instance, are able to probe DM mass for $5\times 10^6\, {\rm GeV}<M_{\rm DM}<1.6\times 10^7$ GeV with a signal-to-noise ratio (SNR) $>10$, along with the observed amount of baryon asymmetry due to gravitational leptogenesis for heavy RHN mass $M_{\cal{N}}$ to be around $8\times 10^{12}$ GeV. Employing Fisher matrix forecast analysis, we identify the parameter space involving non-minimal coupling to gravity $\xi$, reheating temperature of the Universe $T_{\rm rh}$ and DM mass $M_{\rm DM}$ where the GW detector-sensitivities will be the maximum with the least error, along with SNR $>10$. Finally, utilizing mock data for each GW detector, we perform MCMC analysis to find out the combined constraints on the various microphysics parameters. 
We also explore production of other cosmological relics such as QCD axion relic as DM candidate, produced via gravity-portal in early universe. We find that ET, for instance, can probe the decay constant of such DM candidates ($f_{\ax}$) as $10^9\,{\rm GeV}\lesssim f_{\ax}\lesssim 10^{14}\,{\rm GeV}$ for misalignment angle $\theta_i\in[0.1,\pi/\sqrt{3}]$ and $\xi=1$ with SNR $>10$, whereas this range decreases with the increase of non-minimal coupling. Thus the upcoming GW missions will be able to test such non-thermal DM and baryogenesis scenarios involving very high energy scales,  which is otherwise impossible to reach in particle physics experiments in laboratories. 
}
%%%%%%%%%%%%%%%%%%%%%%%%%%%

\maketitle
\newpage
%%%%%%%%%%%%%%%%%%%%%%%%%%%
\section{Introduction}
\label{sec:intro}
%%%%%%%%%%%%%%%%%%%%%%%%%%%

Astrophysical signatures of dark matter (DM) in the Universe first pointed out by Fritz Zwicky in 1933 by measuring its gravitational interactions~\cite{Zwicky:1933gu}, confirmed later on and quantified by cosmological evidences from the measurements of cosmic microwave background radiation (CMBR) anisotropic spectrum ~\cite{Planck:2018vyg} and the large scale structure (LSS) of the Universe~\cite{BOSS:2016wmc}. The standard model (SM) of particle physics does not contain any candidate which could be the possible dark matter. The electrically neutral SM neutrinos or heavy neutral leptons (HNL), are too light to be DM \cite{Gunn:1978gr,Hut:1977zn,Lee:1977ua} while explorations  beyond SM Physics may involve heavy neutrinos with masses around GeV scale or larger, known as sterile neutrinos which could be DM candidates. These DM models often referred to as WIMPs (weakly interacting massive particles) which occur in several extensions of the SM including supersymmetry \cite{Ellis:1983ew}. Classification of WIMP DM candidates includes a Higgs-portal \cite{Silveira:1985rk,McDonald:1993ex,Burgess:2000yq,Davoudiasl:2004be,Cline:2013gha,Han:2015hda,Djouadi:2011aa,Lebedev:2011iq,Mambrini:2011ik,Djouadi:2012zc,Casas:2017jjg}
or a $Z$-portal \cite{Arcadi:2014lta,Escudero:2016gzx}. However, most of these scenarios in their minimal avatar are currently constrained in very stringent manner (see, for example, Refs.~\cite{Roszkowski:2017nbc,Arcadi:2017kky} for reviews), and even extensions to the minimal scenarios like $Z'$-portal~\cite{Mambrini:2010dq,Lebedev:2014bba,Alves:2015pea,Alves:2016cqf,Arcadi:2017jqd} are in significant tension with the electroweak (EW) nature of dark matter due to null results in the DM direct detection experiments.

Unlike WIMPs there could be FIMPs (feebly interacting massive particles) with interactions much weaker than electroweak interactions and are never fully in equilibrium, but are nonetheless produced in the thermal bath after inflation, like the gravitino \cite{Ellis:1983ew,Khlopov:1984pf,Olive:1984bi} among several others ~\cite{McDonald:2001vt,Hall:2009bx,Mambrini:2013iaa} as an alternative to WIMPs (see \cite{Bernal:2017kxu} for a recent review). Due to the very weak interactions involved that they evade the current constraints from experiments unlike WIMPs, and such a non-thermal mechanism for DM production is known as the freeze-in mechanism~\cite{Hall:2009bx,Elahi:2014fsa}. Some candidates for FIMPs  are postulated in unified theories like $SO(10)$ with a heavy $Z'$ gauge boson \cite{Mambrini:2013iaa,Mambrini:2015vna,Bhattacharyya:2018evo}, moduli fields \cite{Chowdhury:2018tzw}, high scale SUSY \cite{Benakli:2017whb,Dudas:2017rpa,Dudas:2017kfz,Dudas:2018npp,Kaneta:2019yjn}
or heavy spin-2 constructions \cite{Bernal:2018qlk}.

Even feebler interactions are possible when the only effective coupling at the UV scale is gravity and here is no other interactions between the particles or between SM and the BSM (beyond standard model) particles. Even in this minimal irreducible interaction that should exist between DM and the SM is mediated by graviton ~\cite{Ema:2015dka,Garny:2015sjg,Garny:2017kha,Tang:2016vch,Tang:2017hvq,Ema:2016hlw,Bernal:2018qlk,Ema:2018ucl,Ema:2019yrd,Chianese:2020yjo,Chianese:2020khl,Redi:2020ffc,Mambrini:2021zpp,Barman:2021ugy,Haque:2021mab,Clery:2021bwz,Clery:2022wib,Ahmed:2022qeh,Ahmed:2022tfm}.
This leads to the observed DM relic through the scattering of
the particles in the thermal bath or via the
the inflaton condensate gravitationaly~\footnote{Gravitational preheating has been considered in Refs.~\cite{Zhang:2024ggn,Zhang:2023xcd,Zhang:2023hjk}}, as already been discussed in detail
Refs.~\cite{Mambrini:2021zpp,Barman:2021ugy,Haque:2021mab,Clery:2021bwz,Clery:2022wib,Haque:2023yra,Barman:2022qgt}. 
For a scalar field, $\phi$, one could naturally extend the interaction sector with a non-minimal coupling term of the form 
\begin{equation}
\label{eq:gen_nonmin_int}
   \mathcal{L} \propto \xi \phi^2 \mathcal{R} \;.
\end{equation}
Here $\mathcal{R}$ is the Ricci scalar and $\xi$ is non-minimal coupling. The particles interact with each other and leads not only to successful reheating of the Universe but also gives rise to the suitable amount of dark matter observed. It is needless to say tat due to the dim-4 nature of these terms the presence of these terms is inevitable, which means that even if there were no such couplings at tree level were present, they are eventually generated by quantum corrections at loop level. 
The non-minimal coupling will naturally influence the particle production in the processes $SS\to XX$, $\phi\phi\to SS$, and $\phi\phi\to XX$ via the s-channel graviton exchange that sets minimal possible production rates which are induced by the non-minimal couplings $\xi_{\phi}$. Here $\phi$ represents the inflaton background oscillating around its minima in the post-inflationary epoch~\cite{Ford:1986sy}, $X$ is the DM candidate and $S$ are SM particles. Needless to say, as mentioned also before that there is no direct interaction between $\phi$, $S$, and $X$.

Other than the DM puzzle of the Universe, the visible or baryonic matter content of the Universe is asymmetric and requires a dynamical mechanism to generate such matter-antimatter asymmetry if one wants avoid fine-tuning of initial conditions of the Universe. One of the simplest way to do this is to produce the baryon asymmetry of the Universe (BAU) is to produce lepton asymmetry and then transfer to baryon asymmetry via $(B+L)$-violating electroweak sphaleron transitions~\cite{Kuzmin:1985mm}, a scenario which is called leptogenesis~\cite{Fukugita:1986hr}. Thermal leptogenesis \cite{Buchmuller:2002rq,Buchmuller:2003gz,Chankowski:2003rr,Giudice:2003jh}, requires right-handed neutrinos (RHNs) masses which are heavy to be produced from the SM plasma, such that it a lower bound on the reheating temperature $\Trh\gtrsim 10^{10}$ GeV (called the Davidson-Ibarra bound)~\cite{Davidson:2002qv}. However this is easily avoided in the case 
non-thermal production of RHNs~\cite{Giudice:1999fb,Asaka:1999yd,Lazarides:1990huy,Campbell:1992hd,Hahn-Woernle:2008tsk} from the inflaton decay. 

Therefore, non-thermal leptogenesis are somewhat model dependent, for instance, needing a coupling between the inflaton and the right-handed neutrino or in some cases the supersymmetric partner of the right-handed neutrino may drive cosmic inflation \cite{Ellis:2005uk,Ghoshal:2022fud}, or in supergravity models there may arise gravitational coupling due to specific forms of K\"ahler potential and superpotential. But in gravity-mediated leptogenesis the the inflaton potential determines the particle production with the absence of any other no assumptions regarding the inflaton and right-handed neutrino sector. Its production from the inflaton condensate can be wholly gravitational in nature \cite{Barman:2021ugy,Co:2022bgh}. The abundance of RHNs is calculated in the same manner as the DM abundance during the gravitational reheating period. For such we need gravitational interactions are described by the Lagrangian~(see \textit{e.g.},)
\begin{equation}
\sqrt{-g}{\cal L}_{\rm int}= -\frac{1}{M_P}h_{\mu \nu}
\left(T^{\mu \nu}_{\rm SM}+T^{\mu \nu}_\phi + T^{\mu \nu}_{\mathcal{N}} \right) \, .
\label{Eq:lagrangian1}
\end{equation}
Here SM represents Standard Model fields, $\phi$ is the inflaton and $\mathcal{N}$ is the right-handed neutrino. The standard form of the stress-energy tensor $T^{\mu \nu}_i$ which depends on the spin of the field, $i = 0$, $1/2$, $1$ is considered.  In Fig. \ref{fig:feynman}, we have shown the $s$-channel exchange of a graviton obtained from the Lagrangian~(\ref{Eq:lagrangian1}) for the production of right-handed neutrinos from  the inflaton condensate.

Just like DM or the RHN, QCD axions can also be produced purely due to gravitational interactions and behave as the DM candidate of the universe, however this parameter space is different than the standard ``vacuum misalignment production" of QCD axion~\footnote{see the Refs.~\cite{Preskill:1982cy,Abbott:1982af,Dine:1982ah} for ``vacuum misalignment production" of QCD axion} as shown in Refs. \cite{Barman:2023icn}.

Naturally such scenarios involving only gravitational interactions is very challenging to test in laboratory experiments or astrophysical observations.
A possible way to probe such production of non-thermal DM and baryogenesis, is via tensor fluctuations of the metric during inflation that naturally sources primordial gravitational waves (GWs). The production of these primordial GWs could get amplified in the pre-BBN (Big Bang nucleosynthesis) era depending upon the inflaton oscillations, thereby acting as a secondary source for Primordial GW on top of the standard vacuum fluctuations.  
The minimal scenario, that is without any non-minimal coupling to gravity, if one tries to reheat the visible universe solely via gravitational reheating then it is known that this scenario is excluded due to overproduction of dark radiation in the form of GWs during BBN \cite{Figueroa:2018twl}. Therefore in this analysis we will concentrate on non-minimal couplings of fields with gravity (or the Ricci scalar ($\mathcal{R}$)). Since these are gravity-mediated production scenarios, there is possibly no other ways (than Primordial Gravitational Waves) of testing such DM productions in early universe as the DM do not interact with the visible sector with any other interaction than gravitational interactions. 

In the present article, with an aim to investigate the prospects of the upcoming GW interferometer missions in detecting such signals of primordial GW via DM/baryogenesis, we first estimate the signal-to-noise ratio (SNR) for a series of such detectors. Next, in order to have an idea of the errors on any such detection, we employ Fisher matrix forecast analysis, which basically estimates the uncertainties on the parameters involving dark matter physics such as the non-minimal coupling $\xi$ and background equation of state $\omega$, by choosing the allowed fiducial values of these parameters and investigate its relative uncertainties for future GW missions. Further, we complement our Fisher analysis  with a rigorous Markov chain Monte Carlo (MCMC) analysis using mock catalogs generated from the instrumental specifications of these GW missions and search for the combined constraints on the relevant parameters  like $\xi$, $\omega$ and tensor spectral tilt $\nt$ and their possible correlations, if any.
As representative future GW detectors, we mostly make use of BBO~\cite{Corbin:2005ny,Harry_2006}, DECIGO~\cite{Yagi:2011yu}, ET~\cite{Punturo_2010,Hild:2010id} and LISA~\cite{amaroseoane2017laser,Baker:2019nia}, although this analysis can be extended to other detectors as well.

Our analysis therefore leads us to identify the suitable regions of the parameter space where not only non-thermal production of DM, baryogenesis could happen in early universe only via gravity-mediated processes but also be tested with future GW missions. This in turn leads to additional constraints on the parameter spaces on top of what can be achieved by the state-of-the-art particle physics and/or cosmological experiments.
Nevertheless, this analysis paves the path of future investigations for primordial GW as a probe of DM and baryogenesis via a brand new experimental window, namely, the interferometer missions, with more realisitc catalogs with other possible sources of errors, either individually or with possible combinations of missions till the real dataset is available to the community.

\textit{The paper is organized as follows:} we start with the discussion of production of GWs during the inflation and the impacts of non-minimal coupling ($\xi$), scale of inflation ($H_{\rm inf}$) and tensor-to-scalar ratio ($\nt$) on primordial GWs in Sec.~\ref{sec:grav-wave}. Then we have discussed about the prospects of detection of the GW signal in the detector by calculating the SNR in Sec.~\ref{sec:prospects_of_GW}. After that to estimate the uncertainties on the parameters for the detectors, we have presented the Fisher matrix analysis in Sec.~\ref{sec:fisher_into} and also shown the MCMC analysis in Sec.~\ref{sec:parameter_inference} to understand the inference on the parameters. At the end we have explored the prospect of probing the gravitational DM and axions, along addressing the baryon asymmetry, at the GW detectors with robust SNR values in Sec.~\ref{sec:production}, before concluding in Sec.~\ref{sec:conclusion}.

%%%%%%%%%%%%%%%%%%%%%%%%%
\section{Production of gravitational waves during inflation}
\label{sec:grav-wave}
%%%%%%%%%%%%%%%%%%%%%%%%%
Let us begin with a discussion on Primordial GWs generated during  inflation and its subsequent propagation during the post-inflationary era. 
To analyze at the reheating epoch, we can express the behavior of the inflationary potential ($V(\phi)$) at the proximity of its minima as~\cite{Barman:2023ktz}
\begin{eqnarray}
    V(\phi) = \lambda\frac{\phi^n}{M_P^{n-4}}; \quad \phi\ll M_P, \label{Eq:potmin}
\end{eqnarray}
where $M_P = \frac{1}{\sqrt{8\pi G}} \simeq 2.4\times 10^{18}$ GeV, is the reduced Planck mass and $n$ is a free parameter. This emphasizes the fact that the whole analysis is mostly independent of the whole form of the inflationary potential. It only depends on the behavior of the potential at the proximity of its minima. The parameter $\lambda$ in Eq.~\eqref{Eq:potmin} determines the overall scale of the potential and is linked to the amplitude of the power spectrum, $A_S$ and number of e-folding at the time of inflation as $\lambda \simeq \frac{18\pi^2 A_s}{6^{k/2}N_e^2}$. Nevertheless, $N_e$ does not have strong dependence on the results, which is also supported by Ref.~\cite{Barman:2022qgt}.  According to Planck 2018 collaboration~\cite{Planck:2018jri}, ${\rm ln}(10^{10} A_s) = 3.044\pm 0.014$ and scalar spectral index, $n_s = 0.9649\pm 0.0042$, which are kept unaltered during our analysis.

In the subsequent post-inflationary reheating phase, we are primarily scrutinizing the gravity-mediated processes for particle production. The equation of state ($\omega$), during the period of reheating for the gravity-mediated processes, depends on $n$ which is given by~\cite{Ford:1986sy}~\footnote{Here we have assumed typical oscillation frequency, $\omega_{\rm eff}\ll \mathcal{H}$ (conformal Hubble parameter)~\cite{Cembranos:2015oya}}
\begin{eqnarray}\label{eq:eos_omega}
    \omega \approx \frac{n - 2}{n + 2}\, .
\end{eqnarray}
The era of gravitational reheating~\footnote{By gravitational reheating, we mean that the particle production from inflaton, occurring during the period of reheating, is mediated via gravity.}, driving particle production, commences around $\omega \sim 0.65$~\cite{Clery:2021bwz,haque2022gravitational,RT_2022}, 
corresponds to $n > 9$. However, this minimal scenario encounters severe constraints from BBN due to over-production of GWs~\cite{Barman:2022qgt,Yeh_2022}.
To reconcile this, an essential coupling between DM and Ricci scalar $\mathcal{R}$ maybe assumed~\cite{Barman:2022qgt}, as described in Appendix~\ref{app:inflaton} in detailed.

The presence of such non-minimal coupling, $\xi$,  relaxes the aforementioned constraint, allowing for the room of $n>4$~\cite{Barman:2022qgt} or even lower ($n>2$)~\cite{Barman:2023opy} in the scenario. 
Therefore, to investigate the role of non-minimal coupling, we keep $\xi$ as a free parameter in our analysis. In presence of such a non-minimal coupling, the reheating temperature ($\Trh$) can be expressed as~\cite{Barman:2022qgt}
\begin{eqnarray}\label{eq:grav-trh}
\Trh^4 = \frac{30}{\pi^2\,g_{\rm RH}}\, M_P^4\,\left(\frac{\rho_{\rm end}}{M_P^4}\right)^{\frac{4n-7}{n-4}}\,\left(\frac{\alpha^{\xi}_n\,\sqrt{3}\,(n+2)}{8n-14}\right)^{\frac{3n}{n-4}}\, .    
\end{eqnarray}
Here $g_{\rm RH}$ is the relativistic degree of freedom during the epoch of reheating and $\alpha^{\xi}_n$ is defined as in  Ref.~\cite{RT_2022,Barman:2022qgt}, discussed in detailed in Appendix~\ref{app:inflaton}. The $\xi$ parameter always appears in square in the $\alpha_n^{\xi}$. Hence positive or negative value of $\xi$ does not alter our result. That is why we have considered modulus values of $\xi$ throughout our analysis.~\footnote{There maybe some upper bounds on $\xi$ based on the choices of inflationary potentials, for instance like Higgs inflation. But that too depends upon assumptions regarding if the formalism is metric or Palatini~\cite{Shaposhnikov:2020fdv}. However, since our study is mostly independent of the choice of inflationary potential, we have not considered them.}
In this equation $\rho_{\rm end}$ indicates the inflation energy density at the end of inflation which can be expressed as $\rho_{\rm end} = 3M_P^2H_{\rm inf}^2$, with $H_{\rm inf}$ being the Hubble parameter during inflation, determining the scale of inflation.
%%%%%%%%%%%%%%%%%%%%%%%%%
\subsection{Primordial GWs in CMB}
\label{subsec:grav-wave}
%%%%%%%%%%%%%%%%%%%%%%%%%
In this section we will mainly discuss the GWs which are produced during the time of inflation due to quantum fluctuations. The produced PGWs  can act as an important probe of the early universe as it bears the imprints of the reheating epoch \textit{e.g.} how DM and SM are produced from inflaton, via gravity portal. 
We know that if the momentum modes of GW re-enter the horizon during the radiation dominated era, GW spectrum will become scale-independent, and the modes re-entering during the inflationary epoch, enhance the GW spectrum by a factor of $\frac{\rho_{\phi}}{\rho_{\rm R}}$~\cite{Barman:2022qgt,Saikawa:2018rcs}, with $\rho_{\rm R}$ being the radiation energy density.
In our study we have assumed gravitational reheating where inflaton is coupled with DM and SM via graviton, during the generation of particles. The reheating phase can be characterized by the background equation of state, $\omega$, of that era. We can consider the scale-dependent tensor power spectrum, $P_{\rm T}(k) = A_{\rm T} \left(\frac{k}{k_{\star}}\right)^{\nt}$ with $\nt$ being the tensor spectral index and $k_{\star} = 0.05\, {\rm Mpc}^{-1}$ being the pivot scale~\cite{Planck:2018jri} and $k$ indicates the modes. Here $A_{\rm T}$ represents the scale independent tensor power spectrum, which can be expressed as~\cite{Langlois:2010xc}
\begin{eqnarray}
    A_{\rm T} = \frac{2H^2_{\rm inf}}{\pi^2 M_P^2},
\end{eqnarray}
with $H_{\rm inf}$ being the Hubble parameter during inflation when the modes re-enter the horizon. This provides us the upper bound on $H_{\rm inf}$ from CMB~\cite{Artymowski:2017pua}
\begin{eqnarray}\label{eq:Hinf}
    H_{\rm inf} \lesssim 8.5\times 10^{13}\; {\rm GeV}.
\end{eqnarray}

Now the energy density of the GWs should be smaller than the limit of the effective number of relativistic species in the Universe, $\Delta N_{\rm eff}$, as~\cite{Maggiore:1999vm}
\begin{eqnarray}\label{eq:omega_delta_nu}
    \int^{\infty}_{f_{\rm min}} \frac{df}{f} \Omega_{\rm GW}(f)h^2\;\leq 5.6\times 10^{-6}\;\Delta N_{\rm eff},
\end{eqnarray}
where $f_{\rm min}$ indicates the lower limit of the integration, which is $\simeq \; 10^{-10}$ Hz for BBN and $\simeq \; 10^{-18}$ Hz for CMB. In our study $h$ is mentioned as reduced Hubble parameter which is defined as $h \equiv \frac{H_0}{100}$, with $H_0$ being the Hubble parameter at present epoch. In this study we have considered $h=0.6766$~\cite{Planck:2018vyg}. Now several current and upcoming observations provide upper bounds on $\Delta N_{\rm eff}$ which read
\begin{eqnarray}\label{eq:neff}
    \Delta N_{\rm eff} <
    \begin{cases}
        0.28 & \ \text{for Planck 2018 + BAO~\cite{Planck:2018vyg}}\\
        0.4 & \ \text{for BBN~\cite{Cyburt:2015mya}}\\
        0.05 & \ \text{for CMB Bharat$^{\star}$~\cite{CMB-bharat}}\\
        0.06 & \ \text{for CMB-S4$^{\star}$~\cite{Abazajian:2019eic,TopicalConvenersKNAbazajianJECarlstromATLee:2013bxd}, PICO$^{\star}$~\cite{NASAPICO:2019thw}}\\
        0.014 & \ \text{for CMB-HD$^{\star}$~\cite{CMB-HD:2022bsz}}\\
        0.12 & \ \text{for COrE$^{\star}$~\cite{CORE:2017oje}, South Pole Telescope$^{\star}$~\cite{SPT-3G:2014dbx}, Simons Observatory$^{\star}$~\cite{SimonsObservatory:2018koc}}.
    \end{cases}
\end{eqnarray}
Here the sign $(\star)$ indicates the forecasts for upcoming experiments. These upper bounds on $\Delta N_{\rm eff}$ from several experiments, lead to a strong upper bound on $\Omega_{\rm GW}(f)h^2$, according to Eq.~\eqref{eq:omega_delta_nu}. 

\subsection{Primordial GWs in interferometer missions}

While the CMB provides crucial insights into the early Universe, our exploration delves into the prospects of detection of PGWs in interferomemeter based GW missions beyond the CMB scale. 
We may express the GW spectrum for today as~\cite{Haque_2021} 
\begin{eqnarray}\label{eq:omegagwh2}
    \Omega_{\rm GW} (k)h^2\simeq 
    \begin{cases}
        \Omega_R h^2 P_{\rm T}(k)\frac{4\mu^2}{\pi}\left[\Gamma\left(\frac{5+ 3\omega}{2+6\omega} \right)\right]^2\left(\frac{k}{2\mu k_{\rm RH}}\right)^{\frac{6\omega - 2}{3\omega + 1}} & \ {\rm for} \ k_{\rm RH}<k \leq k_{\rm end}\\
        \Omega_R h^2 P_{\rm T}(k) & \ {\rm for} \ k \leq k_{\rm RH} ,
    \end{cases}
\end{eqnarray}
where $\mu\,\equiv\,\frac{1}{2}\left(1 + 3\omega\right)$, with $\Omega_R h^2 = 4.15\times 10^{-5}$ being the today's radiation abundance~\cite{Planck:2018vyg}. $\krh$ and $\kend$ represent the modes, corresponding to era of reheating and end of inflation, respectively~\footnote{Here, $\kend$ can be calculated as Eq.~\eqref{eq:fend}. On the other hand $\krh = \frac{a_{\rm RH}}{a_0} H_{\rm RH}$, with $\frac{a_{\rm RH}}{a_0} = \left(\frac{g_{\star}(T_0)}{g_{\star}(\Trh)}\right)^{1/3} \frac{T_0}{\Trh}$, \textit{i.e.} $\krh$ depends on reheating temperature ($\Trh$).}. This transition allows us to bridge our understanding about the early Universe, as obtained from  the CMB, with the ongoing quest to probe through the GW detectors.

With the advancement in the experimental front, several existing and upcoming/proposed interferometer-based GW detectors are in vogue,
%There are several existing and proposed GW experiments, 
which show promise in detecting the high-frequency GWs. These endeavors can be categorized into two distinct groups based on their instrumental and observational methodologies, namely,
\begin{enumerate}[a)]
    \item \textbf{Ground based interferometers:} \textit{Laser Interferomenter Gravitational-wave Observatory} (LIGO)~\cite{LIGOScientific:2016aoc,LIGOScientific:2016sjg,LIGOScientific:2017bnn,LIGOScientific:2017vox,LIGOScientific:2017ycc,LIGOScientific:2017vwq}, \textit{Advanced} LIGO (a-LIGO)~\cite{LIGOScientific:2014pky,LIGOScientific:2019lzm},  \textit{Einstein Telescope} (ET)~\cite{Punturo_2010,Hild:2010id}, \textit{Cosmic Explorer} (CE)~\cite{Reitze:2019iox}, \textit{etc.}
    \item \textbf{Space based interferometers: }\textit{Big-Bang Observer} (BBO)~\cite{Corbin:2005ny,Harry_2006}, \textit{Deci-Hertz Interferometer Gravitaitonal-wave Observatory} (DECIGO)~\cite{Yagi:2011yu}, \textit{Upgraded} DECIGO (U-DECIGO)~\cite{Seto:2001qf,Kawamura_2006,Yagi:2011wg}, \textit{Laser Interferometer Space Antenna} (LISA)~\cite{amaroseoane2017laser,Baker:2019nia}, $\mu$-ARES~\cite{Sesana:2019vho}, \textit{etc.}
\end{enumerate}

\begin{figure*}[!ht]
    \centering
    \begin{subfigure}{0.49\textwidth}
    \includegraphics[width=\textwidth]{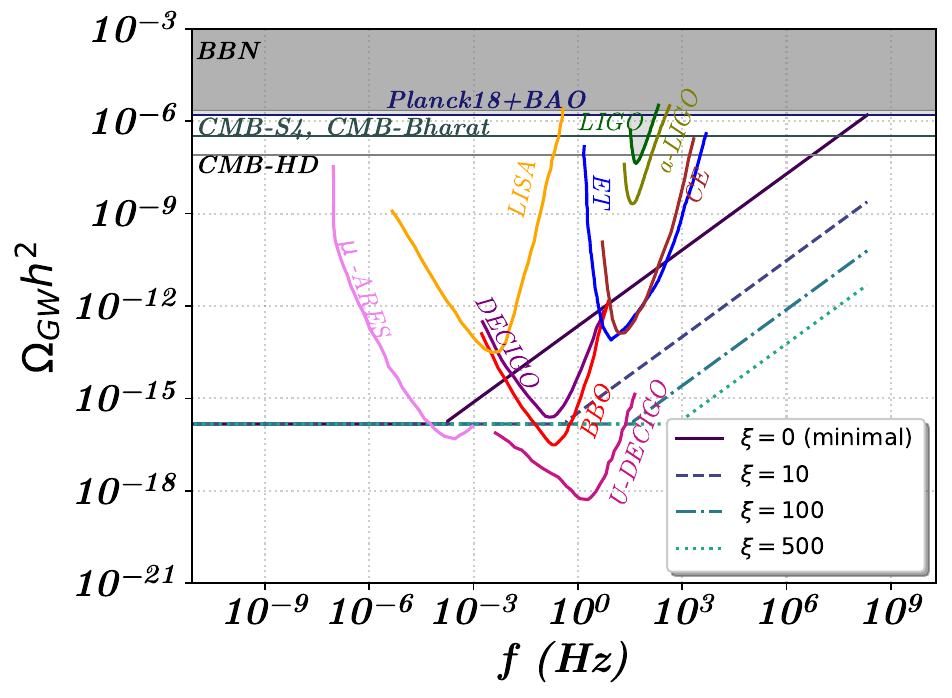}
    \caption{$\omega=0.8$}
    \label{fig:gwnt_fixed_omega}
    \end{subfigure}%
    \begin{subfigure}{0.49\textwidth}
    \includegraphics[width=\textwidth]{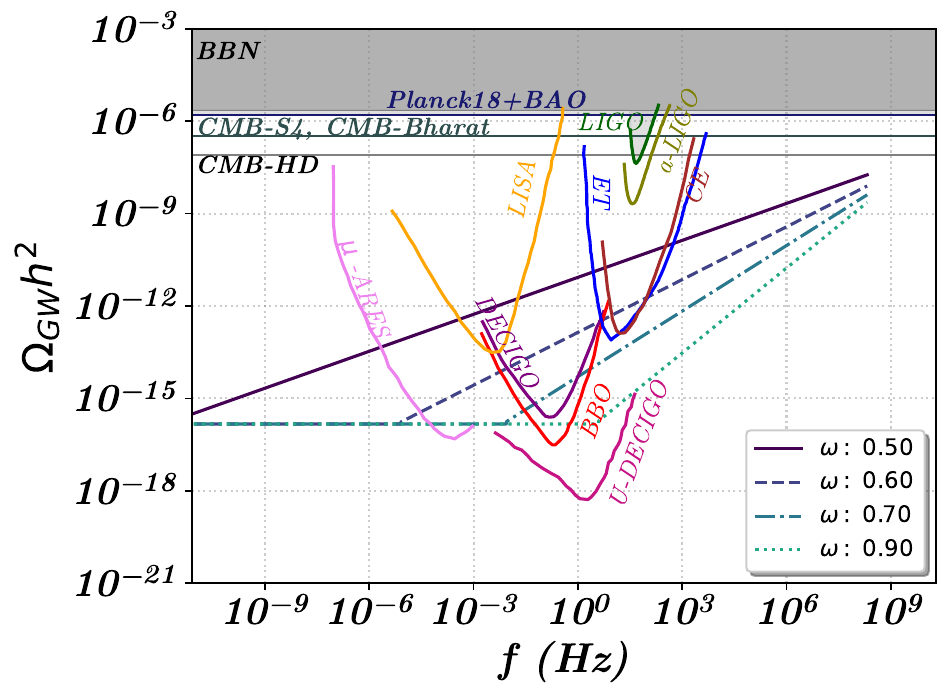}
    \caption{$\xi=10$}
    \label{fig:gwnt_fixed_xi}   
    \end{subfigure} 
    \caption{\it Plots show the variation of GW amplitude with respect to frequency ($f$) for $H_{\rm inf} = 1.0\times 10^{13}$ GeV and $\nt = 0$. \textbf{Left figure} shows the variation of coupling parameter ($\xi$) with $\omega = 0.8$, whereas \textbf{right one} represents the $\Omega_{\rm GW}h^2$ vs. $f$ behavior due to the variation of $\omega$ for $\xi=10$. \textbf{Left figure} depicts that the minimal scenario ($\xi = 0$) is ruled out by BBN and Planck 2018+BAO due to overproduction GWs, whereas $\xi>0$ obeys that bound and shows potential to be detected at BBO, DECIGO, ET, CE, LISA, U-DECIGO and $\mu$-ARES. Also we have shown the projected sensitivities on $\Omega_{GW}h^2$ from CMB-S4, CMB-Bharat and CMB-HD in both of the figures. Their sensitivity reach overlaps with LIGO's sensitivity which is depicted by shaded region. Here we have considered the frequency ($f$) as $f=\frac{k}{2\pi}$.}
    \label{fig:gwnt}
\end{figure*}

The sensitivity curves of these detectors are shown in the Fig.~\ref{fig:gwnt}, \ref{fig:GWs_nt} and \ref{fig:GWhi}. Since $\krh$ depends on the reheating temperature, GW spectrum also depends on the reheating temperature and hence on the coupling parameter ($\xi$) according to Eq.~\eqref{eq:grav-trh}. Now for the minimal scenario ($\xi=0$) we have depicted the GW spectrum, in Fig.~\ref{fig:gwnt_fixed_omega}, which is excluded due to the over-production of the GWs, 
for $H_{\rm inf} = 1.0\times 10^{13}$ GeV and $\nt = 0$ (also pointed out earlier in Ref.~\cite{Saikawa:2018rcs,Barman:2022qgt}). Upper bound for excessive radiation from BBN and Planck 2018 + BAO are shown in the gray shaded regions in the plot. However, with a nonzero $\xi$ , as in the case in the present article, an increase in $\xi$ results in a relaxation of the above constraints for a fixed choice of scale of inflation and tensor spectral index. This happens due to the fact that an increment of $\xi$ increases the reheating temperature resulting a decrement of gravitational energy density. The scale of inflation ($H_{\rm inf}$) also plays an important role in determining the maximum attainable scale, $k_{\rm end}$, which sets the end point of the frequency ($f_{\rm end}$) and also the maximum amplitude of the GW spectrum. 
Now, the modes with comoving wave number, $k$, can be expressed in terms of Hubble parameter ($H_k$) and scale factor ($a_k$) of the Universe when the mode re-enters the cosmic horizon, as $k = \frac{a_k}{a_0} H_k$. $a_0$ being the scale factor of the Universe at the present time which is set to be $1$ during the whole analysis. The frequency corresponds to the wave number can be expressed as $f_k = \frac{k}{2\pi}$, which sets the maximum possible frequency to be 
\begin{eqnarray}\label{eq:fend}
    f_{\rm end} \equiv \frac{H_{\rm inf}}{2\pi} \frac{\aend}{a_0},
\end{eqnarray}
where $\aend$ indicates the scale factor at the end of inflation.
As the enhancement of the GW spectrum occurs for the modes which re-enter the horizon during the post-inflationary reheating period, the enhanced spectrum can be detected at the high frequencies.

\medskip

\subsection{Non-trivial imprints of inflation on sensitivity curves}

In the previous section, we have demonstrated that the minimal scenario is ruled out due to the overproduction of the GWs, that hits the BBN bound. Consequently, it becomes imperative to explore the prospects of non-minimal couplings to gravity in producing sufficient  GWs in those missions, as indicated in the recent studies~\cite{Barman:2022qgt,Yeh_2022,Barman:2023opy}. On top of that, a non-zero value of the tensor spectral index $\nt$ can result in blue tilt in the GWs spectrum making it more prone for detection as we shall discuss next. 
 
As stated earlier, an increase in the  non-minimal coupling parameter $\xi$ leads to a higher value for the maximum temperature during the reheating, resulting an enhanced amount of maximum radiation density at that time. Consequently, the  GW spectrum gets suppressed due to the increment of $\xi$, depicted in Fig.~\ref{fig:gwnt_fixed_omega}. 
However, this scenario becomes particularly intriguing when the tensor spectral index ($\nt$) is also taken into consideration. In cosmological context, $\nt>0$ represents the blue-tilted spectrum, while $\nt<0$ points to the red-tilted spectrum. 
It is important to note that single field slow roll inflation predicts red-tilted $\nt$~\cite{Liddle:1993fq}. However it does not rule out the possibility for the blue-tilted $\nt$ from several motivation beyond simple slow-roll. For instance, string cosmology~\cite{Brandenberger:2006xi}, particle production during inflation~\cite{Cook:2011hg,Mukohyama:2014gba}, G-inflation~\cite{Kobayashi:2010cm}, super inflation modes~\cite{Baldi:2005gk} and others~\cite{Calcagni:2013lya,Kuroyanagi:2020sfw} provide the avenues to explore the blue-tilted $\nt$. In the subsequent analysis we will explore the scenario where $n_T$ is simply a non-zero quantity (as an independent parameter) and its possible consequences in GW missions in the context of gravitational production.

\begin{figure*}[!ht]
    \centering
    \begin{subfigure}{0.49\textwidth}
    \includegraphics[width=\textwidth]{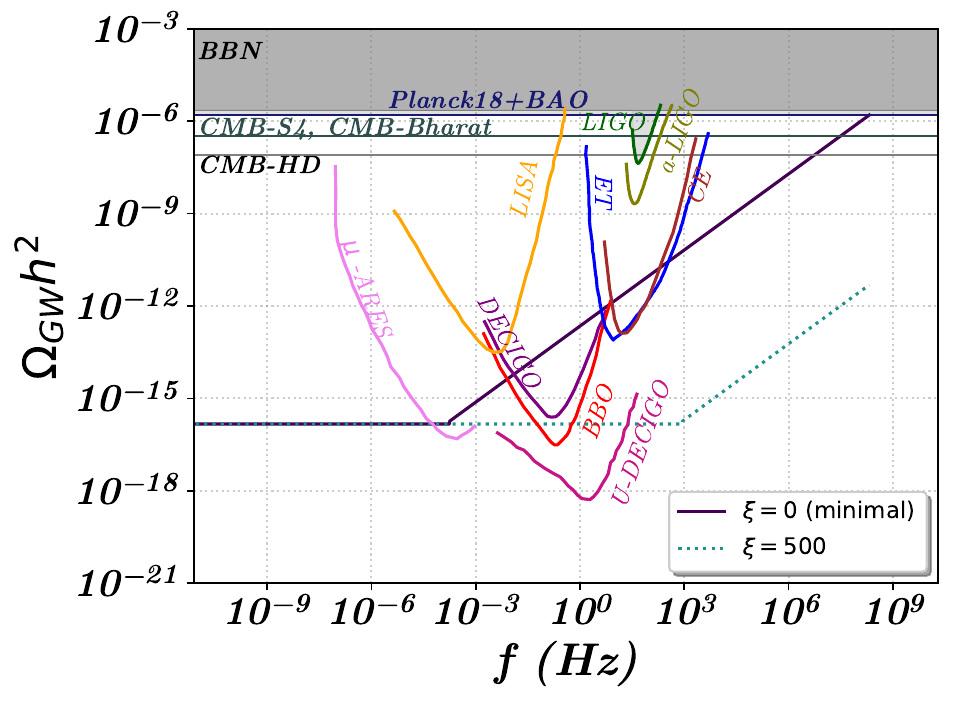}
    \caption{$\nt=0.05$ and $\omega=0.8$}
    \label{fig:gw_nt5}
    \end{subfigure}%
    \begin{subfigure}{0.49\textwidth}
    \includegraphics[width=\textwidth]{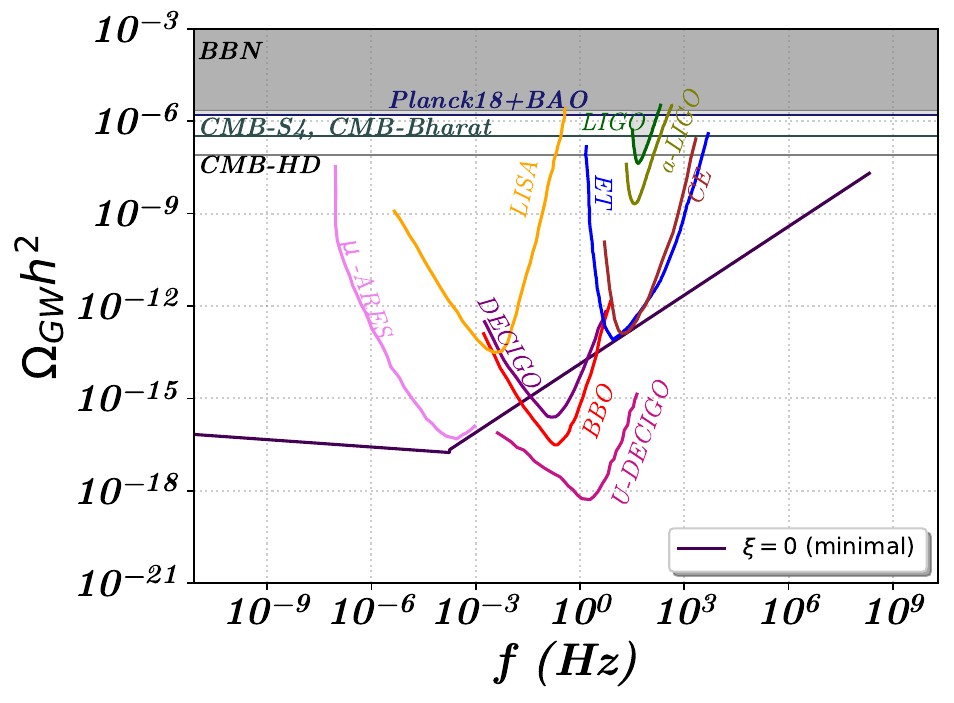}
    \caption{$\nt=-0.08$ and $\omega=0.8$}
    \label{fig:gw_nt_n1}   
    \end{subfigure} 
    \caption{\it \textbf{Left plot} represents the situation where $\nt = 0.05$ while \textbf{right plot} indicates $\nt=-0.08$ scenario. For both of the plots we have set $\omega=0.8$ and $H_{\rm inf} = 1.0\times 10^{13}$ GeV. The shaded regions are excluded due to overproduction of GWs from BBN and CMB era. Also we have shown the projected bounds on $\Omega_{GW}h^2$ from CMB-S4, CMB-Bharat and CMB-HD. All these bound regions overlap with LIGO sensitivity which is depicted by shaded region.  }
    \label{fig:GWs_nt}
\end{figure*}

In Fig.~\ref{fig:GWs_nt} we have displayed the impacts of $\nt$ on GW spectrum for a fixed value of $\omega$ ($\omega=0.8$ here). Fig.~\ref{fig:gw_nt5} depicts that $\nt>0$ allows for larger $\xi$ for which, PGWs could be detected by BBO~\cite{Corbin:2005ny,Harry_2006}, U-DECIGO~\cite{Seto:2001qf,Kawamura_2006,Yagi:2011wg}, DECIGO~\cite{Yagi:2011yu} and $\mu$-ARES~\cite{Sesana:2019vho}
% \ag{where ?}\deba{added.} 
for $\nt=0.05$ and $\xi = 500$. Conversely in Fig.~\ref{fig:gw_nt_n1}, we have depicted that the previously ruled out minimal scenario (\textit{i.e.} $\xi=0$) due to excess GW production (for a particular choice of scale of inflation), can be survived with red-tilted $n_T$ (with $n_T = -0.08$). Under this configuration, the PGWs can be detectable through a wider array of observations, including BBO~\cite{Corbin:2005ny,Harry_2006}, DECIGO~\cite{Yagi:2011yu}, U-DECIGO~\cite{Seto:2001qf,Kawamura_2006,Yagi:2011wg}, ET~\cite{Punturo_2010,Hild:2010id}, CE~\cite{Reitze:2019iox} and $\mu$-ARES~\cite{Sesana:2019vho}. 

\begin{figure*}[!ht]
    \centering
    \begin{subfigure}{.49\textwidth}
    \includegraphics[width=\textwidth]{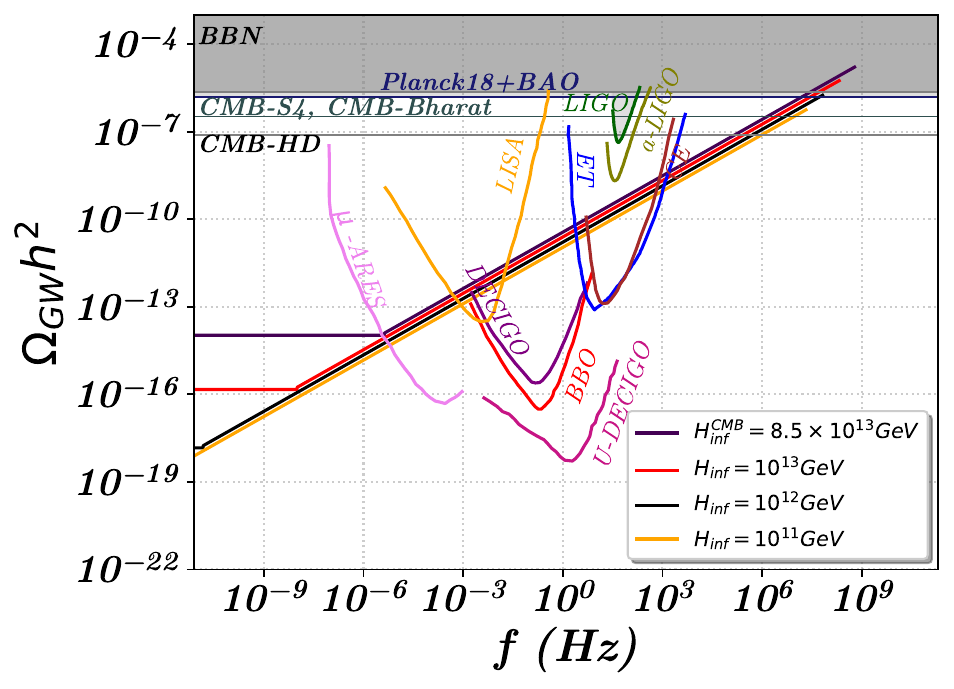}
    \caption{$\xi=0$ (minimal), $\omega=0.65$ and $\nt=0$}
    \label{fig:gw_hi}
    \end{subfigure}
    \begin{subfigure}{.49\textwidth}
    \includegraphics[width=\textwidth]{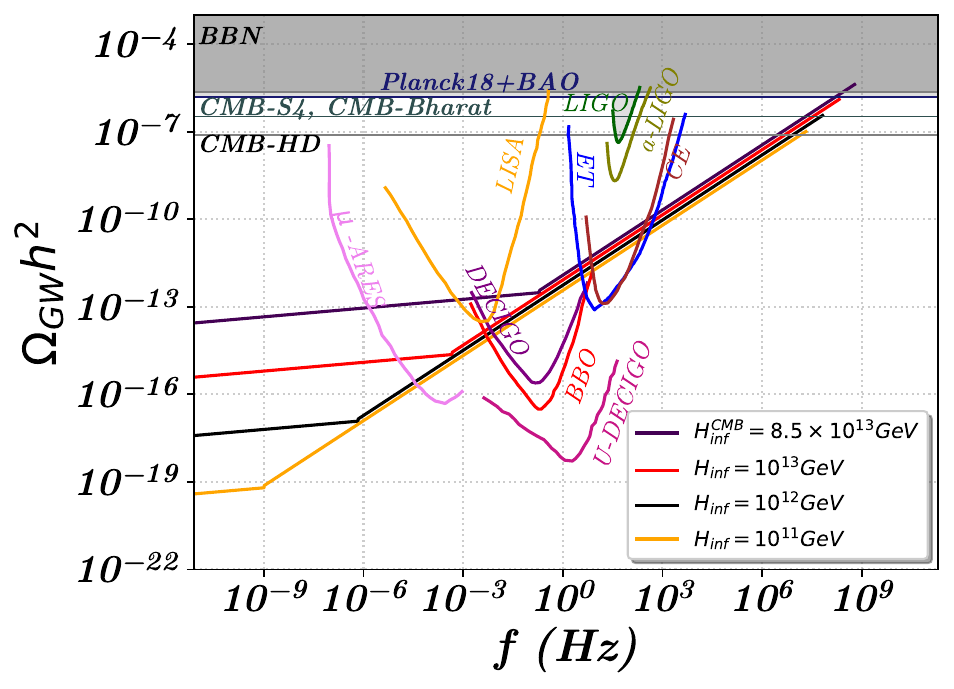}
    \caption{$\xi = 10$, $\omega=0.65$ and $\nt=0.1$}
    \label{fig:gw_hi2}   
    \end{subfigure} 
    \caption{\it These two plots indicate the variation of $H_{\rm inf}$ for fixed values of the other parameter to highlight the effects of $H_{\rm inf}$ on $\Omega_{GW}h^2$. \textbf{Left plot} represents a situation where $\omega=0.65$ and $\nt = 0$ for the minimal scenario ($\xi = 0$), while \textbf{right plot} indicates $\nt=0.1$ with $\omega=0.65$ and $\xi=10$. }
    \label{fig:GWhi}
\end{figure*}

It also transpires from the analysis  that scale of inflation plays a non-trivial role in determining the strength of the GWs spectrum and maximum possible value of frequency ($f_{\rm end}$). We have shown the behavior of GWs spectrum with the variation of $H_{\rm inf}$ in Fig.~\ref{fig:GWhi}. These figures display that with the increment of $H_{\rm inf}$, maximum possible value of frequency ($f_{\rm end}$) increases (evident from Eq.~\eqref{eq:fend}) and the scale invariant part of the GW spectrum also increases (evident from Eq.~\eqref{eq:omegagwh2}). Fig.~\ref{fig:gw_hi2} shows that lower values of $H_{\rm inf}$ are effective to detect the GW spectrum for high non-minimal coupling ($\xi=10$ here) and blue-tilted $\nt$ by $\mu$-ARES~\cite{Sesana:2019vho}, BBO~\cite{Corbin:2005ny,Harry_2006}, DECIGO~\cite{Yagi:2011yu}, U-DECIGO~\cite{Seto:2001qf,Kawamura_2006,Yagi:2011wg}, LISA~\cite{amaroseoane2017laser,Baker:2019nia}, ET~\cite{Punturo_2010,Hild:2010id} and CE~\cite{Reitze:2019iox}. For both of the plots we have fixed $\omega$ at $0.65$.

\medskip

%%%%%%%%%%%%%%%%%%%%%%%%%%%

%%%%%%%%%%%%%%%%%%%%%%%%%%%%%
\section{Prospects of detection in GW detectors}
\label{sec:prospects_of_GW}
%%%%%%%%%%%%%%%%%%%%%%%%%%%%%
For any experimental setup,  the noise from diverse sources (like the instrumental noise, survey area, observational strategy, \textit{etc.}, as well as astrophysical uncertainties and errors due to intervening medium, \textit{etc.}) comes into play along with the signal, and its significant role in the overall data is nearly unavoidable. This needs a good understanding on the noise spectrum. Hence, in order to estimate the prospects of detection of a signal in a particular detector, the quantity signal-to-noise ratio (SNR) serves as an important tool. In the following sections, we will discuss about the noise modeling and SNR for the GW detectors.

%%%%%%%%%%%%%%%%%%%%%%%%%%%%%
\subsection{Noise model for different GW detectors}
\label{sec:noise}
%%%%%%%%%%%%%%%%%%%%%%%%%%%%%
Since the role of noise on the signal is essential for any observation, we should have a good idea about the signal strength over the noise strength is necessary for any detection, 
so as to give us a reliable impression on the chances of detection of the signal. That's why we need to analyze the noise spectral density for each of the detectors. 

For the GW detectors, one 
can construct the effective noise power spectral density in terms of detectors' noise. If $R(f)$ represents the detector response function, effective noise power spectral density, $S(f)$, can be expressed in terms of detector's noise ($N(f)$) as~\cite{Gowling:2021gcy},
\begin{eqnarray}
    S(f)\;\equiv\;\frac{N(f)}{R(f)}.
\end{eqnarray}
Hence the expected sensitivity curves for any GW detectors can be depicted as~\cite{Gowling:2021gcy}
\begin{eqnarray}\label{eq:omegagw}
    \Omega (f) \;=\; \left(\frac{4\pi^2}{3H_0^2}\right)f^3 S(f).
\end{eqnarray}
Below we summarize the noise power spectral density for the GW missions under our consideration, that we are going to make use of in the subsequent analysis.
\begin{itemize}
%% PSD for BBO
\item
\textbf{BBO}

BBO mainly consists of four triangular sets of detectors each of having arm-length $5\times 10^4$ km. The non sky-averaged instrumental noise power spectral density for the BBO can be expressed as~\cite{Yagi:2011yu} %consists
\begin{align}
S_{\rm BBO}(f) &= \left[1.8 \times 10^{-49} (f / 1{\rm Hz})^2+2.9 \times 10^{-49} + 9.2 \times 10^{-52} (f / 1{\rm Hz})^{-4}\right]~{\rm Hz^{-1}}.
\label{eq:SeffBBO}
\end{align}
Hence GW energy spectrum density for BBO, considering non sky-averaged instrumental noise, can be expressed according to Eq.~\eqref{eq:omegagw}.
%% PSD for DECIGO
\item
\textbf{DECIGO}

The basic setup for DECIGO and BBO are more or less same except DECIGO is a Fabry-Perot type interferometer while BBO is a transponder type interferometer. Also its arm-length is $1000$ km, smaller than BBO. The non sky-averaged instrumental noise power spectral density for the DECIGO can be expressed as~\cite{Yagi:2013du} 
\begin{align}
S_{\rm DECIGO}(f) &= 5.3\times 10^{-48}\Big[\left(1 + (f / f_p)^2\right)
\nonumber \\
&~~~~
+ 2.3 \times 10^{-7} \frac{(f / f_p)^{-4}}{1 + (f/f_p)^2} + 2.6 \times 10^{-8} (f / f_p)^{-4} \Big]~{\rm Hz^{-1}},
\label{eq:SeffDECIGO}
\end{align}
with $f_p = 7.36$~Hz. So as earlier, one can easily find out the GW energy spectrum density from Eq.~\eqref{eq:omegagw}.
%% PSD for LISA
\item
\textbf{LISA}

LISA, having a quadrupolar antenna pattern, mainly consists of 3 drag-free spacecraft in which free-falling mirrors are contained. the instrumental noise in LISA is expected to arise mainly due to three reasons~\cite{Gowling:2021gcy}: \textit{(a)} optical metrology noise (omn), \textit{(b)} shot noise (sn) and \textit{(c)} acceleration noise of test mass (accn), due to the disturbance of test mass.

The instrumental noise spectral density for each of the mentioned noise can be expressed respectively as~\cite{Gowling:2021gcy}: 
\begin{align}
S_{\rm acc}(f) &= 9 \times 10^{-30} \frac{1}{(2\pi f / 1{\rm Hz})^4} \left( 1 + \frac{10^{-4}}{f / 1{\rm Hz}} \right)~{\rm m^2Hz^{-1}},\\
S_{\rm sn}(f) &= 2.96 \times 10^{-23}~{\rm m^2Hz^{-1}}, \\
S_{\rm oms}(f) &= 2.65 \times 10^{-23}~{\rm m^2Hz^{-1}}.
\end{align}
Collecting all the noises, the instrumental noise for LISA can be expressed as 
\begin{align}
S_{\rm LISA}(f) &= \frac{40}{3} \frac{4S_{\rm acc}(f) + S_{\rm sn}(f) + S_{\rm oms}(f)}{L^2} \left[1 + \left( \frac{f}{0.41c/2L} \right)^2\right],
\label{eq:SeffLISA}
\end{align}
with $L = 2.5 \times 10^8$~m being the constellation arm-length of the instrument. Hence the GW energy density power spectrum can be calculated from Eq.~\eqref{eq:omegagw}.

\item
\textbf{ET}

For ET we have followed Ref.~\cite{Chowdhury:2022pnv} to generate noise curve.
\end{itemize}

%%%%%%%%%%%%%%%%%%%%%%%%%%%%%
\subsection{Estimation of signal-to-noise ratio}
\label{subsec:snr}
\begin{figure*}[!ht]
    \centering
    \begin{subfigure}{.48\textwidth}
    \includegraphics[width=\textwidth]{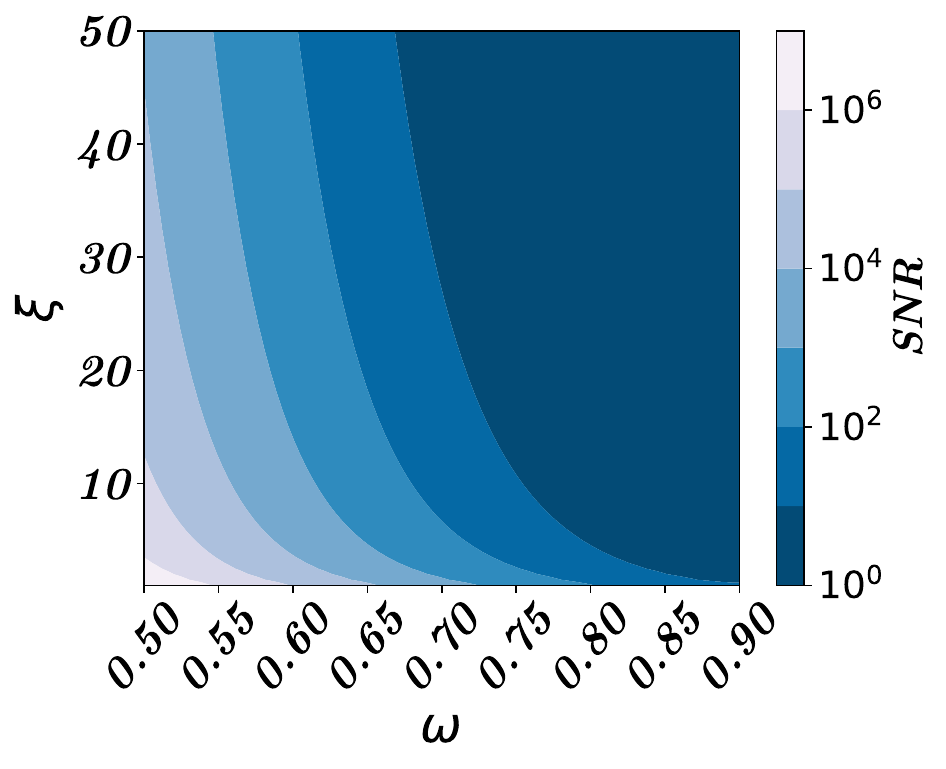}
    \caption{\it BBO}
    \label{fig:snr_bbo1}
    \end{subfigure}
    \begin{subfigure}{.48\textwidth}
    \includegraphics[width=\textwidth]{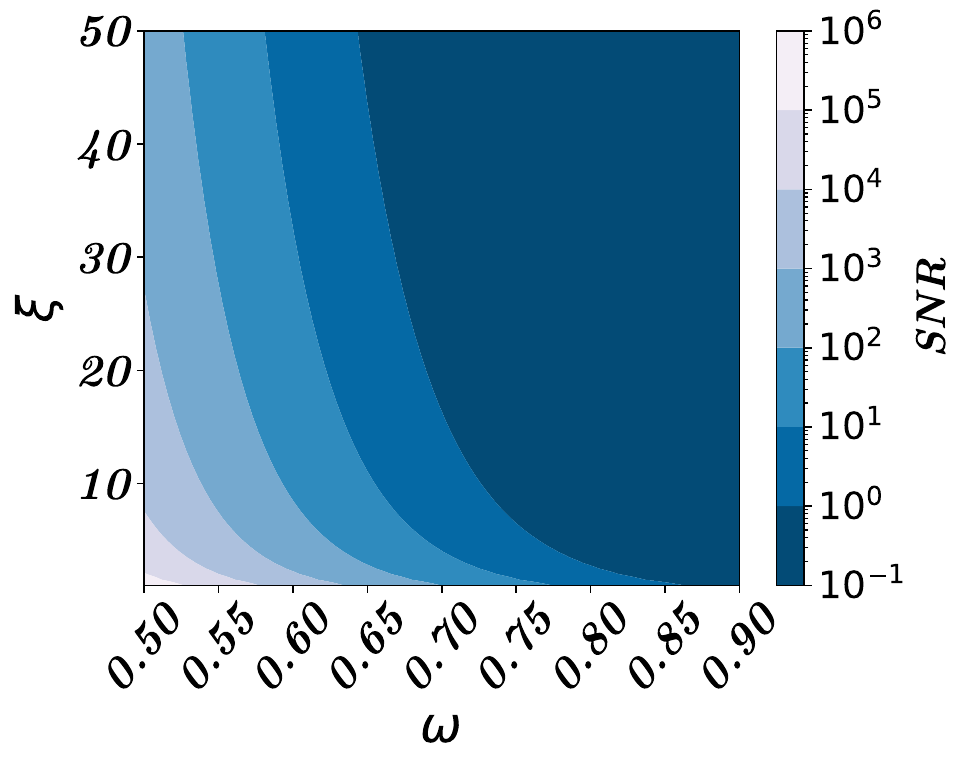}
    \caption{\it DECIGO}
    \label{fig:snr_decigo1}   
    \end{subfigure}
    \begin{subfigure}{.48\textwidth}
    \includegraphics[width=\textwidth]{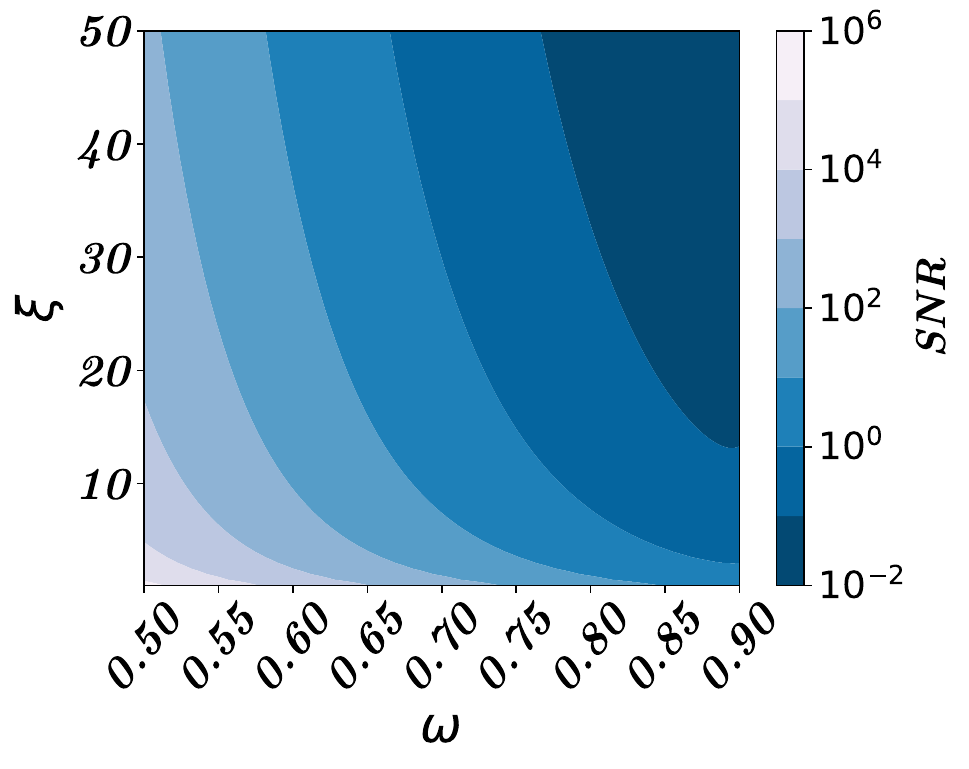}
    \caption{\it ET}
    \label{fig:snr_et1}   
    \end{subfigure}
    \begin{subfigure}{.48\textwidth}
    \includegraphics[width=\textwidth]{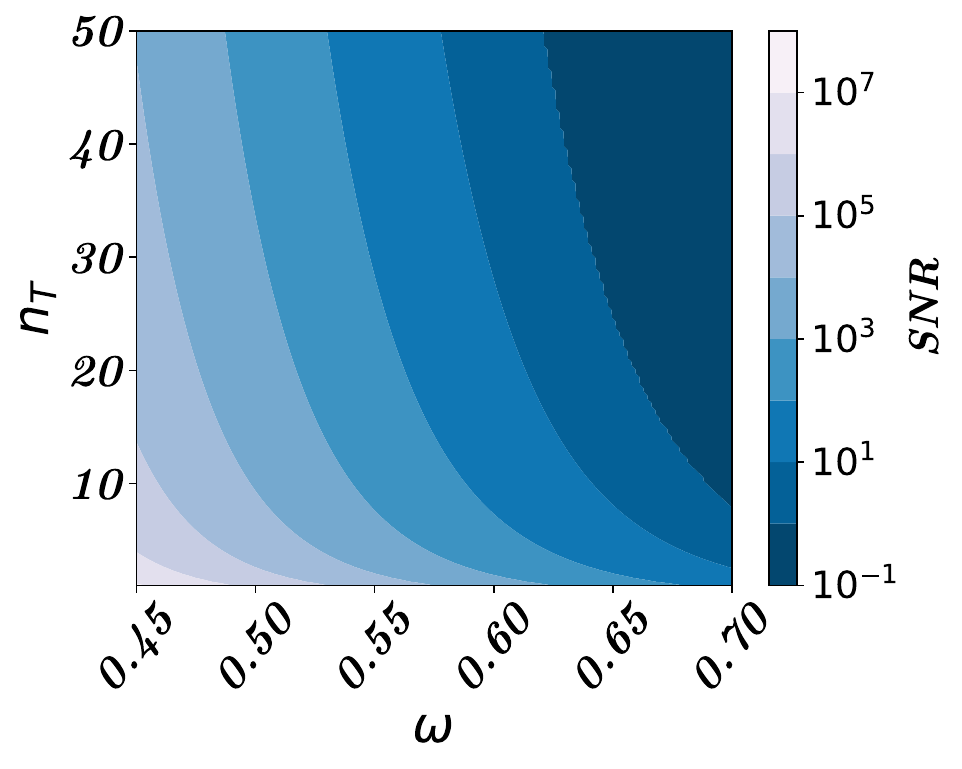}
    \caption{\it LISA}
    \label{fig:snr_LISA1}   
    \end{subfigure}
    \caption{\it The plots represent the dependence of SNR on $\omega$ and $\xi$ for $\nt=0$ for BBO, DECIGO, and ET, and $\nt = 0.5$ for LISA. We have also set $H_{\rm inf} = 5.5\times 10^{12}$ GeV.}
    \label{fig:snr_contour}
\end{figure*}

\begin{figure*}[!ht]
    \centering
    \begin{subfigure}{.48\textwidth}
    \includegraphics[width=\textwidth]{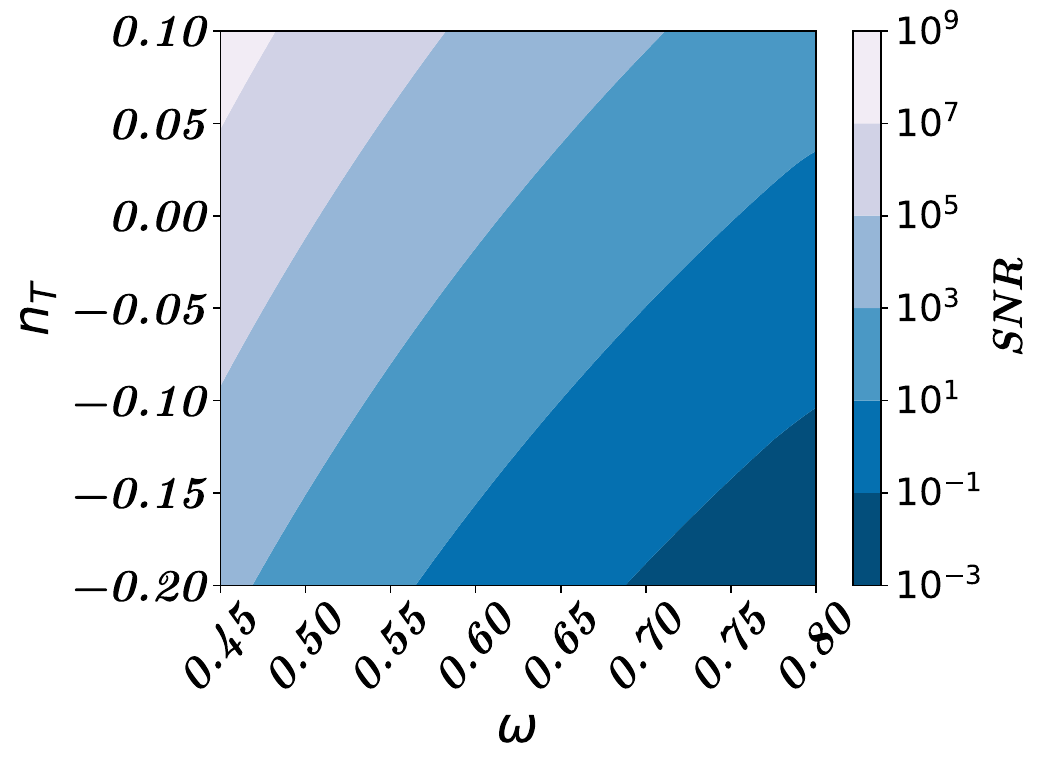}
    \caption{\it BBO}
    \label{fig:snr_bbo2}
    \end{subfigure}
    \begin{subfigure}{.48\textwidth}
    \includegraphics[width=\textwidth]{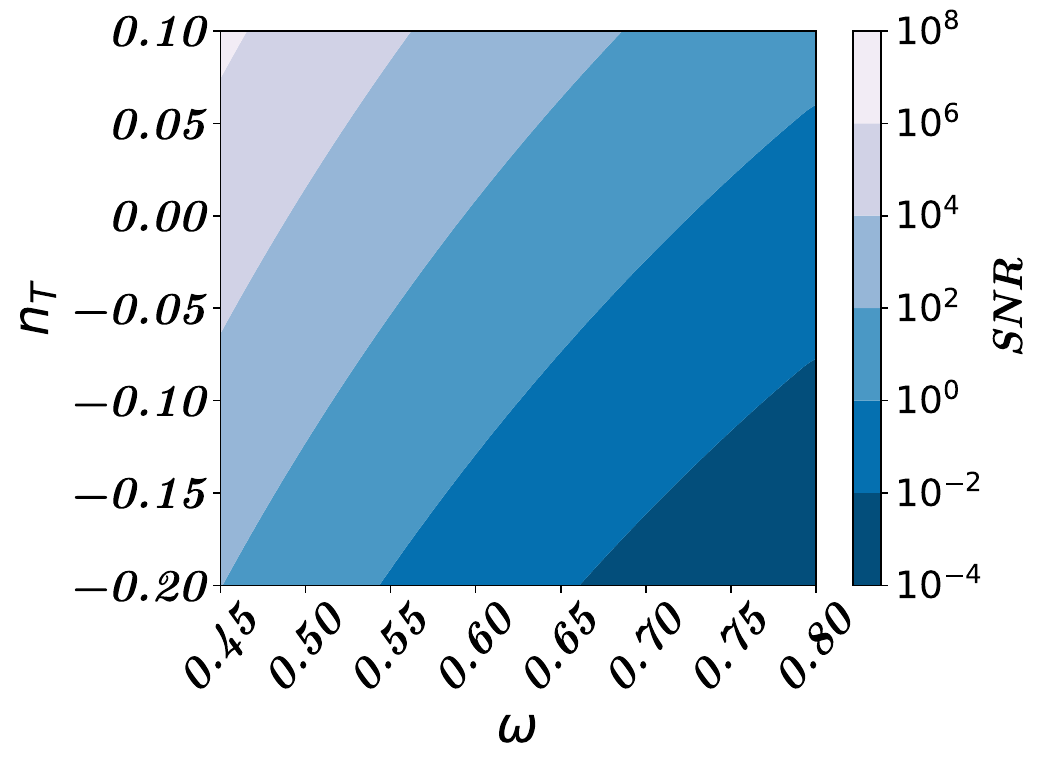}
    \caption{\it DECIGO}
    \label{fig:snr_decigo2}   
    \end{subfigure}
    \begin{subfigure}{.48\textwidth}
    \includegraphics[width=\textwidth]{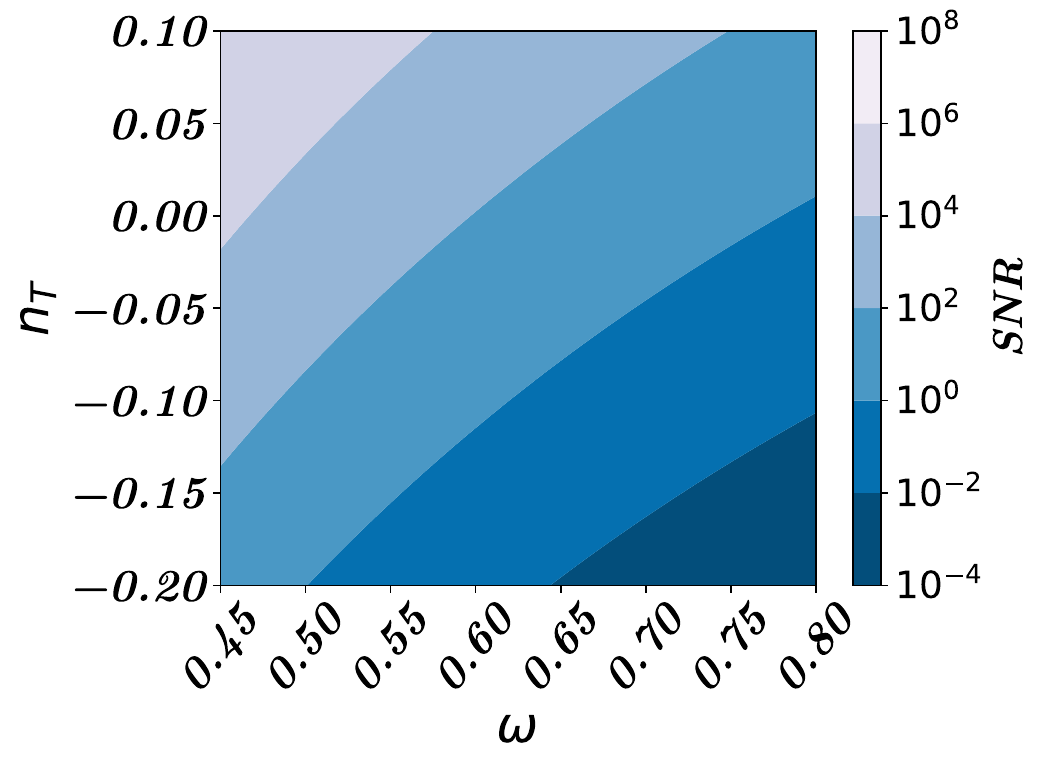}
    \caption{\it ET}
    \label{fig:snr_et2}   
    \end{subfigure}
    \begin{subfigure}{.48\textwidth}
    \includegraphics[width=\textwidth]{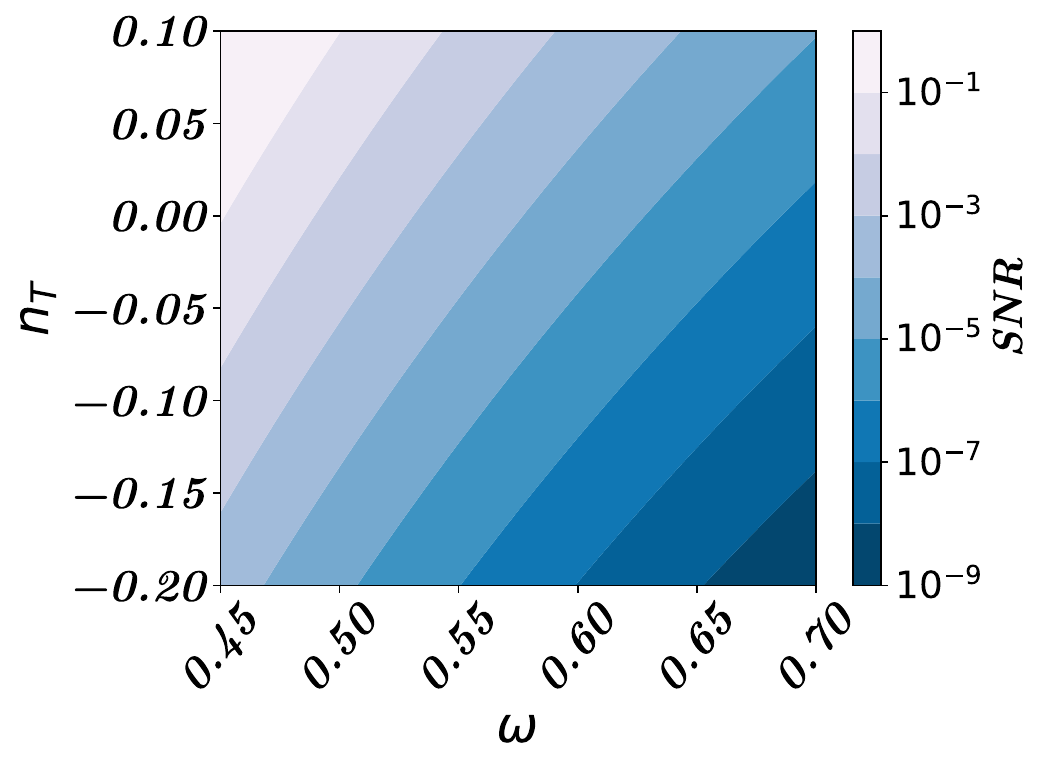}
    \caption{\it LISA}
    \label{fig:snr_LISA2}   
    \end{subfigure}
    \caption{\it The plots represent the dependence of SNR on $\omega$ and $\nt$ for a fixed value of $\xi$ for BBO, DECIGO, ET and LISA. For all the  detectors $\xi$ is set to $10$. We have also set $H_{\rm inf} = 5.5\times 10^{12}$ GeV.}
    \label{fig:snr_contour2}
\end{figure*}

\begin{figure*}[!ht]
    \centering
    \begin{subfigure}{.48\textwidth}
    \includegraphics[width=\textwidth]{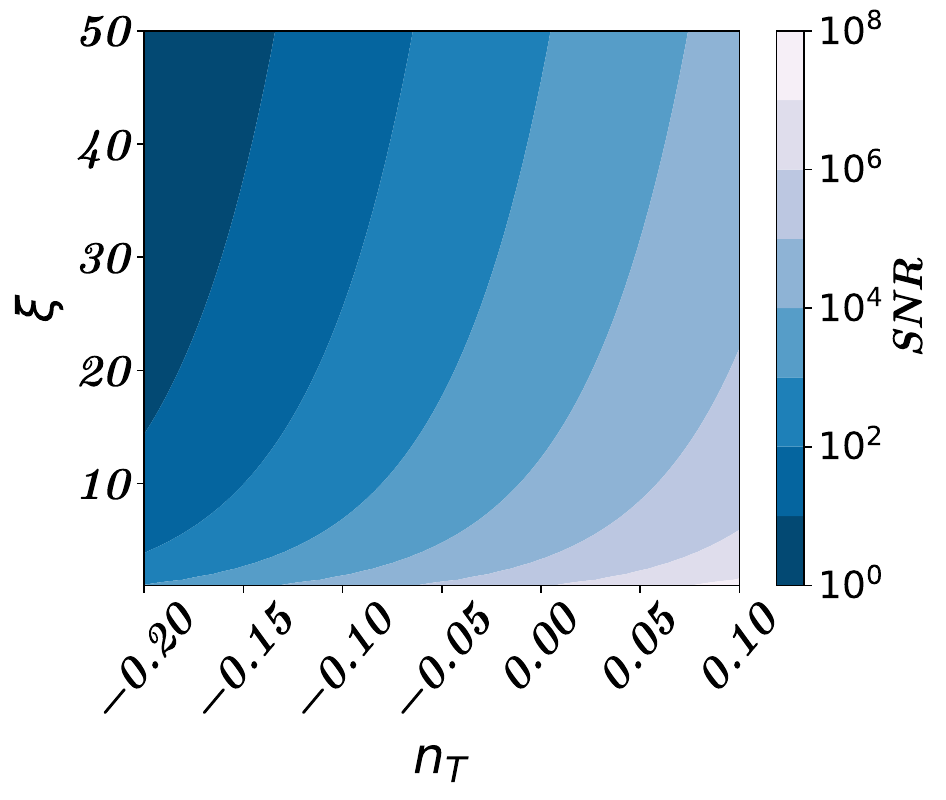}
    \caption{\it BBO}
    \label{fig:snr_bbo3}
    \end{subfigure}
    \begin{subfigure}{.48\textwidth}
    \includegraphics[width=\textwidth]{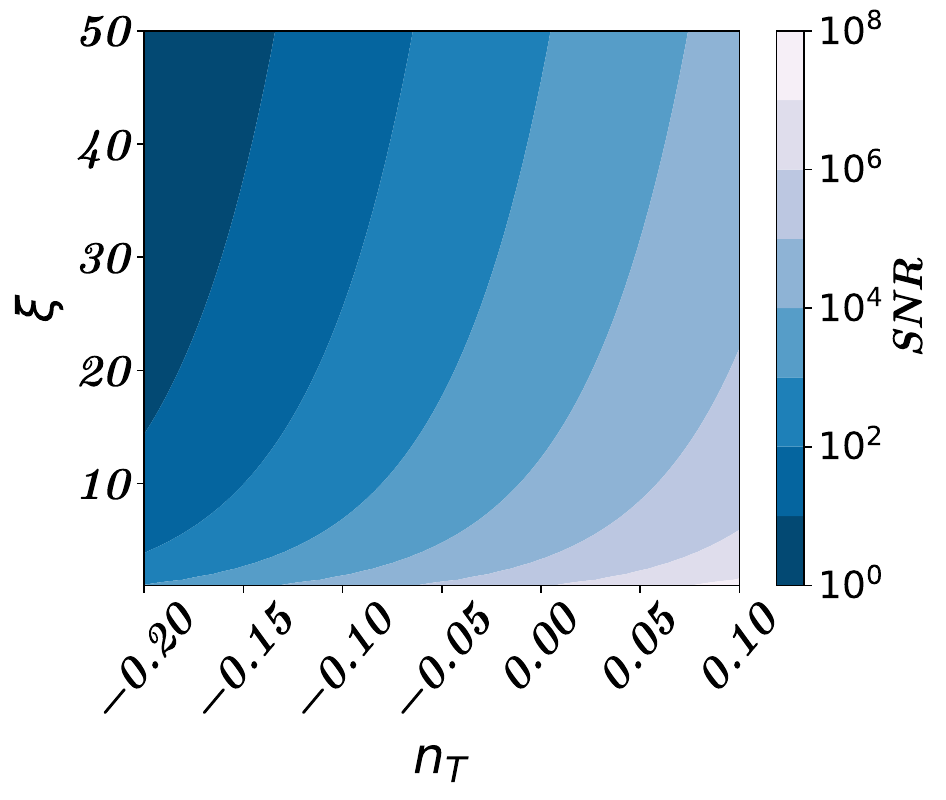}
    \caption{\it DECIGO}
    \label{fig:snr_decigo3}   
    \end{subfigure}
    \begin{subfigure}{.48\textwidth}
    \includegraphics[width=\textwidth]{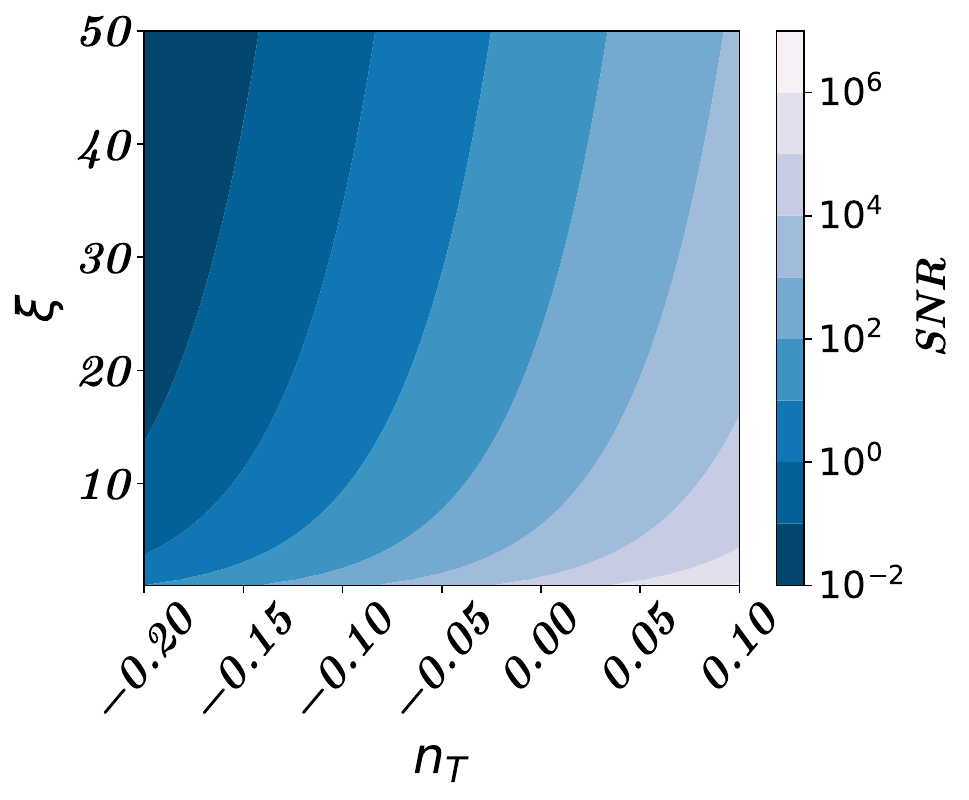}
    \caption{\it ET}
    \label{fig:snr_et3}   
    \end{subfigure}
    \begin{subfigure}{.48\textwidth}
    \includegraphics[width=\textwidth]{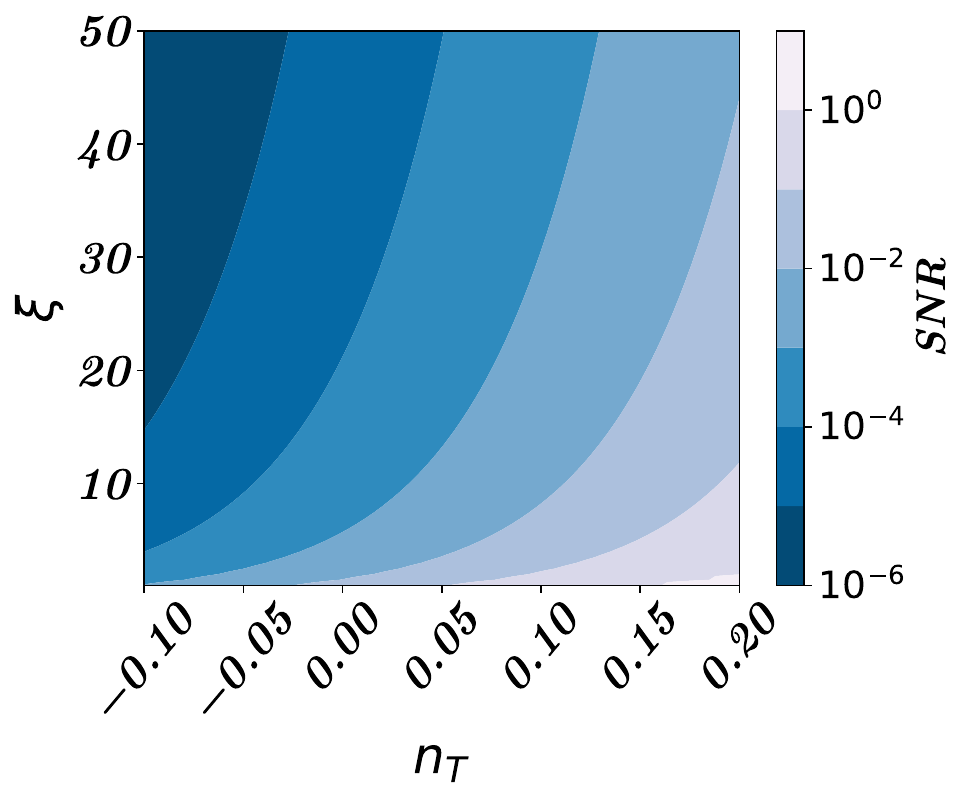}
    \caption{\it LISA}
    \label{fig:snr_LISA3}   
    \end{subfigure}
    \caption{\it The plots represent the dependence of SNR on $\nt$ and $\xi$ for $\omega=0.55$ for BBO, DECIGO, ET and LISA. $H_{\rm inf}$ is set to $5.5\times 10^{12}$ GeV.}
    \label{fig:snr_contour3}
\end{figure*}

%%%%%%%%%%%%%%%%%%%%%%%%%%%%%
As mentioned earlier, SNR serves a crucial tool in assessing the detectional prospect of signal.
SNR from the GW missions can be calculated as~\cite{Thrane:2013oya,Caprini:2015zlo}
\begin{eqnarray}\label{eq:snr}
    \text{SNR}\;\equiv\; \sqrt{\tau\;\int_{f_{\rm min}}^{f_{\rm max}} df \left(\frac{\Omega_{\rm{GW}}(f,\{\theta\})h^2}{\Omega_{\rm n}(f)h^2}\right)^2},
\end{eqnarray}
where $\Omega_{\rm{GW}}(f,\{\theta\})h^2$ represents the GW energy density according to a particular theory, with $\{\theta\}\equiv(\omega,\,\xi,\,n_T$ in our case) being the model parameters and $\Omega_{\rm n}(f)h^2$ carries the information of noise for detector. $\tau$ represents the observation time which is taken to be 4 years in our analysis. We have also set $H_{\rm inf} = 5.5\times 10^{12}$ GeV as it has not only the potential to be detected at the detectors but also it obeys the upper bounds of GW spectrum (higher values of $H_{\rm inf}$ does not obey the upper bound for $\xi=0$ and $w>1/3$, see Fig.~\ref{fig:gw_hi}). $f_{\rm min}$ and $f_{\rm max}$ are the minimum and maximum frequencies for GW detection, respectively, for a given detector. The SNR is plotted as colored contour in Fig.~\ref{fig:snr_contour}, \ref{fig:snr_contour2} and \ref{fig:snr_contour3}, to display the dependency of SNR on the parameters ($\omega$, $\xi$, $\nt$).

Fig.~\ref{fig:snr_contour} represents the SNR on the $\omega$-$\xi$ plane (by simultaneous variation of the parameters $\omega$ and $\xi$) for the four GW detectors under consideration, with a fixed value for $\nt$ for each. For BBO~\cite{Corbin:2005ny,Harry_2006}, DECIGO~\cite{Yagi:2011yu} and ET~\cite{Punturo_2010,Hild:2010id}, we have set $\nt=0$ and for LISA~\cite{amaroseoane2017laser,Baker:2019nia} we set $\nt=0.5$. For each of the detectors, the plots 
indicate that higher values of SNR tends to be concentrated in a region where $\omega$ and $\xi$ both are small. This is intuitive because these regions enhance the GW signal to be detected in the detectors as depicted in Fig.~\ref{fig:gwnt} and also discussed in Sec.~\ref{subsec:grav-wave}.  Fig.~\ref{fig:snr_contour} also indicates that for each of the detectors SNR increases sharply with the decrease of $\xi$ for lower values of $\omega$ which is physical, as decrement the value of $\xi$ results in the increment of the strength of GW spectrum, discussed in Sec.~\ref{subsec:grav-wave}.

Similarly in Figs.~\ref{fig:snr_contour2} and \ref{fig:snr_contour3} we have displayed the SNR on $\omega$-$\nt$ plane and $\xi$-$\nt$ plane, respectively. These plots also lead to a similar conclusion as in Fig.~\ref{fig:snr_contour}. In Fig.~\ref{fig:snr_contour2} we have set $\xi = 10$ for all the four detectors. On the other hand for the contours between $\xi$ and $\nt$ we have set $\omega = 0.55$ for the four detectors, depicted in Fig.~\ref{fig:snr_contour3}. Both of the figures show that the relatively higher SNR values are achieved towards positive values of $\nt$ on top of $\omega$ and $\xi$ are small. Hence these plots favor slightly blue-tilted spectrum \textit{i.e.} $\nt>0$, in order to get high SNR. Notably, LISA~\cite{amaroseoane2017laser,Baker:2019nia} exhibits a stronger inclination towards a blue-tilted spectrum compared to the other detectors, which means that that detection of the signal at LISA~\cite{amaroseoane2017laser,Baker:2019nia} may need the GW spectrum to be more blue-tilted compared to other detectors.

\begin{figure*}[!ht]
    \centering
    \begin{subfigure}{.32\textwidth}
    \includegraphics[width=\textwidth]{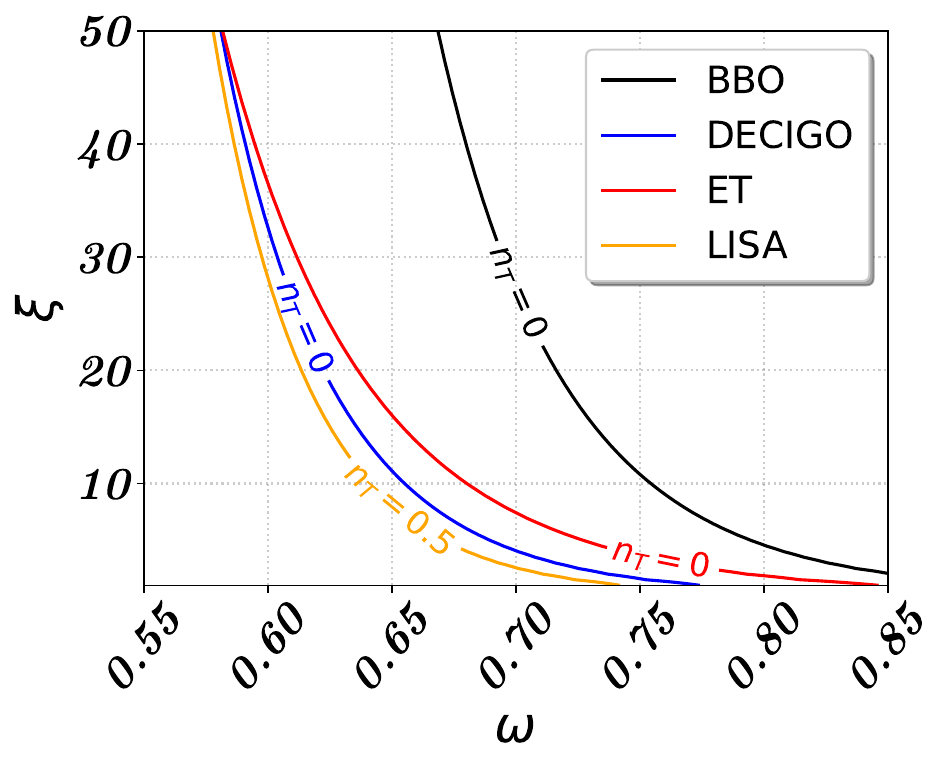}
    \caption{\it For fixed $n_T$}
    \label{fig:snr_xivsw}
    \end{subfigure}
    \begin{subfigure}{.32\textwidth}
    \includegraphics[width=\textwidth]{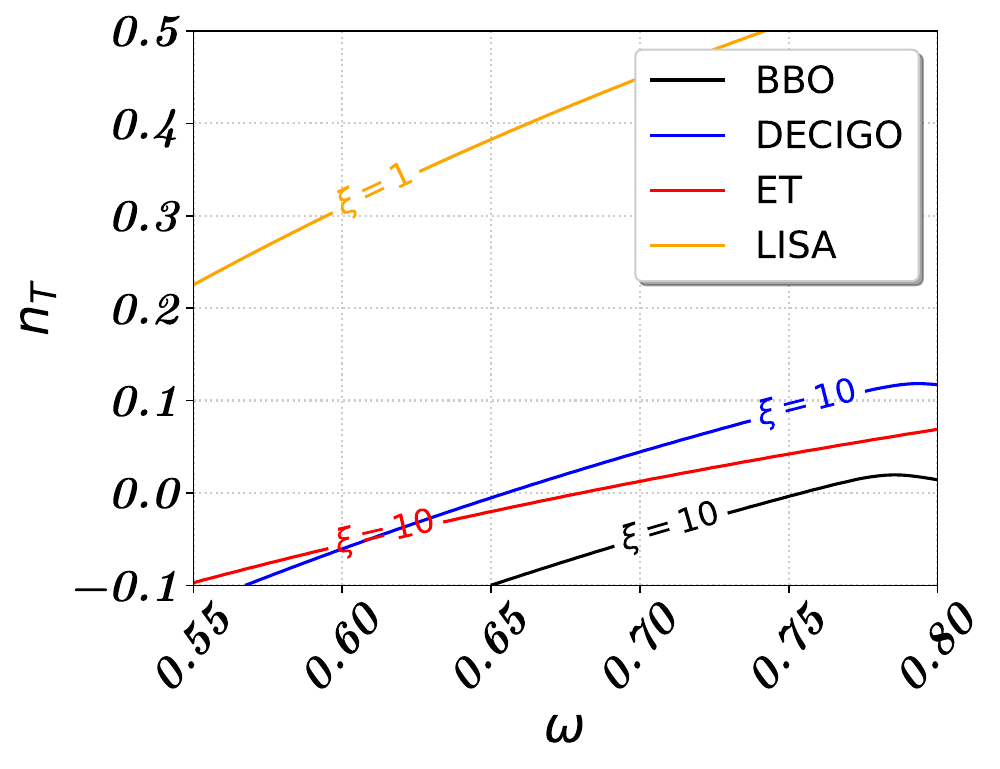}
    \caption{\it For fixed $\xi$}
    \label{fig:snr_ntvsw}   
    \end{subfigure}
    \begin{subfigure}{.32\textwidth}
    \includegraphics[width=\textwidth]{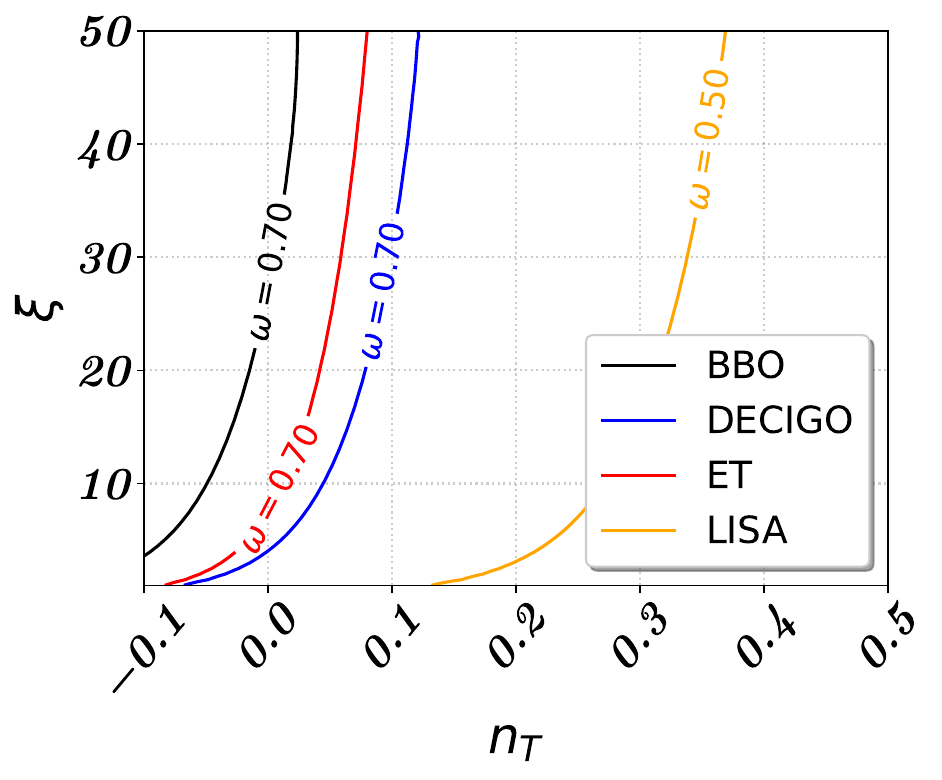}
    \caption{\it For fixed $\omega$}
    \label{fig:snr_ntvsxi}   
    \end{subfigure}
    \caption{\it The figure illustrates the relation between two parameters for a fixed value of the third one, where each of the contours represents SNR=10. In the figures we have compared the scenario among three GW detectors (BBO, DECIGO, ET, LISA) which is indicated by colored line in the figure. For each of the detectors the region below the contour represents SNR $>$10.}
    \label{fig:snr}
\end{figure*}

Further, in Fig.~\ref{fig:snr_xivsw} we have displayed a comparison among the four detectors by taking a benchmark value for SNR $=10$ (which we have set as detection threshold). This comparison depicts that to achieve a considerable amount of SNR for DECIGO~\cite{Yagi:2011yu}, LISA~\cite{amaroseoane2017laser,Baker:2019nia} and ET~\cite{Punturo_2010,Hild:2010id},  $\omega$ and $\xi$ need to be small compared to the BBO~\cite{Corbin:2005ny,Harry_2006}, indicating that BBO~\cite{Corbin:2005ny,Harry_2006} is more sensible to detect the signal. In other words, among all the four detectors, BBO~\cite{Corbin:2005ny,Harry_2006} is more capable to detect the GW signal particular set of parameters. 
Also these plots indicate that when $\omega$ increases $\xi$ needs to decrease in order to remain in the similar detection sensitivity which shows the inter parameter dependence in the context of all of the four detectors. 

\medskip

%%%%%%%%%%%%%%%%%%%%%%%%%%%%%
\section{Fisher matrix forecast analysis}
\label{sec:fisher_into}
%%%%%%%%%%%%%%%%%%%%%%%%%%%%%
Having convinced ourselves of  the high SNR value achievable for certain allowed ranges of the parameters, let us now engage in the estimation of errors of the parameters responsible for gravitational reheating, and gravitational production of DM, as can be obtained by the upcoming GW detectors. To this end, we will employ the standard Fisher forecast analysis.
The Fisher matrix represents the curvature of a Gaussian-shaped approximation used to characterize the posterior likelihood at the proximity of its maximum. This analysis is very important in predicting the uncertainties on estimating the parameters for an experiment for a specific fiducial value. When dealing with a theory comprising $n$ parameters, the corresponding Fisher matrix for an experiment will be an $n\times n$ dimensional matrix and the square root of the diagonal elements of the inverse of the Fisher matrix represent the $1$-$\sigma$ errors on the respective parameters. In the subsequent sections we will delve into the details of this analysis, exploring how it allows us to quantify uncertainties and assess parameter correlations for the PGW signal for the GW detectors.

%%%%%%%%%%%%%%%%%%%%%%%%%%%%%
\subsection{The likelihood  function and Fisher matrix}
\label{subsec:likelihood}
%%%%%%%%%%%%%%%%%%%%%%%%%%%%%
Depending on the noise modeling of the detectors, discussed in Sec.~\ref{sec:noise}, we can generate the mock data for the GW detectors. In this section we will construct the likelihood function which depends on the noise model for each of the detectors (Sec.~\ref{sec:noise}). 

To construct the likelihood function, we have defined the noise  spectral density as $\Omega_{\rm n}$, constructed from the instrumental specifications of the detectors, discussed in Sec.~\ref{sec:noise}. Also we have defined the energy spectral density due to the gravity-mediated processes as signal ($\Omega_{\rm sig}$)(Eq.~\eqref{eq:omegagwh2}) which is specified by model parameters ($\{\theta\}$).

A standard practice in this direction is to split the frequency range of the GW detectors into equally weighted $N_b$ logarithmically spaced positive bins ~\cite{Gowling:2021gcy}, where in each bins, there exist 
\begin{eqnarray}\label{eq:frequencybin}
    n_b \equiv \left[(f_b - f_{b-1})\tau\right],
\end{eqnarray}
with $f_b$ and $f_{b-1}$ denoting the bin boundary where $b$ runs from $0$ to $N_b$ and $\tau$ representing the observation time. The square brackets in the expression of $n_b$ denote the integer value. In this study we have made the estimations considering the signal to be mostly Gaussian. 
For Gaussian distribution the likelihood function for $\Omega_{\rm sig}(f_b,\theta)$ is defined as~\cite{Dodelson:2003ft} 
\begin{eqnarray}
    \mathscr{L} (\theta) = \prod_{b=1}^{N_b} \sqrt{\frac{n_b}{2\pi\Omega_{\rm n}(f_b)^2}}\,\, {\rm exp}\left( - \frac{n_b\left(\Omega_{\rm sig}(f_b,\theta) - \Omega_{fid}(f_b)\right)^2}{\Omega_{\rm n}(f_b)^2}\right),
\end{eqnarray}
with $\Omega_{\rm n}(f_b)^2/n_b$ denoting the variance for particular detectors. In the above equation, we have defined $\Omega_{fid}(f_b)\equiv \Omega_{\rm sig}(f_b,\theta_{\rm fid})$ where  $\theta_{\rm fid}$ represents the fiducial values of the parameters depending on the potential detection of the GW signal by the detectors. In order to maximize the likelihood ($\mathscr{L}$) it is convenient to maximize the logarithm of the likelihood, the so-called log-likelihood function or chi-square distribution, defined as 
\begin{eqnarray}\label{eq:loglikelihood}
    \mathcal{L} (\theta) \equiv ln\,(\mathscr{L} (\theta)).
\end{eqnarray}

%%%%%%%%%%%%%%%%%%%%%%%%%%%%%
\label{subsec:fisher}
%%%%%%%%%%%%%%%%%%%%%%%%%%%%%
As mentioned earlier, we have considered the frequency binning according to Eq.~\eqref{eq:frequencybin}. In terms of the log-likelihood, the Fisher matrix can be expressed as~\cite{Dodelson:2003ft}
\begin{eqnarray}
    F_{ij} = \left\langle-\,\frac{\partial^2 \mathcal{L} (\theta)}{\partial \theta_i \partial \theta_j}\right\rangle,
\end{eqnarray}
where angular brackets indicate the expectation value over the observational data and $\{\theta\}$ are the model parameters. Hence in terms of noise power spectral density ($\Omega_n$), the Fisher matrix can be expressed under Gaussian approximation, which is possible for $2n_b \gtrsim O(10^2)$, as~\cite{Dodelson:2003ft}
\begin{eqnarray}
    F_{ij}\Big|_{\rm Gauss} \;=\; \sum_{b=1}^{N_b}\frac{2n_b}{\Omega_{\rm n}^2}\frac{\partial \Omega_{\rm sig}}{\partial \theta_i}\frac{\partial \Omega_{\rm sig}}{\partial \theta_j}\, .
\end{eqnarray}
In full chi-squared distribution one can expressed the Fisher matrix in terms of Gaussian approximated Fisher matrix as~\cite{Gowling:2021gcy}
\begin{eqnarray}
    F_{ij} = \left(1 + \frac{1}{2n_b}\right) F_{ij}\Big|_{\rm Gauss}\, .
\end{eqnarray}
Note that by considering $N_b=100$, we have set  $2n_b \gtrsim O(10^2)$ that ensures Gaussian distribution and allows us to approximate $F_{ij}$ as $F_{ij}\Big|_{\rm Gauss}$. Hence in terms of GW energy spectrum density, we can express the Fisher matrix under Gaussian approximation as
\begin{eqnarray}\label{eq:fisher}
    F_{ij} \;=\; \tau\sum_{b=1}^{N_b}\frac{2\Delta f_b}{\Omega_{\rm n}^2}\frac{\partial \Omega_{\rm sig}}{\partial \theta_i}\frac{\partial \Omega_{\rm sig}}{\partial \theta_j},
\end{eqnarray}
where $\Delta f_b \equiv f_b - f_{b-1}$.
The reciprocal of the Fisher matrix, $\left[C_{ij}\right] \;\equiv\; \left[F_{ij}\right]^{-1}$,  furnishes the covariance matrix, wherein the square root of the diagonal elements ($C_{ii}$) showcases an uncertainty pertaining to each specific parameter ($\theta_i$).

%%%%%%%%%%%%%%%%%%%%%%%%%%%%%
\subsection{Fisher forecast on the parameters}
\label{subsec:fisher_analysis}
%%%%%%%%%%%%%%%%%%%%%%%%%%%%%
We are now in a position to calculate the Fisher matrix and estimate the errors on the three parameters of our interest, namely, ($\omega,\xi,n_T$) for four different GW experiments: BBO~\cite{Corbin:2005ny,Harry_2006}, DECIGO~\cite{Yagi:2011yu}, LISA~\cite{amaroseoane2017laser,Baker:2019nia} and ET~\cite{Punturo_2010,Hild:2010id}. The analysis are done for the operating duration of the GW experiments of 4 years (\textit{i.e.} $\tau=4$ years) and we have fixed $H_{\rm inf} = 5.5\times 10^{12}$ GeV, as discussed in Sec.~\ref{subsec:snr}, in detail.
We have used publicly available Fisher forecast analysis code, \texttt{CosmicFish}~\cite{Raveri:2016xof,Raveri:2016leq} to plot the results. 

\begin{figure*}[]
    \centering
    \begin{subfigure}{.40\textwidth}
    \includegraphics[width=\textwidth]{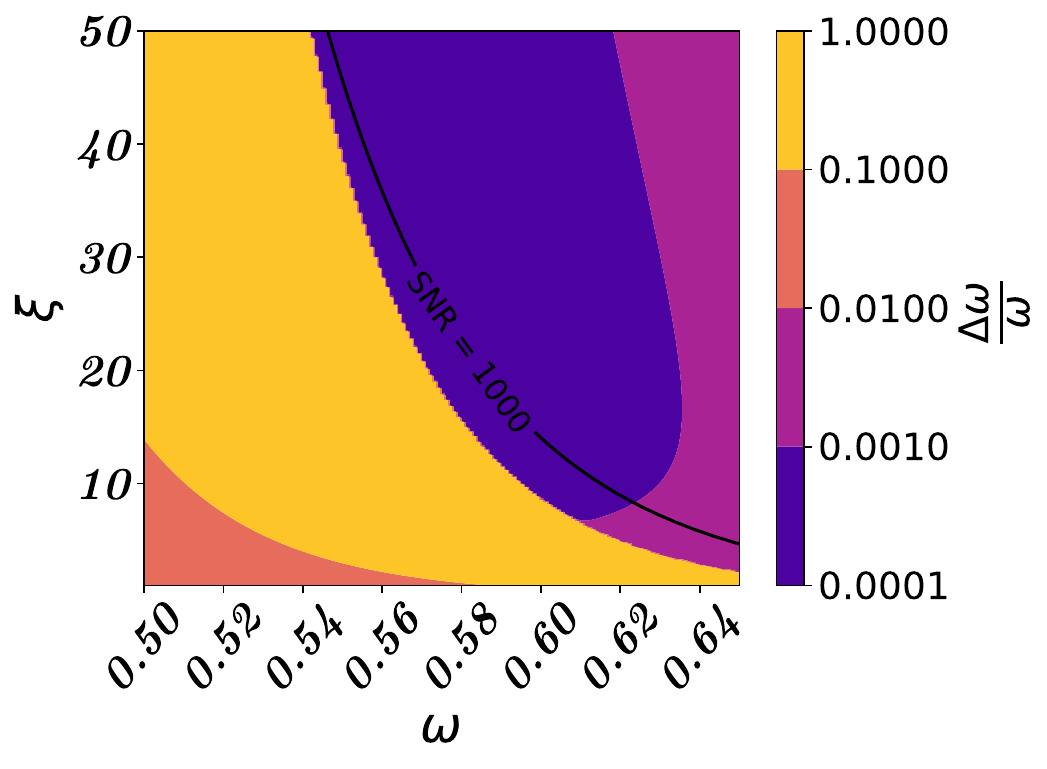}
    \caption{\it BBO (for $\nt = 0$)}
    \label{fig:fisher_deltaw_BBO}
    \end{subfigure}
    \hfill
    \begin{subfigure}{.40\textwidth}
    \includegraphics[width=\textwidth]{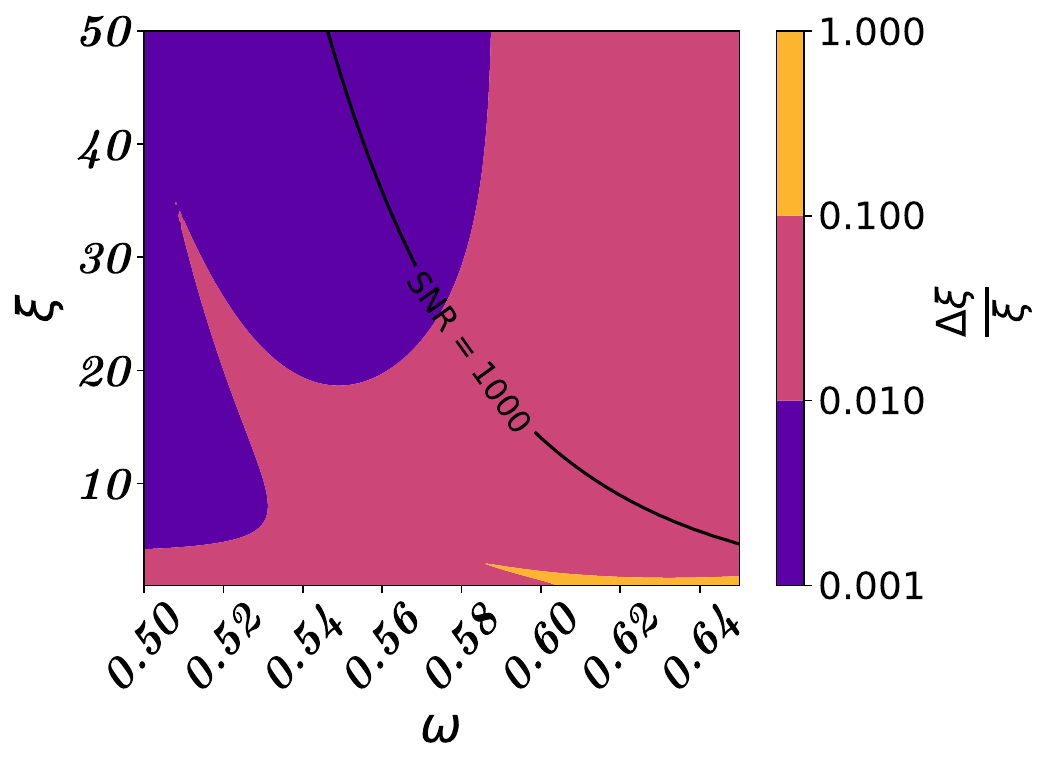}
    \caption{\it BBO (for $\nt = 0$)}
    \label{fig:fisher_deltaxi_BBO}   
    \end{subfigure}
    \begin{subfigure}{.40\textwidth}
    \includegraphics[width=\textwidth]{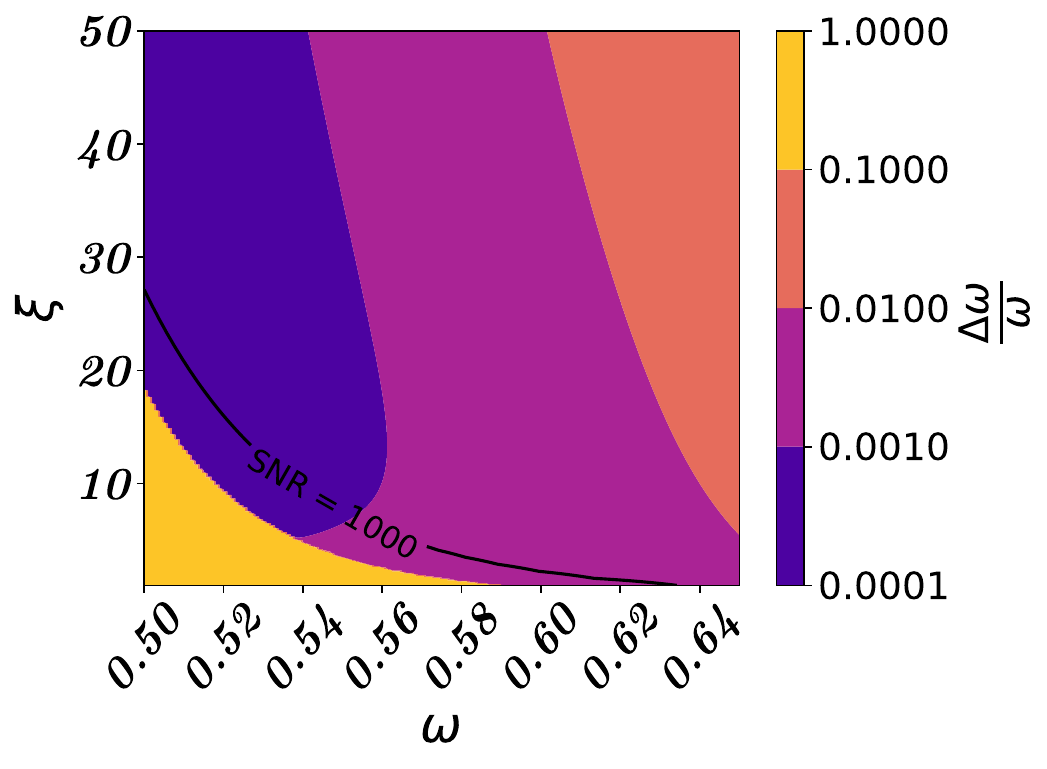}
    \caption{\it DECIGO (for $\nt = 0$)}
    \label{fig:fisher_deltaw_DECIGO}
    \end{subfigure}
    \hfill
    \begin{subfigure}{.40\textwidth}
    \includegraphics[width=\textwidth]{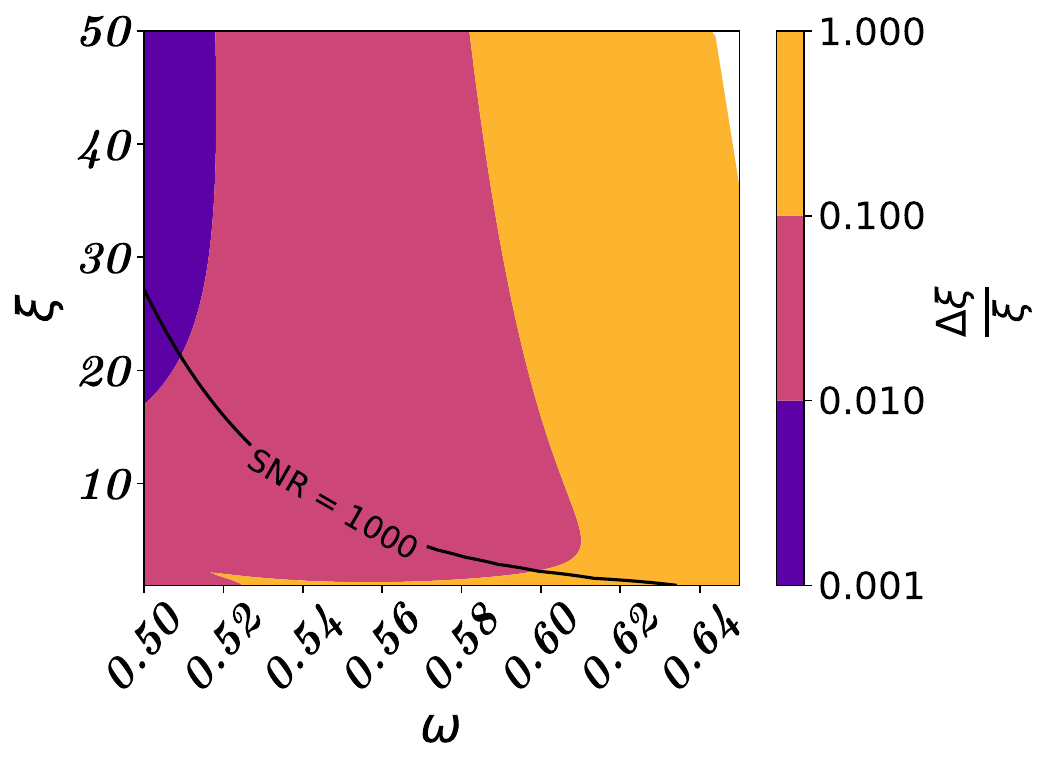}
    \caption{\it DECIGO (for $\nt = 0$)}
    \label{fig:fisher_deltaxi_DECIGO}   
    \end{subfigure}
    \begin{subfigure}{.40\textwidth}
    \includegraphics[width=\textwidth]{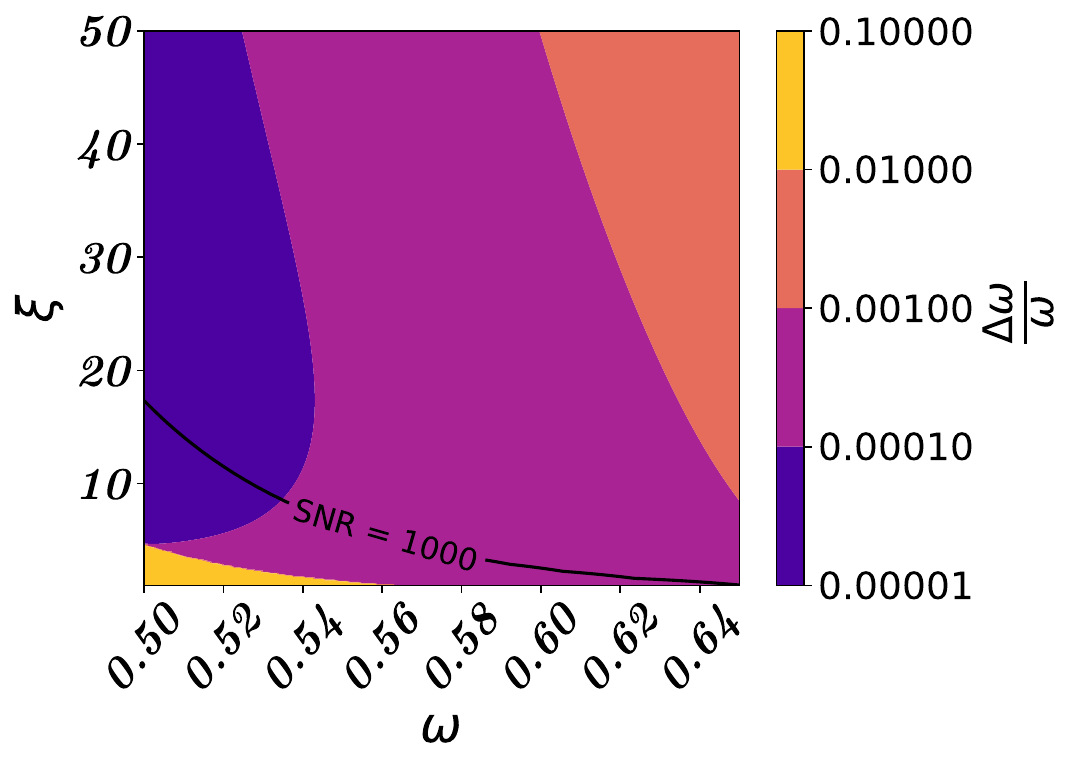}
    \caption{\it ET (for $\nt = 0$)}
    \label{fig:fisher_deltaw_ET}
    \end{subfigure}
    \hfill
    \begin{subfigure}{.40\textwidth}
    \includegraphics[width=\textwidth]{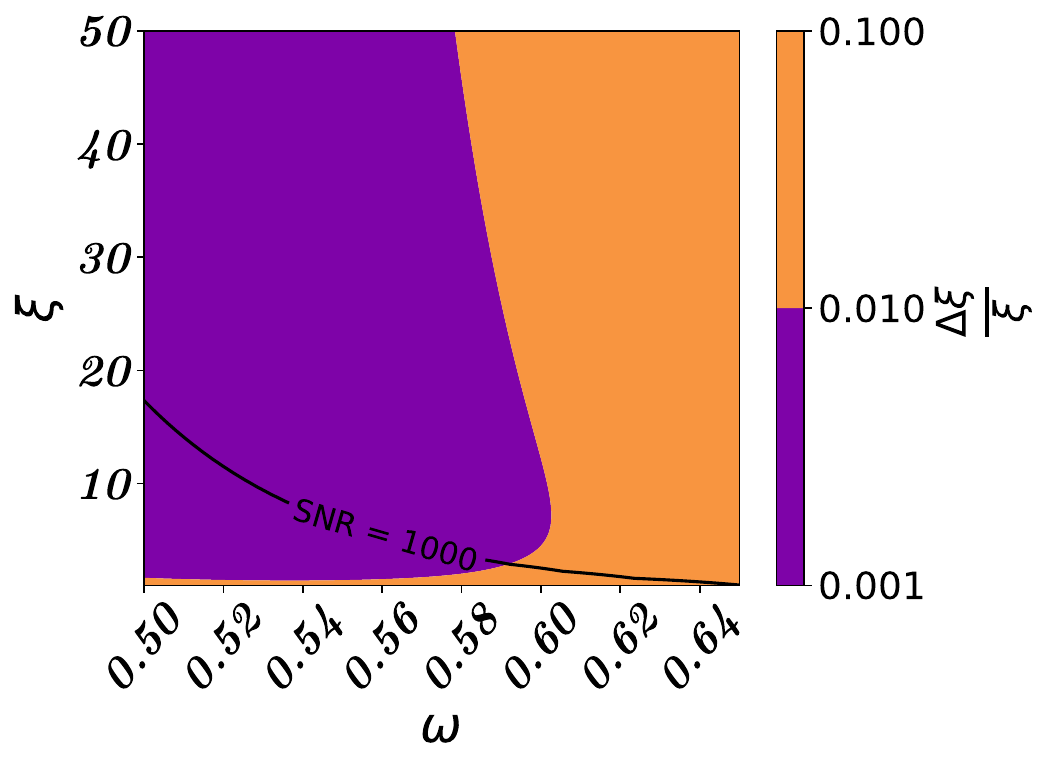}
    \caption{\it ET (for $\nt = 0$)}
    \label{fig:fisher_deltaxi_ET}   
    \end{subfigure}
    \begin{subfigure}{.40\textwidth}
    \includegraphics[width=\textwidth]{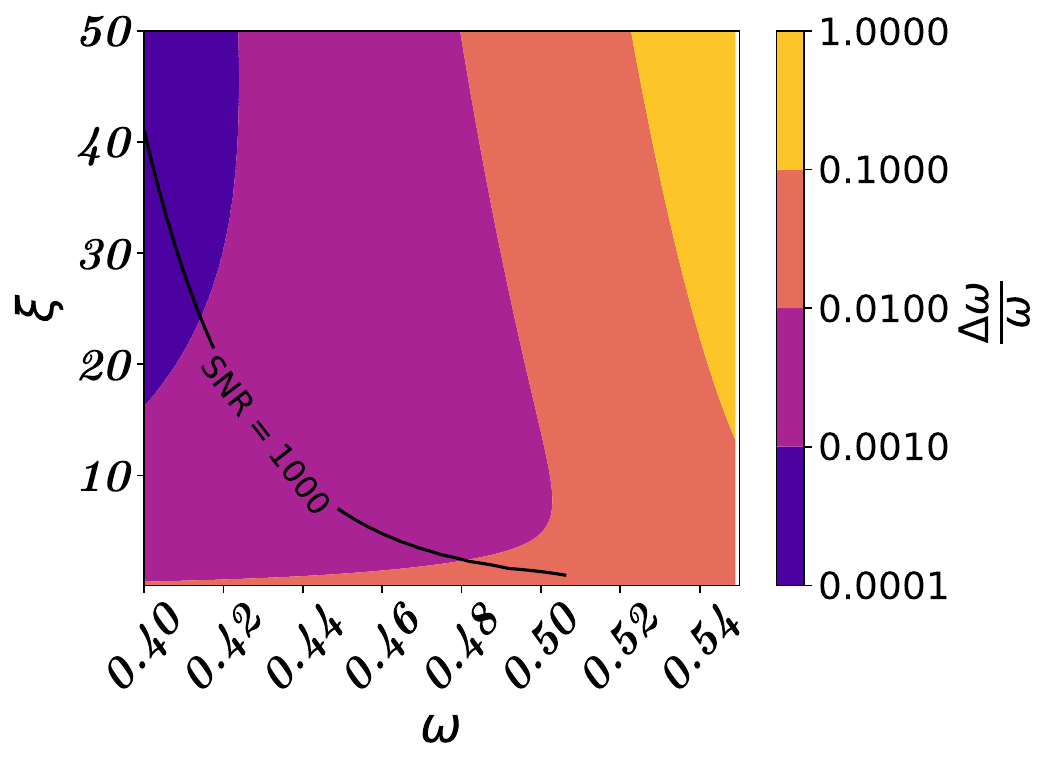}
    \caption{\it LISA (for $\nt = 0.3$)}
    \label{fig:fisher_deltaw_lisa}
    \end{subfigure}
    \hfill
    \begin{subfigure}{.40\textwidth}
    \includegraphics[width=\textwidth]{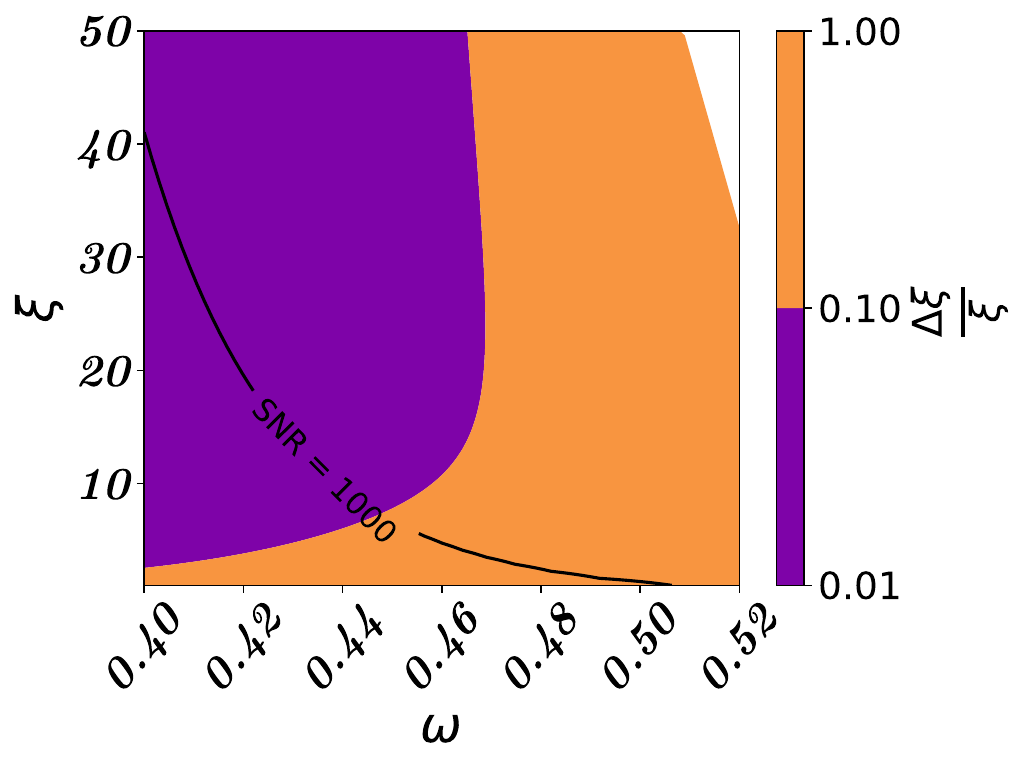}
    \caption{\it LISA (for $\nt = 0.3$)}
    \label{fig:fisher_deltaxi_lisa}   
    \end{subfigure}
    \caption{\it The contours represent the relative uncertainty for the parameters for $4$ years of observations, calculated by Fisher matrix analysis. Each of the pair displays impacts of each of the detectors namely, BBO, DECIGO, ET and LISA respectively. In this analysis we have set $H_{\rm inf} = 5.5\times 10^{12}$ GeV. The black solid line represents the contours for SNR $=1000$ for BBO, DECIGO, ET and LISA.}
    \label{fig:fisher}
\end{figure*}

In Fig.~\ref{fig:fisher} we have represented the relative uncertainties on $\omega$ and $\xi$ for a specific value of $\nt$, chosen based on its detection at the detectors. The figure consists of \textit{left} and \textit{right} panel, illustrating the relative uncertainties on $\omega$ and $\xi$, respectively, in the $\omega$-$\xi$ plane. For BBO~\cite{Corbin:2005ny,Harry_2006}, DECIGO~\cite{Yagi:2011yu} and ET~\cite{Punturo_2010,Hild:2010id}, we have set $\nt=0$ and for  LISA~\cite{amaroseoane2017laser,Baker:2019nia}, it is set to $0.3$. In the subplots of Fig.~\ref{fig:fisher}, we have overlaid the SNR $=1000$ contour using a black solid line. Comparing with Fig.~\ref{fig:snr}, we have found that the underlying portion of the contour represents SNR $>1000$ region (follow Fig.~\ref{fig:snr_contour}).
Also Fig.~\ref{fig:snr} depicts that all the region, depicted in Fig.~\ref{fig:fisher}, are greater than the detection threshold (\textit{i.e.} SNR $>10$). 
At the same time it is interesting to note that the figure depicts the sensitive regions, in terms of SNR value, retains the relative uncertainty on the parameters which is less than unity, indicating high degree of confidence in estimating the parameters. For an example, considering BBO~\cite{Corbin:2005ny,Harry_2006}, for $\omega = 0.50$ and $\xi = 10$, relative uncertainty on $\omega$, i.e.,($\frac{\Delta\omega}{\omega}$) is less than $10\%$ and on $\xi$, i.e., ($\frac{\Delta\xi}{\xi}$) is less than $1\%$.

\begin{figure*}
    \centering
    \begin{subfigure}[!ht]{.48\textwidth}
        \includegraphics[width=\textwidth]{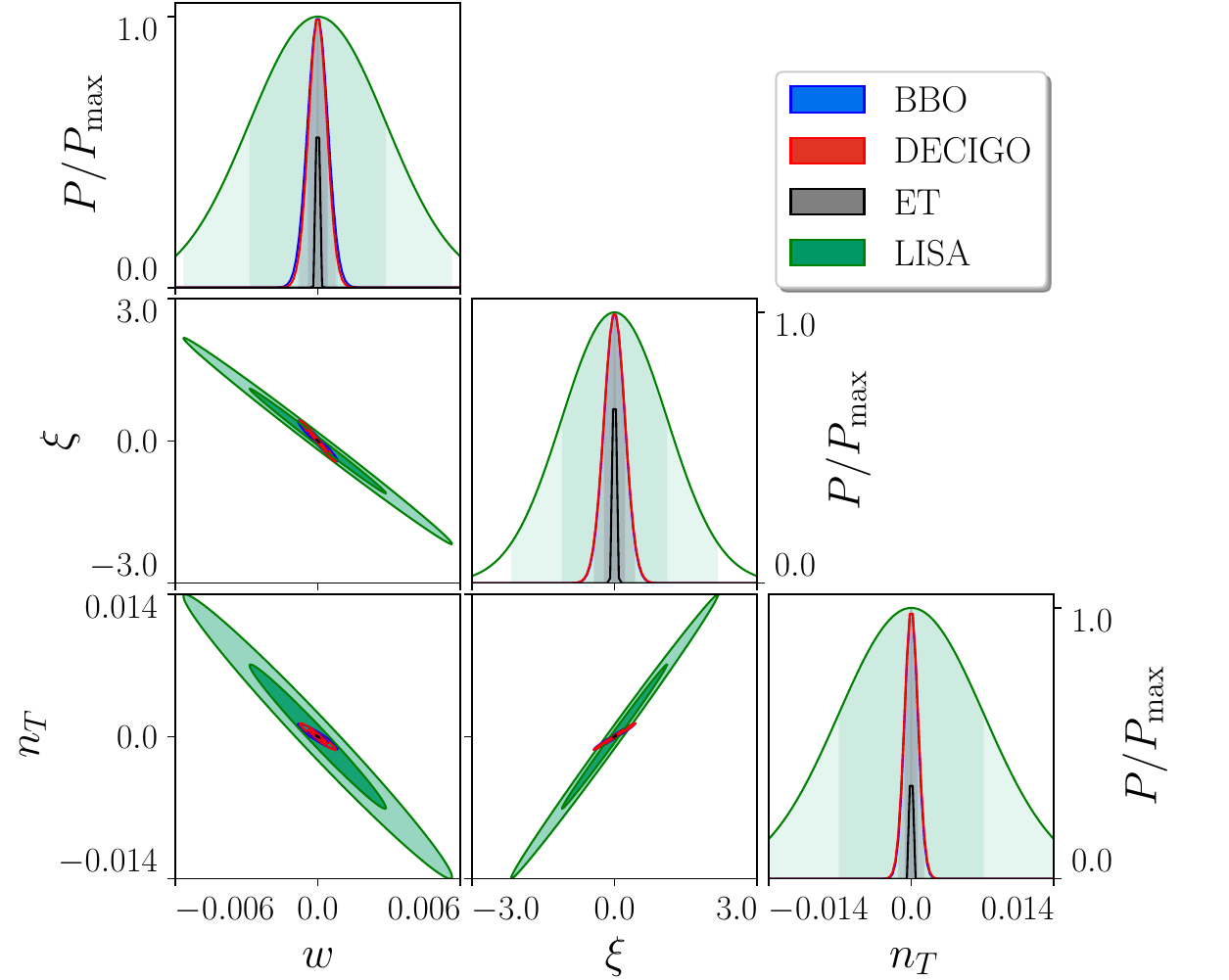}
        \caption{\it 1D and 2D distribution from Fisher matrix analysis for BBO, DECIGO, ET and LISA, \textbf{without displaying the fiducial positions}.} 
        \label{fig:fisher_all}
    \end{subfigure}%
    \hfill
    \begin{subtable}[!ht]{0.42\textwidth}
        \centering
        \renewcommand{\arraystretch}{1.1}
        \begin{tabular}{c| c| c| c}
        \hline \hline
        \textit{Param} & \textit{Detectors} & \textit{Mean} & 1-$\sigma$ \\
        \hline
        \multirow{4}{*}{$\omega$} & BBO & 0.60 & 0.00043 \\
         & DECIGO & 0.54 & 0.00038 \\
         & ET & 0.54 & 0.00006 \\
         & LISA & 0.40 & 0.00286 \\
        \hline
        \multirow{4}{*}{$\xi$} & BBO & 10 & 0.2183 \\
         & DECIGO & 10 & 0.2218 \\
         & ET & 10 & 0.0320 \\
         & LISA & 5 & 1.1035 \\
        \hline
        \multirow{4}{*}{$\nt$} & BBO & 0 & 0.00064 \\
         & DECIGO & 0 & 0.00067 \\
         & ET & 0 & 0.00010 \\
         & LISA & 0.1 & 0.00710 \\
        \hline
        \hline
       \end{tabular}
       \caption{\it Mean and 1-$\sigma$ values for the parameters from Fisher matrix analysis for BBO, DECIGO, ET and LISA.}
       \label{tab:fisher_tab}
    \end{subtable}%
    \vspace{0.7cm}
    \begin{subfigure}[!ht]{0.85\textwidth}
        \includegraphics[width=\textwidth]{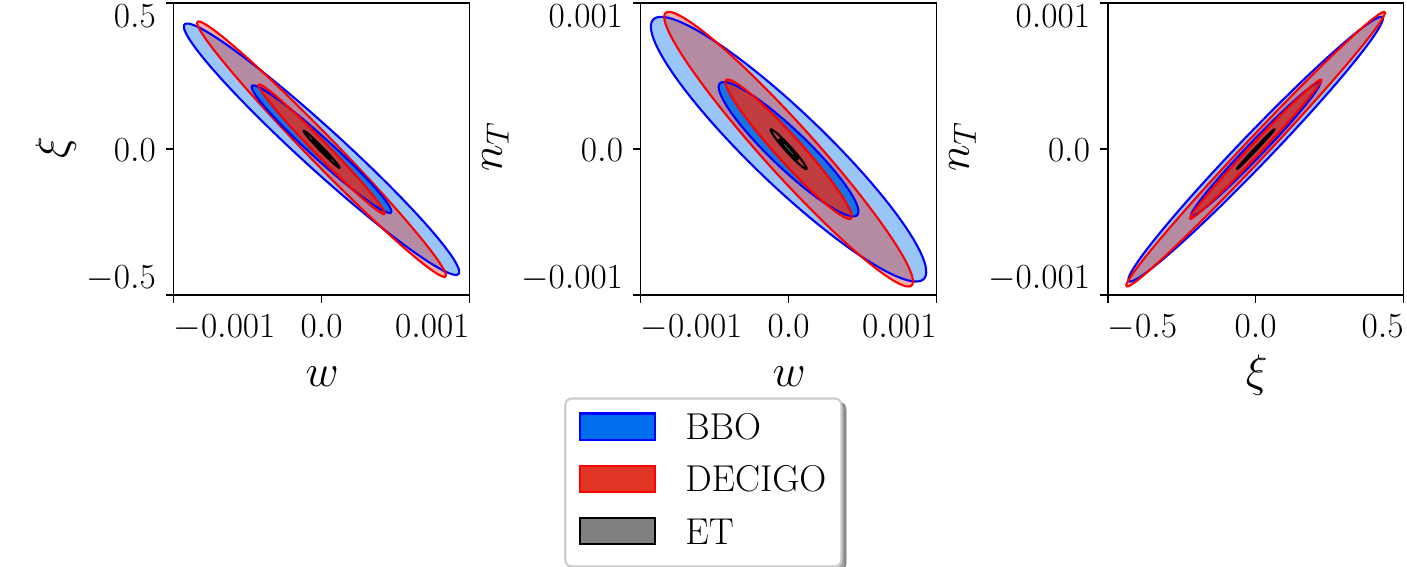}
        \caption{\it 2D distribution from Fisher matrix analysis only for BBO, DECIGO and ET \textbf{without displaying the fiducial positions} to particularly depict the correlation among them. As the 1-$\sigma$ error is very high for LISA, we have only displayed the three detectors to show the correlations.}
        \label{fig:fisher_three}
    \end{subfigure}%
    \caption{\it The figure illustrates the marginalized 1-$\sigma$ and 2-$\sigma$ contours, along with 1-dimensional posterior distribution, from Fisher matrix analysis for BBO, DECIGO, ET and LISA. We compare the correlations among the microphysics parameters for each detectors. The accompanying table on the right displays the mean and 1-$\sigma$ values for the corresponding parameters, for each GW detectors.}
    \label{fig:fisher_triplot}
\end{figure*}

Also in Fig.~\ref{fig:fisher_triplot} we have demonstrated the 1-$\sigma$ and 2-$\sigma$ estimation on the errors in measurements of the parameters for all the detectors. To highlight only on the correlations among the parameters, we have not displayed the fiducial positions in the figure, rather we have displayed the fiducials and corresponding uncertainties in the associated table of Fig.~\ref{fig:fisher_triplot}. Since Fisher analysis only deals with the uncertainties, we need to do a rigorous MCMC analysis to get the combined inference on the parameters.

%%%%%%%%%%%%%%%%%%%%%%%%%%%%%
\section{Inference on the parameters}
\label{sec:parameter_inference}
%%%%%%%%%%%%%%%%%%%%%%%%%%%%%
As mentioned in the previous section, Fisher matrix does not give us a good idea on the combined inference on the parameters under consideration. In order to do this, we need to do a rigorous MCMC analysis for GW detectors, BBO~\cite{Corbin:2005ny,Harry_2006}, DECIGO~\cite{Yagi:2011yu}, LISA~\cite{amaroseoane2017laser,Baker:2019nia} and ET~\cite{Punturo_2010,Hild:2010id}. To analyze the parameters for each of the detectors, we have considered the log-likelihood function as described in Eq.~\eqref{eq:loglikelihood}. By combining the log-likelihood with different priors (to be discussed in the following section), we have sampled the parameter space using MCMC algorithm, in order to obtain estimations on the parameter values with uncertainties associated to them and correlations among them. To perform this analysis, we have considered the equally weighted $100$ logarithmically spaced bins for the frequency range of the GW detectors, with observation time, $\tau = 4$ years (discussed in detailed in Sec.~\ref{subsec:likelihood}). For all of the detectors, we have set the scale of inflation ($H_{\rm inf}$) to be $5.5\times 10^{12}$ GeV, as discussed in Sec.~\ref{subsec:snr}, in detail. To perform this analysis we have used publicly available MCMC analysis code \texttt{emcee}~\cite{2013PASP..125..306F} and the resulting chains are analyzed using \texttt{GetDist}~\cite{Lewis:2019xzd} package to get the posterior distributions of the parameters.

%%%%%%%%%%%%%%%%%%%%%%%%%%%%%
\subsection{MCMC analysis for GW detectors}
\label{subsec:prior}
%%%%%%%%%%%%%%%%%%%%%%%%%%%%%
In the present analysis we have three parameters in our hand: equation of state during the period of reheating ($\omega$), non-minimal coupling parameter ($\xi$) and tensor spectral index ($\nt$). The priors on the parameters are chosen mainly based on the two things: \textit{(i)} detectability of the signal at the detectors and \textit{(ii)} the corresponding SNR $>10$.

For all the detectors, we have chosen flat priors on the parameters which are tabulated in Table~\ref{tab:prior}. We have chosen the same priors of DECIGO~\cite{Yagi:2011yu} and ET~\cite{Punturo_2010,Hild:2010id} because the parameter space, for which the signal can be detected by the detectors with reasonable SNR, are very close (discussed in Sec.~\ref{subsec:snr})~\footnote{In this analysis we have included only these detectors, due to their high capabilities in detecting the signals as the signal is mostly noise dominated for LISA~\cite{amaroseoane2017laser,Baker:2019nia}.}. 

\begin{table}[!ht]
    \centering
    \renewcommand{\arraystretch}{1.2}
    \begin{tabular}{c|c|c}
    \hline
    \hline
        \multirow{2}{*}{\textit{Parameter}} & \multicolumn{2}{|c}{\textit{Prior}}\\
        \cline{2-3}
        & BBO & DECIGO \textit{and} ET\\
        \hline
        $\omega$ & Flat, 0.45 $\rightarrow$ 0.70 & Flat, 0.4 $\rightarrow$ 0.60 \\
        $\xi$ &  Flat, 1 $\rightarrow$ 30 &  Flat, 1 $\rightarrow$ 30\\
        $n_T$ &  Flat, -0.2 $\rightarrow$ 0.1 &  Flat, -0.2 $\rightarrow$ 0.15\\
    \hline
    \hline
    \end{tabular}
    \caption{\it Prior range for the 3 model parameters for for each of the detectors}
    \label{tab:prior}
\end{table}

\begin{figure*}
    \centering
    \begin{subfigure}{.43\textwidth}
        \includegraphics[width=\textwidth]{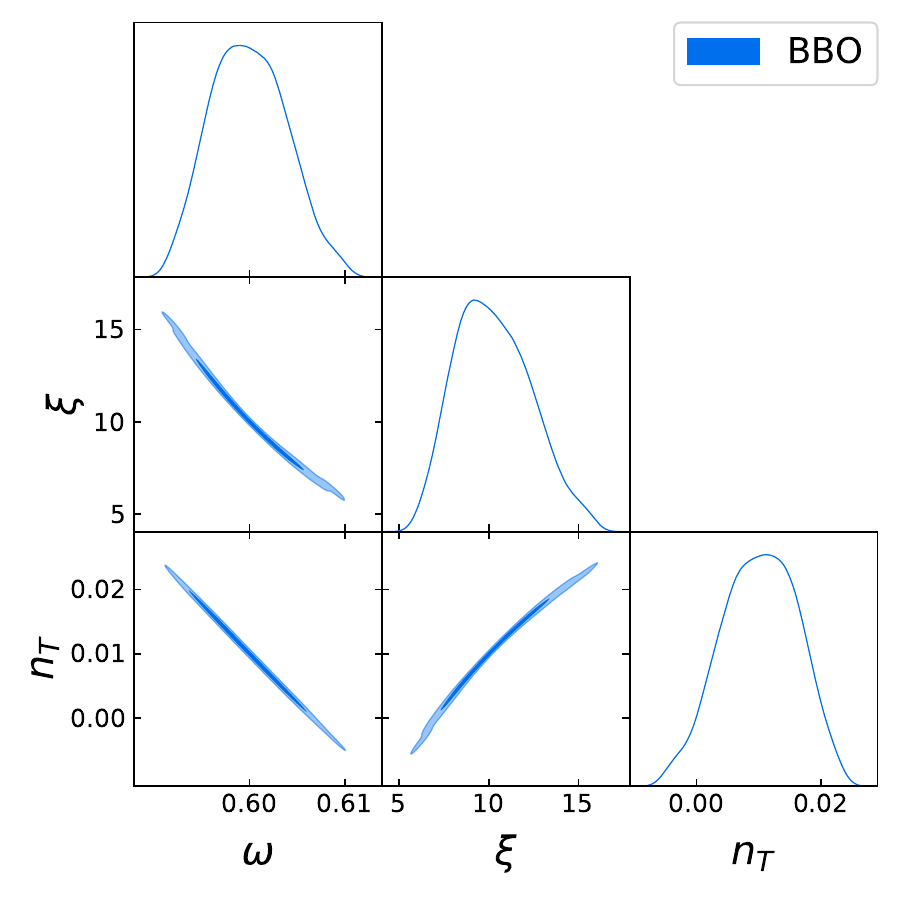}
        \caption{\it For BBO}
        \label{fig:mcmc_BBO}
    \end{subfigure}
    \begin{subfigure}{.43\textwidth}
        \includegraphics[width=\textwidth]{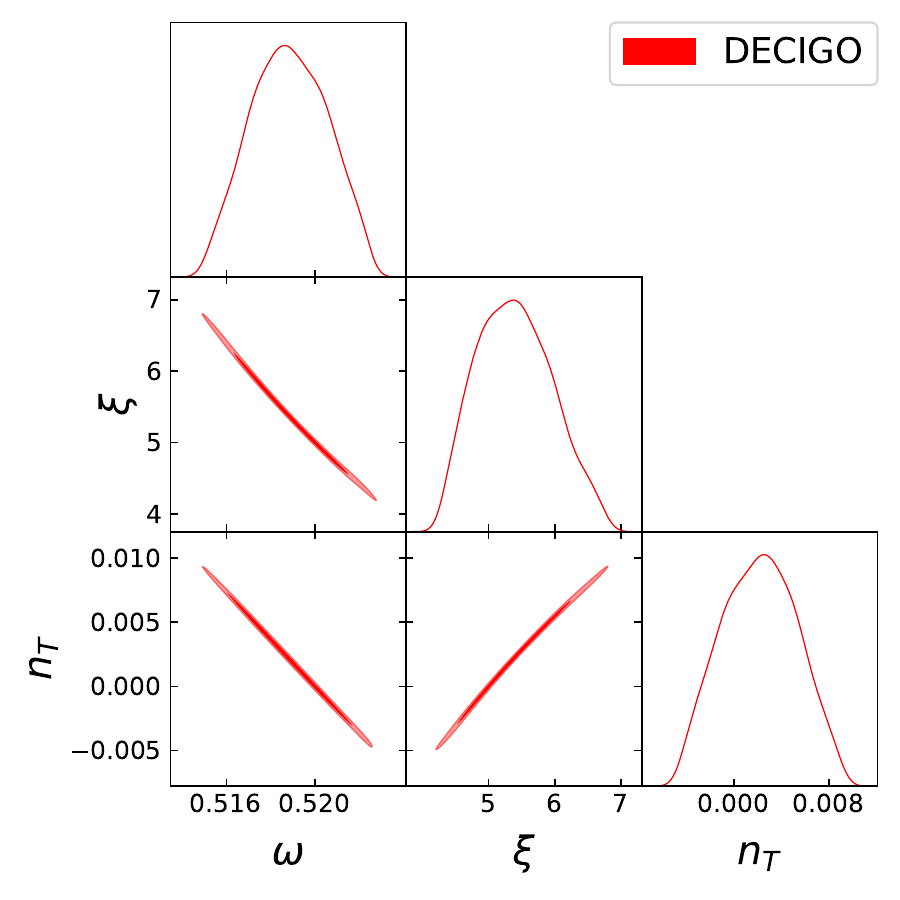}
        \caption{\it For DECIGO}
        \label{fig:mcmc_DECIGO}
    \end{subfigure}
    \begin{subfigure}{.43\textwidth}
        \includegraphics[width=\textwidth]{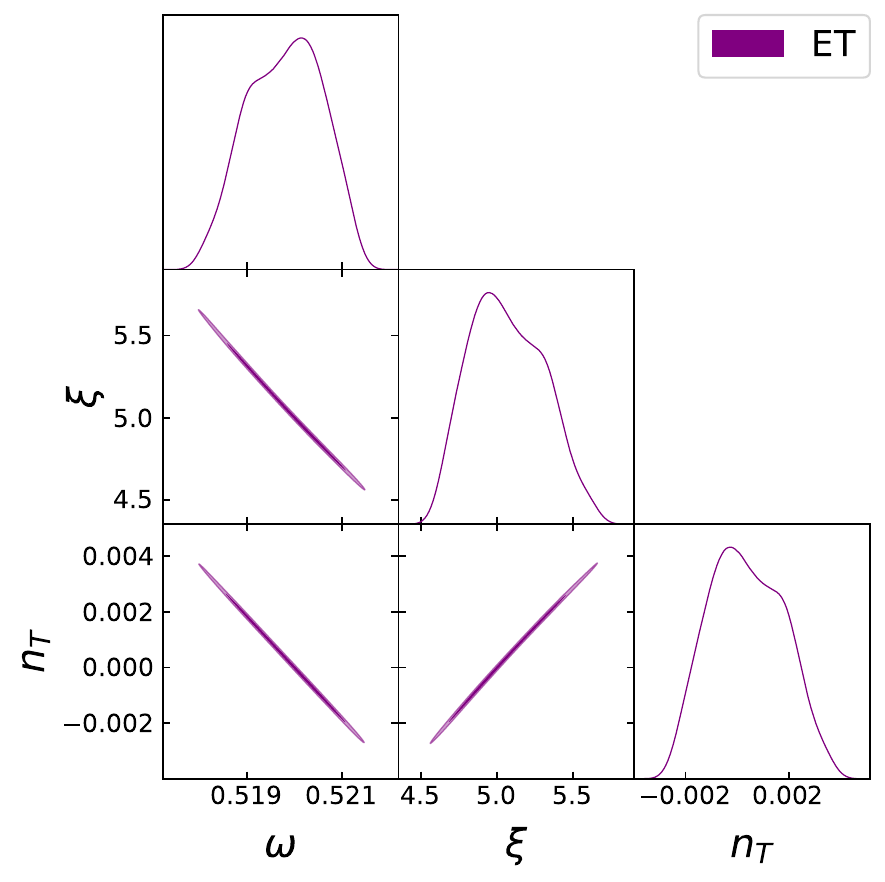}
        \caption{\it For ET}
        \label{fig:mcmc_ET}
    \end{subfigure}
    \caption{\it 2-D posterior distributions for the 3 model parameters ($\omega,\xi,n_T$) considering BBO, DECIGO and ET. LISA has not shown over here as the signal is mostly noise dominated for LISA.}
    \label{fig:mcmc}
\end{figure*}

%%%%%%%%%%%%%%%%%%%%%%%%%%%%%
\subsection{Constraints on the parameters and discussions}
\label{subsec:mcmc_discussion}
%%%%%%%%%%%%%%%%%%%%%%%%%%%%%
In Fig.~\ref{fig:mcmc} we have presented the 2-dimensional posterior distribution of the parameters of the model under consideration using the mock data generated for different detectors. MCMC analysis helps us to understand the combined inference on the parameters. The uncertainties on the parameters thus estimated have been summarized in Table~\ref{tab:mcmc_results}.  The 2-dimensional posterior distribution (Fig.~\ref{fig:mcmc}) depicts that $\omega$ is negatively correlated with $\xi$ and $\nt$, whereas $\xi$ and $\nt$ are in positive correlation, for all the three detectors. This is physical which can be interpreted as the individual behaviors of the parameters. Fig.~\ref{fig:gwnt} depicts that the GW signal enhances for the smaller values of $\omega$ whereas the signal dampens due to the larger values of $\xi$. Hence for the detection of the signal at any detector, higher values of $\omega$ needs smaller $\xi$ and vice versa, which justifies the reason for the correlation between them. On the other hand, positive $\nt$ enhances the spectrum (Fig.~\ref{fig:gw_nt5}), which depicts that large $\xi$ needs higher $\nt$ in order to detect the signal by any detectors. This explains that these two parameters should be in positive correlation for all the detectors which we have got from MCMC analysis (in Fig.~\ref{fig:mcmc}).

Nevertheless, from the results it appears that among the four interferometer missions under consideration, that ET~\cite{Punturo_2010,Hild:2010id} has the potential to constrain the  parameters better compared to the other detectors. Of course, in order to comment conclusively on that, one needs to wait for the real data to arrive. However the present analysis, done with the help of realistic mock catalogs, indeed shows the early trend that serves as a pointer in this topic of research.

\begin{table}[!ht]
    \centering
    \renewcommand{\arraystretch}{1.4}
    \begin{tabular}{c|c|c|c}
    \hline
    \hline
        \multirow{2}{*}{\textit{Parameter}} & \multicolumn{3}{|c}{\textit{mean$\pm\,\sigma$}}\\
        \cline{2-4}
        & BBO & DECIGO & ET\\
        \hline
        $\omega$ & 0.6000$^{+0.0047}_{-0.0042}$ & 0.5188$^{+0.0019}_{-0.0018}$ & 0.5198$^{+.0008}_{-0.0009}$ \\
        $\xi$ &  10.040$^{+2.446}_{-1.940}$ &  5.377$^{+0.590}_{-0.633}$ &  5.049$^{+0.288}_{-0.241}$\\
        $n_T$ &  0.0106$^{+0.0066}_{-0.0068}$ & 0.0022$^{+0.0033}_{-0.0035}$ &  0.0003$^{+0.0017}_{-0.0015}$\\
    \hline
    \hline
    \end{tabular}
    \caption{\it Statistical results for the three parameters for each of the detectors}
    \label{tab:mcmc_results}
\end{table}

\medskip

%%%%%%%%%%%%%%%%%%%%%%%%%
\section{Gravitational Production of cosmological relics}
\label{sec:production}
%%%%%%%%%%%%%%%%%%%%%%%%%

The aforementioned sections are mainly focused on the production of PGWs during the period of inflation, including the calculation of SNR and the estimation of uncertainties associated with the parameters governing the generation of the GWs signal and on gravitational reheating. In this section we will delve into the intricate processes, governing the production of DM via gravity-mediated mechanism as well as gravitational leptogenesis. Our discussion will encompass various scenarios, including the production of RHN DM~\cite{Barman:2022qgt}, axionic DM~\cite{Barman:2023icn} and gravitational leptogenesis processes~\cite{RT_2022,Barman:2022qgt}. Our aim is to elucidate the dependence of the parameters, responsible for sufficient relic abundance of DM as well as sufficient leptogenesis with high SNR. In particular, we will highlight how parameter-regions, yielding high SNR for the respective detectors, are capable of probing both DM and leptogenesis.

%%%%%%%%%%%%%%%%%%%%%%%%%%%
\subsection*{Model framework}
\label{sec:framework}
%%%%%%%%%%%%%%%%%%%%%%%%%%%
In the linearized gravity, the background space-time metric can be characterized by $g_{\mu\nu} \simeq \eta_{\mu\nu} + 2\frac{h_{\mu\nu}}{M_P}$, where $\eta_{\mu\nu}$ is the Minkowski metric ($=diag[1,-1,-1,-1]$) and $h_{\mu\nu}$ is the canonically normalized tensor perturbation. The ensuing gravitational interactions can be succinctly expressed as~\cite{Clery:2021bwz}
\begin{eqnarray}
    \mathcal{L}_{int} = -\frac{1}{M_P}\frac{h_{\mu\nu}}{\sqrt{-g}}\left(T^{\mu\nu}_{SM} + T^{\mu\nu}_{\mathcal{N}_i} + T^{\mu\nu}_{\phi}\right), \label{Eq:lagrangian}
\end{eqnarray}
where $SM, \, \mathcal{N}_i\, \text{and} ~\phi$ represent the standard model field, RHNs and inflaton field, respectively. Here the stress-energy tensor, $T_j^{\mu\nu}$, depends on the spin of the fields ($j=0,\frac{1}{2},1$) and can be expressed as following~\cite{Clery:2021bwz}
\begin{dgroup}[compact]
\begin{dmath}
    T^{\mu\nu}_0 = \partial^\mu S \partial^\nu S - g^{\mu\nu}\left[\frac{1}{2}\partial^\alpha S \partial_\alpha S - V(S)\right],
\end{dmath}
\begin{dmath}
    T^{\mu\nu}_{1/2} = \frac{i}{8}\left[\Bar{\chi}\gamma^\mu{\overleftrightarrow {\partial^\nu}}\chi + \Bar{\chi}\gamma^\nu{\overleftrightarrow {\partial^\mu}}\chi \right] - g^{\mu\nu}\left[\frac{i}{4}\Bar{\chi}\gamma^\alpha{\overleftrightarrow {\partial_\alpha}}\chi - \frac{m_\chi}{2}\Bar{\chi}\chi\right],
\end{dmath}
\begin{dmath}
    T^{\mu\nu}_1 = \frac{1}{2}\left[F_{\alpha}^{\mu}F^{\nu\alpha} + F_{\alpha}^{\nu}F^{\mu\alpha} - \frac{1}{2}g^{\mu\nu}F^{\alpha\beta}F_{\alpha\beta}\right],
\end{dmath}
\end{dgroup}
where $\chi$ is the fermionic field and $F_{\mu\nu} = \partial_{\mu} A_{\nu} - \partial_{\nu} A_{\mu}$, with $A_{\mu}$ being the vector field. $V(\phi)$ is the potential for the scalar field, $S$. Our study is quite general from the point of view of $V(\phi)$ as our study does not depend on the exact form of the potential but the behavior at the proximity of the minima of the potential which is $\sim \phi^n$ (Eq.~\eqref{Eq:potmin}).

\begin{figure}[ht!]
\centering
\includegraphics[scale=0.6]{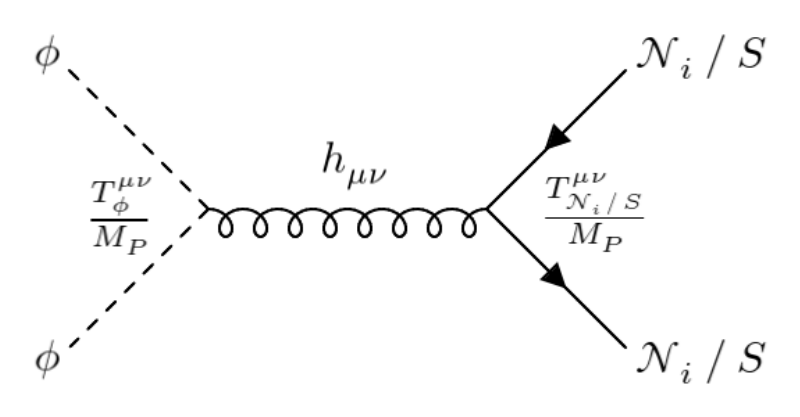}
\caption{\it \em Feynman diagram for the production of RHNs and SM particles from the scattering of the inflaton condensate via s-channel graviton exchange. Here, $\phi$, $\mathcal{N}_i$ and $S$ represent inflaton, RHNs and SM particles, respectively.}
\label{fig:feynman}
\end{figure}

Our study is mainly focused on the $2\to 2$ gravitational scattering, the sole channel for the DM production via the scattering of the inflaton condensate and from the SM thermal bath here. 
Fig.~\ref{fig:feynman} illustrates the production of DM and SM from inflaton via s-channel graviton exchange, corresponding to the Lagrangian in Eq.~\ref{Eq:lagrangian}. 

%%%%%%%%%%%%%%%%%%%%%%%%%%%%%%%%%%%%%%%
%%%%%%%%%%%%%%%%%%%%%%%%%%%%%%%%%

In our study, to address DM and gravitational leptogenesis simultaneously, we have assumed three RHNs, following Ref.~\cite{Barman:2022qgt}. We have considered the most minimal extension of the SM of particle physics where the SM is augmented with 3 RHNs which participates in seesaw type generate SM neutrino masses~\cite{10.1143/PTP.64.1103,PhysRevLett.44.912,FUKUGITA198645}. In order to address the DM puzle, we have considered that the lightest RHN decouples from the other two RHNs and has a vanishing Yukawa coupling, due to $Z_2$ odd symmetry, making it a stable DM candidate. The other two heavy RHNs generate the mass for the SM neutrinos via the type-I vanilla seesaw mechanism. These decay of RHNs into the active neutrino can address baryon asymmetry in the Universe via lepton asymmetry~\footnote{This baryogenesis can also be addressed considering gravitaional neutrino reheating, pointed out earlier in Ref.~\cite{Haque:2023zhb}}.

%%%%%%%%%%%%%%%%%%%%%%%%%%%%%%
\subsection{Gravitational production of dark matter}
\label{subsec:grav-prod-dm}
%%%%%%%%%%%%%%%%%%%%%%%%%%%%%
For gravitational production of DM, we have considered that the least massive RHN which is stable to be a DM candidate, as stated earlier. Now the DM (the least massive RHN) can be produced in two possible ways : \textit{(i)} from the thermal bath of SM and \textit{(ii)} from the direct scattering of inflaton. Both the processes are mediated via graviton. If the mass of the DM candidate be $M_{\rm DM}$, we can express the DM number density at the end of reheating ($\arh$ being the scale factor of the Universe at this time), generated due to the gravitational scattering of the SM particles in the thermal bath for $\arh \gg \aend$, as~\cite{Barman:2022qgt}
\begin{eqnarray}
\label{eq:n_T}
    n_{\rm DM}^T (\arh) \simeq \frac{\beta_{1/2}\,\rho_{\rm RH}^{3/2}}{(\frac{\pi^2 g_{\rm RH}}{30})^2}\,\frac{(7-4n)^2\, (n+2)}{6\,\sqrt{3}\,(n+5)(n-1)(5n-2)}\,\left(\frac{\arh}{\aend}\right)^{\frac{10+2n}{n+2}},
\end{eqnarray}
where $\beta_{1/2} = 3.4\times 10^{-2}$~\cite{Cl_ry_2022}, with $\aend$ being the scale factor at the end of inflation. The radiation energy density of the Universe at the end of reheating, $\rho_{\rm RH}$, can be expressed as $\rho_{\rm RH} = \frac{\pi^2  g_{\rm RH}}{30}\, \Trh^4$. Hence the relic density of the DM, generated from the thermal bath, can be expressed as~\cite{Barman:2022qgt}
\begin{eqnarray}
\label{eq:DM_plasma}
\Omega_{\rm DM}^T\,h^2     & \simeq 1.6\times 10^8\,\frac{g_0\,\beta_{1/2}}{g_{\rm RH}}\,\frac{M_{\rm DM}}{\rm GeV}\, \frac{(\frac{\pi^2 g_{\rm RH}}{30})^{-\frac{5}{6}-\frac{5}{3n}}\,(7-4n)^2\, (n+2)}{6\,\sqrt{3}\,(n+5)(n-1)(5n-2)}\left(\frac{\Trh}{M_P}\right)^{\frac{5n-20}{3n}}\left(\frac{\rho_{\rm end}}{M_P^4}\right)^{\frac{n+5}{3n}}\,,
\end{eqnarray}
with $g$ being the number of relativistic degree of freedom, where $g_0 \equiv g(T_0) = \frac{43}{11}$ and $g_{\rm RH} \equiv g(\Trh) = \frac{427}{4}$ represent the same at present-time and time of reheating, respectively. On the other hand number density of the DM at the end of reheating, generated due to the inflaton scattering, mediated via graviton, can be represented as~\cite{Barman:2022qgt}
\begin{eqnarray}
\label{eq:n_phi}
    n_{\rm DM}^{\phi} (\arh) \simeq \frac{M_{\rm DM}^2 \sqrt{3} (n+2) \rho_{\rm RH}^{\frac{1}{2} + \frac{2}{n}}}{24\,\pi\,n(n-1)\lambda^{2/n}\, M_P^{1+\frac{8}{n}}}\,\left(\frac{\rho_{\rm end}}{\rho_{\rm RH}}\right)^{\frac{1}{n}}\,\Sigma_{\rm DM}^{(n)},
\end{eqnarray}
where $\rho_{\rm RH}$ represents the energy density of radiation at the end of reheating which can be expressed as $\rho_{\rm end} = \frac{\pi^2  g_{\rm RH}}{30} \Trh^4$. The definition of $\Sigma_{\rm DM}^{(n)}$ follows Ref.~\cite{RT_2022,Cl_ry_2022,Barman:2022qgt}, tabulated in Table~\ref{tab:sigma_dm}. Hence the relic abundance of the DM for the inflaton scattering can be expressed as~\cite{Barman:2022qgt,Cl_ry_2022}
\begin{eqnarray}
\label{eq:DM_phi}
\frac{\Omega_{\rm DM}^{\phi} h^2}{0.12}&=&
\frac{\Sigma_{\rm DM}^{(n)}}{2.4^{\frac{8}{n}}}\frac{(n+2)}{n(n-1)}
\left(\frac{10^{-11}}{\lambda}\right)^{\frac{2}{n}}
\left(\frac{10^{40} {\rm GeV}^4}{\rho_{\rm RH}}\right)^{\frac{1}{4}-\frac{1}{n}}
\nonumber
\\
&\times&
\left(\frac{\rho_{\rm end}}{10^{64} {\rm GeV}^4}\right)^{\frac{1}{n}}
\left(\frac{M_{\rm DM}}{1.1\times 10^{7^+\frac{6}{n}} {\rm GeV}}\right)^3\,.
\end{eqnarray}
Therefore the total DM abundance ($\Omega_{\rm DM}^{tot}$) is basically the sum of $\Omega_{\rm DM}^T$ and $\Omega_{\rm DM}^{\phi}$.

\begin{figure*}[!ht]
    \centering
    \begin{subfigure}{.48\textwidth}
    \includegraphics[width=\textwidth]{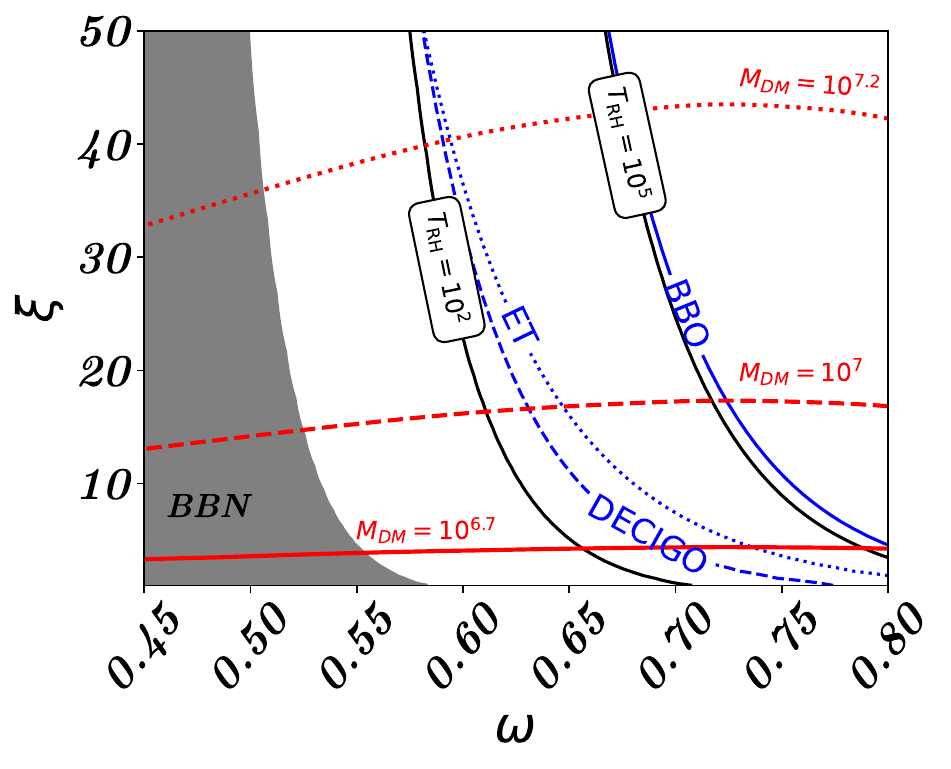}
    \caption{\it $\nt = 0$ for BBO, DECIGO and ET. All the units of $M_{\rm DM}$ and $T_{\rm RH}$ are in GeV.}
    \label{fig:snr_dm_tr_three}
    \end{subfigure}
    \hfill
    \begin{subfigure}{.48\textwidth}
    \includegraphics[width=\textwidth]{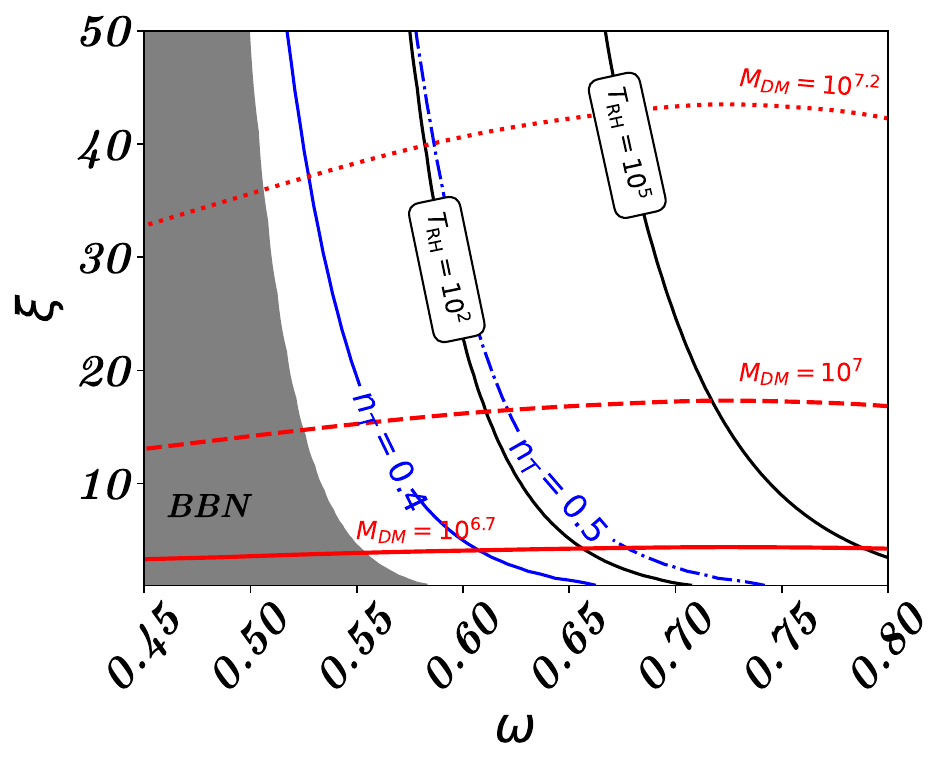}
    \caption{\it Considering $\nt=0.4$ and $\nt=0.5$ for LISA. All the units of $M_{\rm DM}$ and $T_{\rm RH}$ are in GeV.}
    \label{fig:snr_dm_tr_lisa}   
    \end{subfigure}
    \caption{\it The plots represent contours of reheating temperatures (in black) and the observed relic abundance of DM ($\Omega_{\rm DM}^{tot} = 0.12$) for different masses of DM (in red), both are in GeV. The gray shaded region, in both of the plots, indicates $\Trh < 1$ MeV which is ruled out by BBN. In \textbf{left plot} we have shown contours for SNR$=10$, for BBO (blue solid line), DECIGO (blue dashed line), ET (blue dotted line) where we have set $\nt=0$. \textbf{Right plot} showcases the SNR $=10$ contours for LISA, for $\nt=0.4$ (blue solid line) and $\nt=0.5$ (blue dash-dot line). Also we have set $H_{\rm inf} = 5.5\times 10^{12}$ GeV.}
   \label{fig:snr_dm_contour}
\end{figure*}

\begin{table}[h!]
    \centering
    \renewcommand{\arraystretch}{1.3}
    \begin{tabular}{|c||c|c|c|c|c|c|c|c|c|c|}
    \hline
       $n$ & $2$ & $4$ & $6$ & $8$ & $10$ & $12$ & $14$ & $16$ & $18$ & $20$ \\
       \hline
       $\Sigma_{\rm DM}^{(n)}$ & $\frac{1}{64}$ & $0.061$ & $0.101$ & $0.133$ & $0.157$ & $0.177$ & $0.192$ & $0.205$ & $0.216$ & $0.225$ \\
       \hline
    \end{tabular}
    \caption{Coefficient of $\Sigma_{\rm DM}^{(n)}$ for RHNs}
    \label{tab:sigma_dm}
\end{table}

\medskip

%%%%%%%%%%%%%%%%%%%%%%%%%%%%%%
\subsection{Gravitational leptogenesis}
\label{subsec:grav-lepto}
%%%%%%%%%%%%%%%%%%%%%%%%%%%%%
When the Majorana neutrinos ($\mathcal{N}$) are produced, they are unstable, they decay to left-handed neutrinos ($L$) in the following channels:
\begin{dgroup}[compact]
\begin{dmath}
    \mathcal{N} \rightarrow L + h
\end{dmath}
\begin{dmath}
    \mathcal{N} \rightarrow \Bar{L} + \Bar{h},
\end{dmath}
\end{dgroup}
with $h$ being the Higgs doublets. As discussed in the earlier section, least massive RHN provides stable DM candidate whereas the other two decay in the above channels. Here we have assumed the mass of one of the two RHNs is $M_{\mathcal{N}}$. According to Ref.~\cite{Barman:2022qgt} we have considered non-thermal leptogenesis, indicating too weak coupling between the RHNs and SM particles to reach the process in thermal equilibrium. This allows us to consider only the inflaton scattering process (mediated via graviton) to produce the RHNs for leptogenesis.

If CP is violated due to the decay of RHNs, lepton asymmetry ($Y_L$) can be expressed in terms of number density of RHNs ($n_{\mathcal{N}}$) as~\cite{Co:2022bgh,Buchmuller:2004nz}
\begin{eqnarray}
    Y_L \equiv \epsilon \frac{n_{\mathcal{N}}}{s},
\end{eqnarray}
where $s$ represents the entropy density which can be expressed as $s(T)=\frac{2}{45}\pi^2\,g(T)\,T^3$. The CP asymmetry term ($\epsilon$) in the above equation, can be expressed as~\cite{PhysRevD.45.455,Flanz:1994yx,Covi:1996wh,Buchmuller:2004nz,Davidson:2008bu}
\begin{align}\label{eq:cp}
\epsilon = \frac{\Gamma({\mathcal{N}} \rightarrow L+h)-\Gamma({\mathcal{N}} \rightarrow \Bar{L}+\Bar{h})}{\Gamma({\mathcal{N}} \rightarrow L +h)+\Gamma({\mathcal{N}} \rightarrow \Bar{L} +\Bar{h})}\, .
\end{align}
Now in terms of the produced SM neutrino mass, $m_{\nu}$, the CP asymmetry term can be expressed as~\cite{Kaneta:2019yjn,Co:2022bgh,Buchmuller:2004nz}
\begin{align}
\epsilon \simeq \frac{3 \delta_{\rm eff}}{16\,\pi}\,\frac{M_{\mathcal{N}}\,m_{\nu}}{v^2}\,,  
\end{align}
with $v\equiv \langle h \rangle \simeq 174$ GeV being vacuum expectation value of Higgs doublet. $\delta_{\rm eff}$, in the above equation, is the effective CP violating  phase change in the neutrino mass matrix, where $0\leq\delta_{\rm eff}\leq 1$. Now this lepton asymmetry can be converted into baryon asymmetry via electroweak sphaleron processes as~\cite{Co:2022bgh,Kaneta:2019yjn,Khlebnikov:1988sr,
PhysRevD.42.3344}
\begin{eqnarray}
    Y_B = \frac{28}{79} Y_L\, ,
\end{eqnarray}
which leads to~\cite{Co:2022bgh,Barman:2022qgt}
\begin{align}\label{eq:yb}
Y_B &= \frac{28}{79} \epsilon\,  \frac{n^\phi_{\mathcal{N}}(\Trh)}{s} \\
&\simeq 3.5\times 10^{-4}\, \delta_{\rm eff} \left(\frac{m_{\nu}}{0.05 \,\text{eV}}\right) \left(\frac{M_{\mathcal{N}}}{10^{13}\,\text{GeV}}\right)\,\frac{n_{\mathcal{N}}^\phi (\Trh)}{s (\Trh)}\,,
\end{align}
with $m_{\nu}$ being the SM neutrino mass, while the observed value of $Y_B$ is $8.7\times 10^{-11}$~\cite{Planck:2018vyg}. $n_{\mathcal{N}}^{\phi} (\Trh)$ represents the number density of the RHNs at the end of reheating which will follow the same expression as Eq.~\eqref{eq:n_phi}, replacing $M_{\rm DM}$ by $M_{\mathcal{N}}$.

\begin{figure*}[!ht]
    \centering
    \begin{subfigure}{.48\textwidth}
    \includegraphics[width=\textwidth]{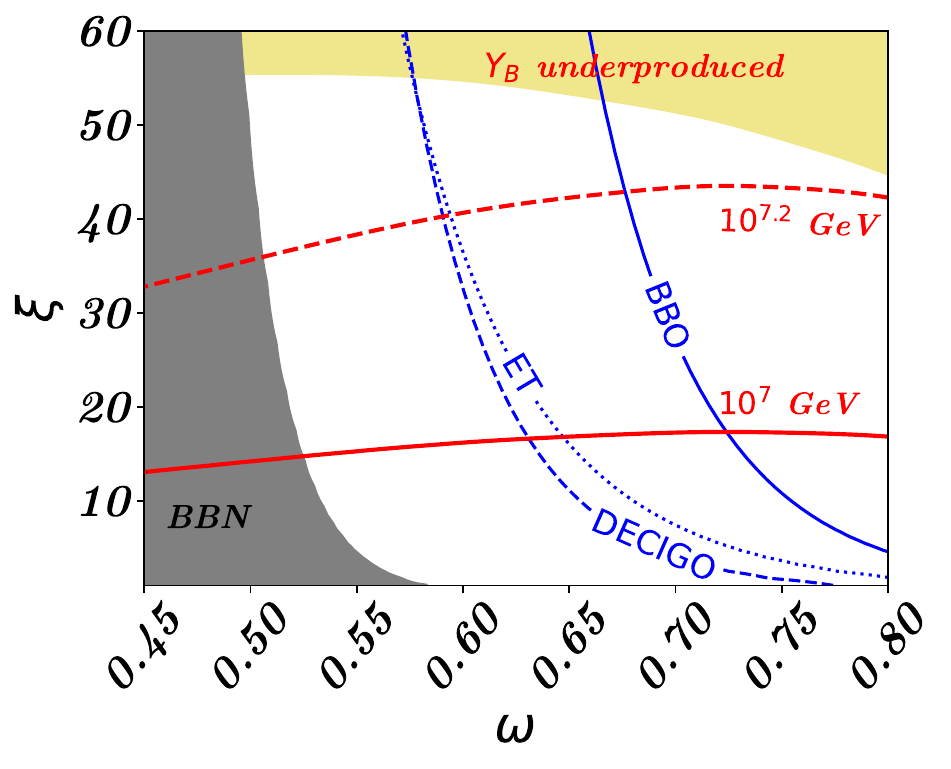}
    \caption{\it $\nt = 0$ for BBO, DECIGO and ET}
    \label{fig:snr_dm_lepto_three}
    \end{subfigure}
    \begin{subfigure}{.48\textwidth}
    \includegraphics[width=\textwidth]{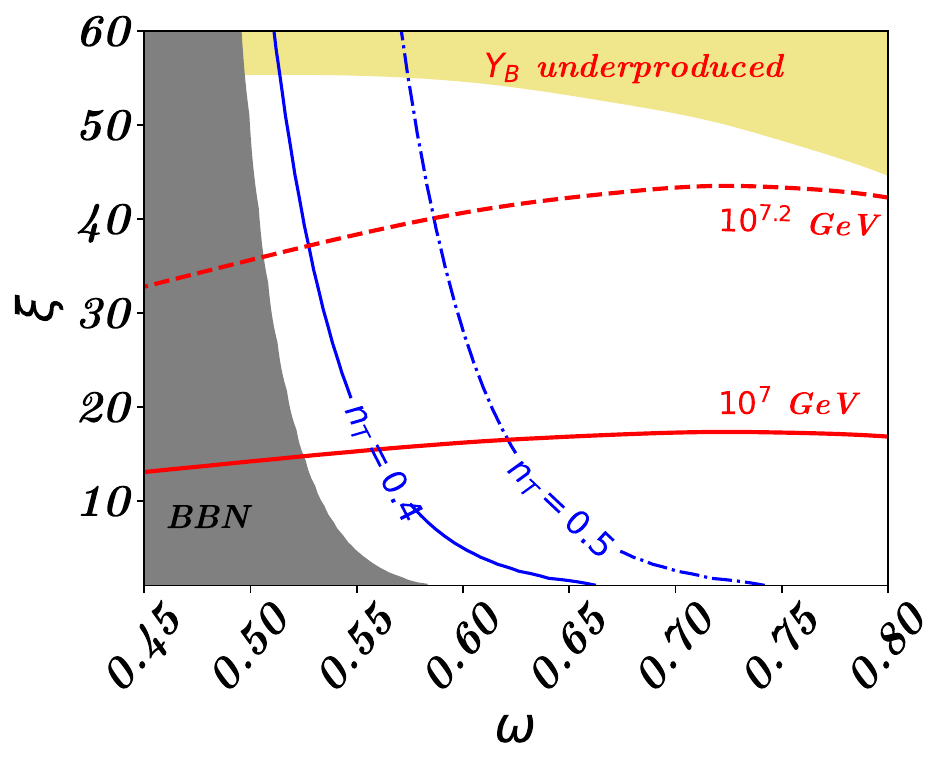}
    \caption{\it considering $\nt=0.4$ and $\nt=0.5$ for LISA}
    \label{fig:snr_dm_lepto_lisa}   
    \end{subfigure}
    \caption{\it The plots represent region which can produce sufficient amount of leptogenesis and baryogenesis for $M_{\mathcal{N}} = 8\times 10^{12}$ GeV. The gray shaded region indicates $\Trh<1$ MeV which is ruled out by BBN. In the \textbf{left plot}, we have shown the contours for SNR$=10$, for BBO (blue solid line), DECIGO (blue dashed line), ET (blue dotted line) where we have set $\nt = 0$. Right plot showcases the contours for SNR $=10$ for LISA, for $\nt=0.4$ and $\nt=0.5$. Also we have set $H_{\rm inf} = 5.5\times 10^{12}$ GeV.}
    \label{fig:snr_tr_contour}
\end{figure*}

%%%%%%%%%%%%%%%%%%%%%%%%%%%%%%
\subsection{Production of QCD axions as dark matter during gravitational reheating}
\label{subsec:grav-prod-axion}
%%%%%%%%%%%%%%%%%%%%%%%%%%%%%
This section is primarily focused on the production of axions (in presence of gravitational reheating) and their detectional aspects at the GW detectors. We have considered the production of QCD axions via standard misalignment mechanism which occurs during the epoch of reheating which is governed by gravitational interaction, as indicated in Ref.~\cite{Barman:2023icn}. We intend to demonstrate the aspect of detection of the QCD axions via GW detectors. To delve into the scenario, we have considered the reheating temperature, $\Trh$, as mentioned in Eq.~\eqref{eq:grav-trh}. Before discussing the detectional aspects of the axion, we discuss the detail of production of QCD axion via the well-known misalignment mechanism.

For any axion field ($\textsl{a}(t)$), the Lagrangian density can be expressed as~\cite{GrillidiCortona:2015jxo}
\begin{eqnarray}\label{eq:axion_lag}
    \mathcal{L}_{\ax} \supset \frac{1}{2} \partial^{\mu}\ax \partial_{\mu}\ax - m_{\ax}^2(T) f_{\ax}^2\left[1 - \cos{\left(\frac{{\ax (t)}}{f_{\ax}}\right)}\right],
\end{eqnarray}
which can be reduced to the equation of motion of the zero modes of axions as
\begin{eqnarray}\label{eq:axion_eom}
    \ddot{\theta}(t) + 3H\dot{\theta}(t) + \Tilde{m}_{\ax}^2(T) \sin{\theta(t)} = 0,
\end{eqnarray}
where $\theta(t) \equiv \frac{\ax (t)}{f_{\ax}}$ with $f_{\ax}$ being the decay constant. $\Tilde{m}_{\ax}$ is the temperature dependent mass of the axions which can be expressed as~\cite{Borsanyi:2016ksw} 
\begin{eqnarray}\label{eq:axion_mass_time}
    \Tilde{m}_{\ax} \simeq m_{\ax} \times
    \begin{cases}
        \left(\frac{T_{\rm qcd}}{T}\right)^4 & \ \text{for}\, T \ge T_{\rm qcd} \\
        \, 1 & \ \text{for}\, T \le T_{\rm qcd}.
    \end{cases}
\end{eqnarray}
Here~\cite{DiLuzio:2020wdo}
\begin{eqnarray}\label{eq:axion_mass}
    m_{\ax}\simeq 5.7\times 10^{-6} \left(\frac{10^{12} \text{GeV}}{f_{\ax}}\right)\, \text{eV},
\end{eqnarray}
is the zero temperature axion mass, with $T_{\rm qcd}$ being the QCD temperature. Now depending on the onset of the oscillation of axions ($T_{\rm osc}$), oscillation temperature can be expressed as following~\cite{Barman:2023icn}:
\begin{enumerate}[i)]
    \item if oscillation of axion starts after the reheating \textit{i.e.} $T_{\rm osc}<\Trh$,
    \begin{eqnarray}\label{eq:tosc1}
    T_{\rm osc} \simeq
    \begin{cases}
        \sqrt{\frac{1}{\pi}\sqrt{\frac{10}{g_{\star}(T_{\rm osc})}}m_{\ax}M_P T_{\rm qcd}^4} & \ \text{for}\, T_{\rm osc} \ge T_{\rm qcd} \\
        \sqrt{\frac{1}{\pi}\sqrt{\frac{10}{g_{\star}(T_{\rm osc})}}m_{\ax}M_P} & \ \text{for}\, T_{\rm osc} \le T_{\rm qcd}.
    \end{cases}
\end{eqnarray}
    \item if oscillation of axion starts before the reheating \textit{i.e.} $T_{\rm osc}>\Trh$,
    \begin{eqnarray}\label{eq:tosc2}
    T_{\rm osc} \simeq
    \begin{cases}
        \sqrt{\frac{1}{\pi}\sqrt{\frac{10}{g_{\star}(T_{\rm osc})}}\frac{m_{\ax} M_P T_{\rm qcd}^4}{\Trh^6}} & \ \text{for}\, T_{\rm osc} > T_{\rm qcd} \\
        \sqrt{\frac{1}{\pi}\sqrt{\frac{10}{g_{\star}(T_{\rm osc})}}\frac{m_{\ax} M_P}{\Trh^2}} & \ \text{for}\, T_{\rm osc} < T_{\rm qcd}.
    \end{cases}
\end{eqnarray}
\end{enumerate}
Subsequently, based on the oscillation temperature, axion energy density at present day can be evaluated as~\cite{Drees:2015exa}
\begin{eqnarray}\label{eq:rho_axion}
    \rho_{\ax}(T_0) = \rho_{\ax}(T_{\rm osc}) \frac{m_{\ax}}{\Tilde{m}_{\ax}(T_{\rm osc})} \frac{s(T_0)}{s(T_{\rm osc})} \times
    \begin{cases}
        1 & \ \text{for}\, T_{\rm osc} < \Trh \\
        \frac{S(T_{\rm osc})}{S(\Trh)} & \ \text{for}\, T_{\rm osc} > \Trh,
    \end{cases}
\end{eqnarray}
where $s(T)$ indicates the SM entropy density~\footnote{The SM entropy density can be expressed as $s(T) = \frac{2\pi^2}{45} g_{\star s}(T)\, T^3$. The present day entropy density, $s(T_0) \simeq 2.69\times10^{-3} \text{cm}^{-3}$~\cite{Planck:2018vyg}.} and the $\frac{S(T_{\rm osc})}{S(\Trh)}$ being dilution factor which can be estimated as $\frac{S(T)}{S(\Trh)} = \frac{g_{\star s}(T)}{g_{\star s}(\Trh)}$~\cite{Barman:2023icn}. The axion energy density during the oscillation temperature, $\rho_{\ax}(T_{\rm osc})$, as $\simeq \,\frac{1}{2}\, \Tilde{m}_{\ax}^2\, f_{\ax}^2\, \theta_i^2$, with $\theta_i$ being the initial misalignment angle~\cite{DiLuzio:2020wdo,kolb1990early,Drees:2015exa,Hertzberg:2008wr}. Hence the relic abundance for the axionic DM, for present day, can be expressed as 
\begin{eqnarray}
    \Omega_{\ax} h^2 \equiv \frac{\rho_{\ax} (T_0)}{\rho_c} h^2,
\end{eqnarray}
with $\rho_c/h^2 = 8.54\times 10^{-47}\, \text{GeV}^4$ being the critical density of the Universe at the present epoch.

\begin{figure*}[!ht]
    \centering
    \begin{subfigure}{.48\textwidth}
    \includegraphics[width=\textwidth]{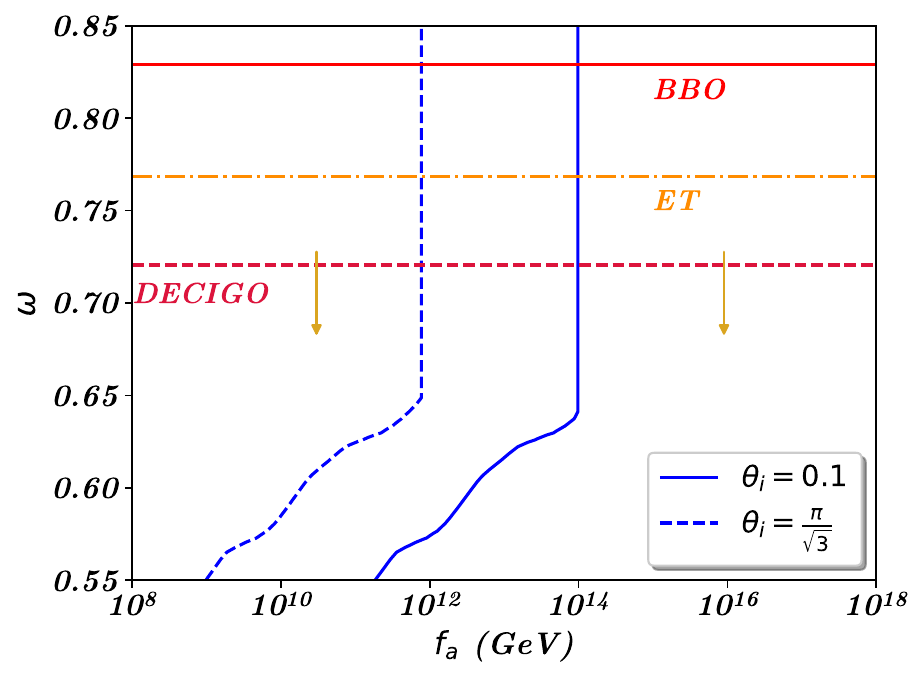}
    \caption{\it For BBO, DECIGO and ET considering $\xi = 1$ and $\nt=0$}
    \label{fig:axion_xi1_three}
    \end{subfigure}
    \hfill
    \begin{subfigure}{.48\textwidth}
    \includegraphics[width=\textwidth]{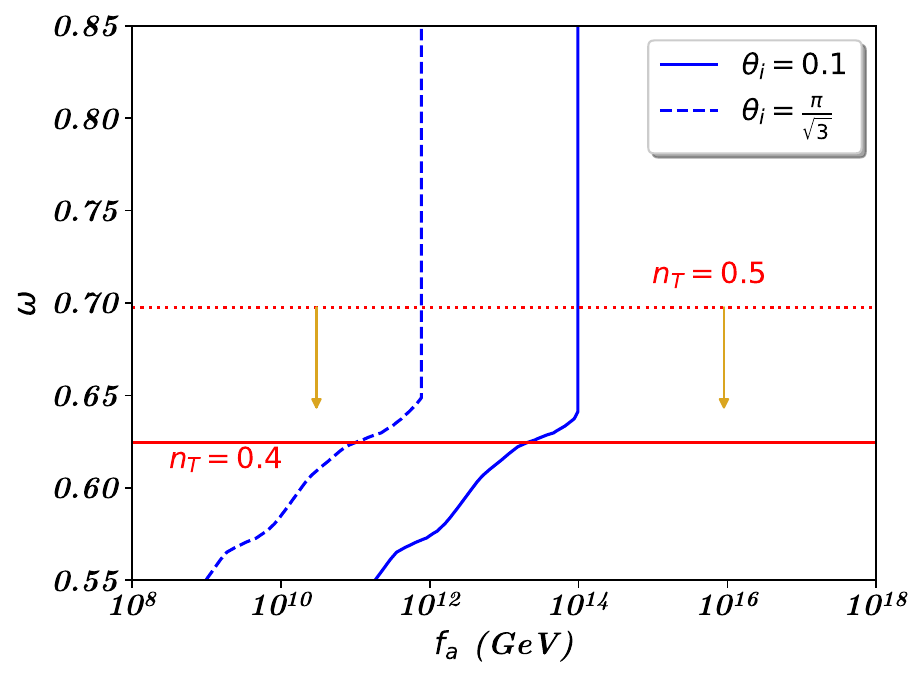}
    \caption{\it For LISA considering $\xi = 1$ for $\nt=0.4$ and $\nt=0.5$}
    \label{fig:axion_xi1_lisa}
    \end{subfigure}
    \begin{subfigure}{.48\textwidth}
    \includegraphics[width=\textwidth]{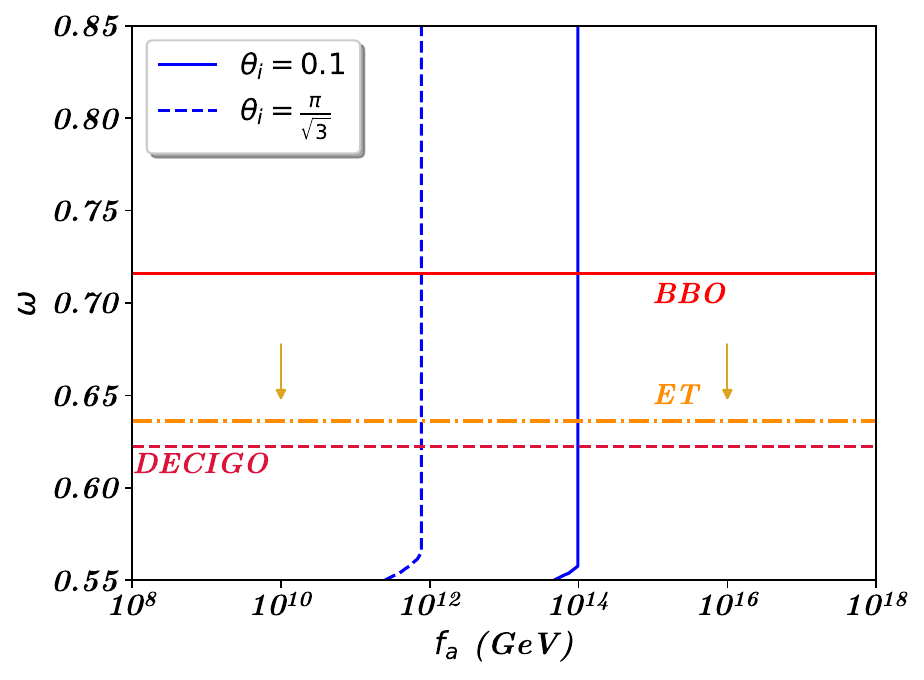}
    \caption{\it For BBO, DECIGO and ET considering $\xi = 20$ and $\nt=0$}
    \label{fig:axion_xi20_three}   
    \end{subfigure}
    \hfill
    \begin{subfigure}{.48\textwidth}
    \includegraphics[width=\textwidth]{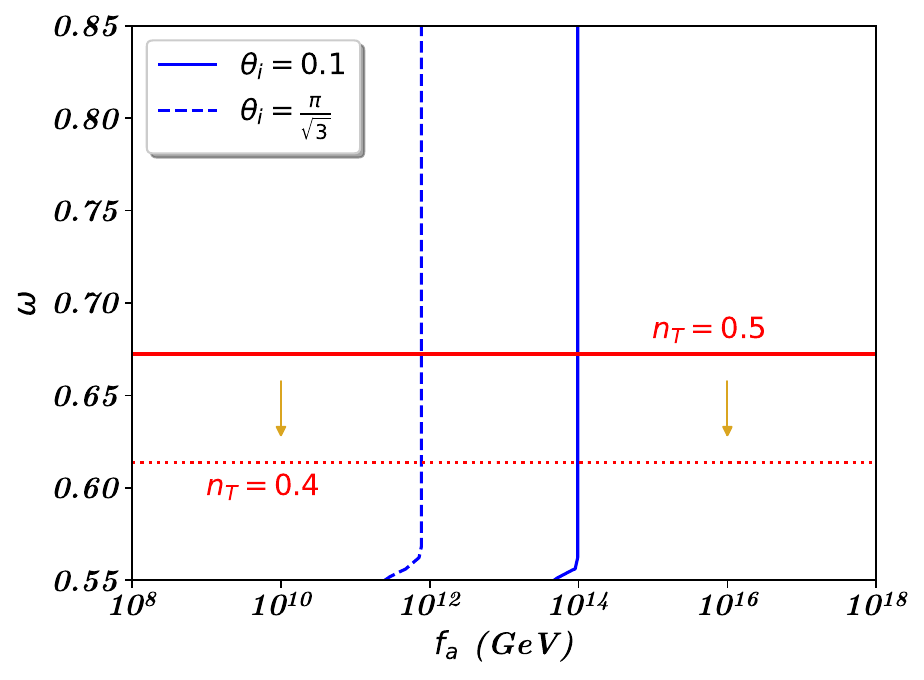}
    \caption{\it For LISA considering $\xi = 20$ for $\nt=0.4$ and $\nt=0.5$}
    \label{fig:axion_xi20_lisa}
    \end{subfigure}
    \caption{\it The figures represent the relic abundance of axions as well as its detectional aspects with respect to the GW detectors. The horizontal lines in the figures represent the SNR $=50$ contours for the mentioned detectors, where the arrows indicate SNR $>50$ regions for all the detectors. For the SNR contours we have set $\nt=0$ for BBO, DECIGO and ET, whereas $\nt=0.45$ is considered for LISA. In the \textbf{left plot} we have depicted the relic abundance for axions, considering $\xi=1$ whereas, \textbf{right plot} represents the same but $\xi=20$. For both the cases $H_{\rm int}$ is set to be $5.5\times 10^{12}$ GeV. }
    \label{fig:axion_contour}
\end{figure*}

%%%%%%%%%%%%%%%%%%%%%%%%%%%%%%
\subsection{Signal-to-noise ratio}
\label{subsec:result}
%%%%%%%%%%%%%%%%%%%%%%%%%%%%%
In Figs. \ref{fig:snr_dm_contour} and \ref{fig:snr_tr_contour}, we present the parameter space defined by $\omega$ and $\xi$, which facilitates the exploration of DM properties while simultaneously addressing the requirements for adequate baryogenesis via leptogenesis. Achieving sufficient baryon asymmetry necessitates that the baryon asymmetry ($Y_B$) attains a value of $8.7\times 10^{-11}$ \cite{Planck:2018vyg}. We denote regions in the parameter space where baryogenesis under-produces as light-orange shaded, and those where the reheating temperature ($\Trh$) falls below 1 MeV, hence being inconsistent with BBN, as gray-shaded.

The expression for $Y_B$ (Eq.~\eqref{eq:yb}) underscores its strong dependence on the mass of the right-handed neutrinos ($M_{\mathcal{N}}$), crucial for generating the requisite baryon asymmetry. In our analysis, we have fixed $M_{\mathcal{N}}=8\times 10^{12}$ GeV to expand the accessible parameter space for DM exploration. Notably, we have marked dashed contours representing DM masses ($5.0\times 10^6$ GeV, $10^7$ GeV and $1.6\times 10^7$ GeV)
consistent with the observed relic abundance. Consequently, the viable parameter space is able to probe DM candidates within the mass range of $5.0\times 10^6$ GeV to $1.6\times 10^7$ GeV while satisfying the conditions for baryogenesis.

In the above scenario we have set $\nt=0$ for BBO~\cite{Corbin:2005ny,Harry_2006}, DECIGO~\cite{Yagi:2011yu} and ET~\cite{Punturo_2010,Hild:2010id}, and $\nt=0.4$ for LISA~\cite{amaroseoane2017laser,Baker:2019nia}. Also we have set $H_{\rm inf} = 5.5\times 10^{12}$ GeV. Fig.~\ref{fig:snr_tr_contour} depicts that out of all the detectors,  BBO~\cite{Corbin:2005ny,Harry_2006} stands out as particularly promising for probing both DM properties and achieving sufficient baryon asymmetry, boasting relatively high SNR values we get. Fig.~\ref{fig:axion_contour}  depicts the parameter space in the $\omega$-$f_{\ax}$ plane, for a specific value of $\xi$, which displays the relic abundance of axions. \textit{Upper} and \textit{lower} row of Fig.~\ref{fig:axion_contour}, represent the same for $\xi = 1$ and $\xi = 20$, respectively. For $\xi = 1$, we have shown that the detectors can probe axions for $10^9\, {\rm GeV}\lesssim f_{\ax} \lesssim 10^{14}\, {\rm GeV}$ for $\theta_i\in [0.1,\pi/\sqrt{3}]$, whereas the range of $f_{\ax}$ shrinks as we increase $\xi$ (Fig.~\ref{fig:axion_xi20_three} and \ref{fig:axion_xi20_lisa}). On top of these, in the figures we have depicted the SNR$\,=50$ contours by the horizontal lines for BBO~\cite{Corbin:2005ny,Harry_2006}, DECIGO~\cite{Yagi:2011yu} and ET~\cite{Punturo_2010,Hild:2010id} for $\nt=0$, and for LISA~\cite{amaroseoane2017laser,Baker:2019nia} for $\nt=0.4$ \& $0.5$ where the portion, underneath each of the horizontal line, represents the SNR$\,>50$ region for the respective detectors (indicated by arrow in the figure). This indicates the capability of the detectors to probe the axion with high SNR values (\textit{i.e.} SNR $>10$). Out of all the detectors BBO~\cite{Corbin:2005ny,Harry_2006} comes out as a more promising detector which can probe the axion with a considerable SNR value for relatively wide range of $\omega$. Now the mass of axions depends on $f_{\ax}$ according to Eq.~\eqref{eq:axion_mass}, which depicts that GW detectors can probe a wide range of axion mass, which can contribute as DM, for lower values of $\xi$.

 \medskip

%%%%%%%%%%%%%%%%%%%%%%%%%%%%%
\section{Discussion and Conclusion}
\label{sec:conclusion}
%%%%%%%%%%%%%%%%%%%%%%%%%%%%%

Various cosmological sources of GWs such as first-order phase transitions, topological defects such as cosmic strings or domain walls, inflationary reheating and preheating, graviton bremmstrahlung \textit{etc.} are believed to produce detectable stochastic gravitational wave background signals from early Universe, the GW spectrum of which are completely different from each other cosmic sources described above. The detection of such a signal along with GW spectral shape will open up a compelling new possibility into the pre-BBN Universe. In fact it can even help probing new physics beyond the SM, as for example GUT-scale physics, high scale baryogenesis and leptogenesis physics and axion physics \cite{Dasgupta:2022isg,Bhaumik:2022pil,Barman:2022yos,Ghoshal:2022jdt,Dunsky:2021tih,Bernal:2020ywq,Ghoshal:2020vud,Chen:2024roo,Borboruah:2024eal} which are otherwise beyond the reach of LHC or any other laboratory or astrophysical searches for new physics due to either heavy scales or very weak couplings to SM that are involved. In this work, we studied gravity-mediated production of RHNs, lightest of which is stable and is the DM candidate of the universe and the heavier ones decay leading to baryogenesis via leptogenesis. Within this  cosmological framework involving such gravitational production, we find large GW signals arising from first-order inflationary tensor perturbations \footnote{In context to non-thermal DM search with GW from preheating, see \cite{Ghoshal:2022jdt}}.   
Similarly we have also discussed the production of QCD axion relic as well, in presence of gravitational reheating. Assuming a maximal amplitude of scale-invariant tensor modes produced during inflation, as allowed by the current CMBR measurements, we have studied the effect of those different parameters on the shape of the GW spectrum at present time. particularly, we have exhibited the GW spectral shapes that may be observed in LISA~\cite{amaroseoane2017laser,Baker:2019nia}, ET~\cite{Punturo_2010,Hild:2010id}, BBO~\cite{Corbin:2005ny,Harry_2006}, DECIGO~\cite{Yagi:2011yu}, U-DECIGO~\cite{Seto:2001qf,Kawamura_2006,Yagi:2011wg}.

Below are some of the salient features of our the result of our analysis:
\begin{itemize}
    \item Previous studies~\cite{Barman:2022qgt,Yeh_2022} have demonstrated that the presence of non-minimal coupling is inconsistent with BBN~\cite{Cyburt:2015mya} due to overproduction GWs, prompting the introduction of non-minimal coupling to satisfy BBN constraints. In our investigation, we have presented compelling evidence that the minimal scenario can indeed be reconciled within the framework of primordial gravitational waves (PGWs) generation without resorting to non-minimal coupling. We have illustrated that for red-tilted spectrum $\nt \leq -0.08$ the minimal scenario compiles with the BBN bound for PGWs generation, as depicted in Fig.~\ref{fig:gw_nt_n1}. At the same time we have allowed the scale of inflation ($H_{\rm inf}$) to be a free parameter. This shows that $H_{\rm inf} <\, 10^{12}$ GeV allows the minimal scenario (\textit{i.e.} $\xi = 0$), restricting the overproduction of the PGWs and shows the potential to be detected at BBO~\cite{Corbin:2005ny,Harry_2006}, DECIGO~\cite{Yagi:2011yu}, U-DECIGO~\cite{Seto:2001qf,Kawamura_2006,Yagi:2011wg} (Fig.~\ref{fig:gw_hi}). Hence, we have described that minimal scenario (\textit{i.e.} $\xi = 0$) can survive if we either consider a slightly red-tilted scenario for the GW spectrum or allow the scale of inflation to be less than $10^{12}$ GeV.
    
    \item Although the non-minimal coupling effectively curtails the overproduction of PGWs, it simultaneously dampens the PGWs for larger coupling values, thereby diminishing the chances of detectability of the signal. A blue-tilted GW spectrum not only can enhance the signal to be detected but also be detected at the detectors at relatively smaller frequency-range like LISA~\cite{amaroseoane2017laser,Baker:2019nia}, as presented in Fig.~\ref{fig:gw_nt5}.

    \item In assessing the detection capability of any detection of signal, noise in the instrument can hamper its detection. Hence SNR provides us a good assessment about the detection of the signal by any detectors. Among all of the detectors, we have depicted, as in Fig.~\ref{fig:snr}, that BBO~\cite{Corbin:2005ny,Harry_2006}
    exhibits notably a high SNR value, indicative to its heightened sensitivity for detecting the signal. For LISA~\cite{amaroseoane2017laser,Baker:2019nia}, we have observed a blue-tilted GW spectra, as evidenced by the elevated SNR values depicted in Fig.~\ref{fig:snr}. Furthermore our investigation, as delineated in Fig.\ref{fig:snr_contour}, reveals a negative correlation between the parameters $\omega$ and $\xi$ across all the detectors. This aligns with physical expectations since lower the values of $\omega$ more are the chances to shoot the BBN bound (see Fig.~\ref{fig:gwnt_fixed_xi}). Hence, in order to comply with the bounds, smaller values of $\omega$ need more dampening, implying larger value of $\xi$ is needed to be allowed within BBN limits.
    
    \item To represent the ability of the detectors for the detection of the signal, we have employed Fisher forecast analysis, which basically estimates the uncertainties on the parameters. In Fig.~\ref{fig:fisher} we have displayed region in the $\omega$-$\xi$ plane for which the relative uncertainties on the parameters are less than unity, across all the detectors. Additionally in the same figure, we have also illustrated a region, where relative uncertainties on the parameters are less than $10\%$, which is greater than the detection threshold \textit{i.e.} which we have considered SNR $>10$. We have also observed that the correlations among the uncertainties on the  parameters in Fig.~\ref{fig:fisher_triplot}. We have displayed that $\omega$-$\xi$ are in strongly negative correlation for ET~\cite{Punturo_2010,Hild:2010id} whereas it is least for LISA~\cite{amaroseoane2017laser,Baker:2019nia}, the reason for which is that at the high frequencies, detectors are more sensitive to $\omega$ and $\xi$. The Fisher analysis however only provides the relative uncertainties on the parameters.

    \item We have studied the MCMC analysis for the detectors in Fig.~\ref{fig:mcmc} which is important to understand the combined inferences on the parameters. This analysis reveals that $\omega$ are in negative correlation with both, $\xi$ and $\nt$, whereas $\xi$ and $\nt$ are in positive correlation. These correlations are physical which are mentioned in Sec.~\ref{subsec:mcmc_discussion}. Although MCMC analysis is time consuming compared to Fisher forecast analysis, Fisher forecast analysis does not take into account the variation of fiducial values of the parameters whereas MCMC analysis contains both the information about the shifting of fiducial values and their uncertainties. This study is also essential to understand the situation when the real data comes and how it will estimate the parameters. Here, the uncertainties on the parameters are unveiled as the standard deviation of the marginalized posterior distributions. 
    
    \item At the end we have explored the DM, generated via gravitational interaction, as well as the gravitational leptogenesis. To address this we have considered three RHNs, where the lightest one acts as DM, whereas the other two contribute to the leptogensis (according to Ref.~\cite{Barman:2022qgt}). Our findings indicate that the fermionic DM (RHNs in our case), produced from gravitational interaction, can be detected at the detectors at high SNR for $5.0\times 10^6$ GeV $<M_{\rm DM}<1.6\times 10^7$ GeV with a considerable amount of gravitational leptogensis, for heavy RHN mass, $M_{\mathcal{N}} = 8\times 10^{12}$ GeV, as illustrated in Fig.~\ref{fig:snr_dm_contour} and \ref{fig:snr_tr_contour}. Notably, among all of the detectors, BBO~\cite{Corbin:2005ny,Harry_2006} is more reliable as it can probe the DM with high SNR. On top this, we have extended our analysis to include considerations of the parameter space responsible for addressing the baryon asymmetry as well as gravitational leptogenesis, as presented in Fig.~\ref{fig:snr_tr_contour}. This indicates that we have depicted the parameter space which will be able to probe the fermionic DM as well as address the baryon asymmetry at a high SNR value for all the detectors and demonstrate that BBO~\cite{Corbin:2005ny,Harry_2006} is more reliable to probe. 

    \item Additionally, we have illustrated the detectability of QCD axions. Our analysis is focused on axions produced via standard misalignment mechanism, with the reheating period governed by purely gravitational effects, thereby impacting the relic abundance of the axions. Our findings demonstrate that the QCD axions can be detected via various GW detectors BBO~\cite{Corbin:2005ny,Harry_2006}, DECIGO~\cite{Yagi:2011yu}, ET~\cite{Punturo_2010,Hild:2010id} and LISA~\cite{amaroseoane2017laser,Baker:2019nia} with SNR$\,>50$. We have shown that the detectors can probe $10^9\,{\rm GeV}\lesssim f_{\ax} \lesssim 10^{14}\,{\rm GeV}$ for QCD axions for $\xi=1$. Additionally, we have depicted in Fig.~\ref{fig:axion_contour} that the range of $f_{\ax}$ shrinks as we increase $\xi$. Out of them BBO~\cite{Corbin:2005ny,Harry_2006} is more promising in detecting the signal with SNR$\,>50$ for relatively higher values of $\omega$ compared to other GW detectors. According to Eq.~\eqref{eq:axion_mass}, $f_{\ax}$ is related with axion mass which underscores the capability of the GW detectors to explore the broad range of axion masses, particularly evident for lower values of $\xi$, as depicted in Fig.~\ref{fig:axion_contour}.
\end{itemize}

The spectral features proposed in our study if observed in the GW experiments, additional observations will be needed to distinguish between gravitational DM production, axion and RHN production. It is worth noting that RHNs generate other signals that could be explored in the future, particularly GeV-TeV mass RHN may involve HNL searches~\cite{Bolton:2019pcu} or superheavy DM searches at direct detection experiments~\cite{Kavanagh:2017cru,Bramante:2018qbc,Clark:2020mna} and gravitational DM tests in laboratories \cite{Windchime:2022whs} at the laboratories thus complementing with primordial GW searches we propose here. Other indicators of the presence of RHNs, such as neutrino-less double beta decay or lepton number violating processes in labs provide various ways to independently verify their existence. Such detection channels would not only provide valuable independent confirmations of our results but also offer a unique opportunity for synergies between DM, GW searches with laboratory, astrophysical and CMB searches.

The detection of GW in LIGO~\cite{LIGOScientific:2016aoc,LIGOScientific:2016sjg,LIGOScientific:2017bnn,LIGOScientific:2017vox,LIGOScientific:2017ycc,LIGOScientific:2017vwq}, and more recently stochastic GW background in PTA has opened up new windows for studying the early universe, which is complementary to other methods for the same. Our analysis shows that inflationary GW can serve as excellent standard candles for probing the pre-BBN Universe where gravity-mediated dark matter, baryogenesis and QCD axion relic production can be tested, in upcoming GW observatories like LISA~\cite{amaroseoane2017laser,Baker:2019nia}, ET~\cite{Punturo_2010,Hild:2010id}, BBO~\cite{Corbin:2005ny,Harry_2006}, DECIGO~\cite{Yagi:2011yu}, \textit{etc.}. The current and planned GW detectors give us the opportunity to probably in near future to have a worldwide network of such GW detectors using which the future measurements of such GW spectral shapes could we may dare to imagine and explore the the early Universe down to $10^{-16}~\rm s$ after the Big Bang or even beyond which was otherwise impossible to consider in the past decades. It is remarkable that the existence features in the GW spectrum can probe a given microscopic particle physics scenario. Notably, this post-inflationary timeline may also leave imprints in the CMB spectrum itself as it affects the number of $e$-folds of inflation. Such studies are beyond the scope of the current work and will be taken up in future.

%%%%%%%%%%%%%%%%%%%%%%%%%%%

\section*{Acknowledgements}
Authors gratefully acknowledge the use of publicly available codes \texttt{CosmicFish}~\cite{Raveri:2016xof,Raveri:2016leq}, \texttt{emcee}~\cite{2013PASP..125..306F} and \texttt{GetDist}~\cite{Lewis:2019xzd} and thank the computational facilities of the Technology Innovation
Hub, ISI Kolkata. Authors specially thank  Basabendu Barman for his help and discussion related to axions. Authors also thank the anonymous referee for the valuable suggestions in the manuscript. DP thanks ISI Kolkata for financial support through Senior Research Fellowship. DP also thanks Arko Bhaumik, Purba Mukherjee, Rahul Shah and Sourav Pal for fruitful discussions and computational helps.
SP thanks the Department of Science and Technology, Govt. of India
for partial support through Grant No. NMICPS/006/MD/2020-21.

%%%%%%%%%%%%%%%%%%%%%%%%%%%%%%%%

%%%%%%%%%%%%%%%%%%%%%%%%%%%%%%%%
\appendix
\section{Discussion on non-minimal coupling and impact on reheating}\label{app:inflaton}

\subsection*{Introduction to non-minimal interaction}
As the minimal scenario ($\xi = 0$) overproduces the GW spectrum, we need to consider minimal coupling between DM and inflaton, and DM and Higgs doublets, discussed in Sec.~\ref{sec:grav-wave}. Hence considering the non-minimal coupling, we can express the whole action in Jordan frame as~\cite{Barman:2022qgt}
\begin{eqnarray}
    \mathcal{S}_J = \int d^4 x \sqrt{-g} \left[-\frac{M_P^2}{2}\,\Omega^2\, \mathcal{R} + \mathcal{L}_{\phi} + \mathcal{L}_h + \mathcal{L}_{\mathcal{N}_i}\right],
\end{eqnarray}
where
\begin{eqnarray}
    \mathcal{L}_{\phi} &=& \frac{1}{2}\partial_{\mu}\phi\, \partial^{\mu}\phi - V(\phi) \\
    \mathcal{L}_{h} &=& \partial_{\mu}h\, \partial^{\mu}h^{\dagger} - V(hh^{\dagger}) \\
    \mathcal{L}_{\mathcal{N}_i} &=& \frac{i}{2}\,\overline{\mathcal{N}_i}\,\overleftrightarrow{\slashed{\nabla}}\,\mathcal{N}_i  - \frac{1}{2}\,M_{\mathcal{N}_i}\,\overline{(\mathcal{N})^c}_i\,\mathcal{N}_i + \mathcal{L}_{\rm yuk},
\end{eqnarray}
with 
\begin{eqnarray}
    \mathcal{L}_{\rm yuk} = -y_{\mathcal{N}_i} \overline{\mathcal{N}_i}\, h^{\dagger}\,\mathbb{L}\, + \, {\rm h.c.}\,. 
\end{eqnarray}
Here $h$, $\mathcal{N}$ and $\mathbb{L}$ are the Higgs doublet, right-handed neutrino and standard model lepton doublet, respectively. Also the conformal factor is expressed as 
\begin{eqnarray}
    \Omega^2 \equiv 1 + \frac{\xi_{\phi}\phi^2}{M_P^2} + \frac{\xi_{h}|h|^2}{M_P^2},
\end{eqnarray}
where $\xi_{\phi}$ and $\xi_{h}$ are the coupling parameters between DM and inflaton, and DM and Higgs doublet, respectively. Now in our context, we are interested in small-filed limit \textit{i.e.} 
\begin{eqnarray}\label{eq:small-field_limit}
    \frac{\xi_{\phi}\phi^2}{M_P^2},\, \frac{\xi_{h}|h|^2}{M_P^2}\, \ll\, 1.
\end{eqnarray}
As the inflaton field value can be large enough during the time of inflation, $\xi_{\phi}\, \ll\, 1$, which brings strong constraint on $\xi_{\phi}$. On the other hand, there is no such constrains on $\xi_h$, which allows us to consider $\xi_h$ to be free parameter. In our whole analysis, we have used this to be free parameter and for notational simplicity, we have used $\xi_h = \xi$.

\subsection*{Decay of inflaton during the period of reheating}
During the period of reheating the Boltzmann equation for the inflaton energy density ($\rho_{\phi}$) can be expressed as 
\begin{eqnarray}
    \frac{d \rho_{\phi}}{dt} + 3\,H\,(1 + \omega)\rho_{\phi} = - (1 + \omega)\Gamma_{\phi} \rho_{\phi}.
\end{eqnarray}
Hence the decay rate of inflaton, in presence of non-minimal coupling, can be expressed as~\cite{RT_2022,Barman:2022qgt}
\begin{eqnarray}
    (1+\omega)\,\Gamma_\phi\, \rho_\phi \simeq \, \alpha_n^{\xi} M_P^5 \left(\frac{\rho_{\phi}}{M_P^4}\right)^{\frac{5n-2}{2n}} \, ,
\end{eqnarray}
where $\alpha^{\xi}_n$ is presented in a tabular form (Table-~\ref{tab:alphak}) according to the definition of Ref.~\cite{RT_2022,Barman:2022qgt}.
%%%%%%%%%%%%%%%%%%%%
\begin{table}[htb!]
\centering
\renewcommand{\arraystretch}{1.4}
\begin{minipage}{0.45\textwidth}
    \centering
    \begin{tabular}{|c|c|}
    \hline
    \hline
    $k$ & $\alpha_n^{\xi}$ \\
    \hline
    6 & $0.000193 + 0.00766 \, \xi^2$ \\
    8 & $0.000528 + 0.0205 \, \xi^2$ \\
    10 & $0.000966 + 0.0367\, \xi^2$ \\
    12 & $0.00144  +  0.0537 \, \xi^2$ \\
\hline
\hline
\end{tabular}
\end{minipage}%
\hspace{1mm}
\begin{minipage}{0.45\textwidth}
    \centering
    \begin{tabular}{|c|c|}
    \hline
    \hline
    $k$ & $\alpha_n^{\xi}$ \\
    \hline
    14 & $0.00192  +  0.0702 \, \xi^2$ \\
    16 & $0.00239  +  0.0855 \, \xi^2$ \\
    18 & $0.00282  +  0.0995 \, \xi^2$ \\
    20 & $0.00322  +  0.112 \, \xi^2$ \\
\hline
\hline
\end{tabular}
\end{minipage}
\caption{Values of $\alpha_n^\xi$ for the gravitational reheating.}
\label{tab:alphak}
\end{table}

%%%%%%%%%%%%%%%%%%%%%%%%%%%%%%%%

\bibliographystyle{bibi}
\bibliography{biblio.bib}
%%%%%%%%%%%%%%%%%%%%%%%%%%%%%%%%

\end{document}